\documentclass[11pt,english,vecarrow]{extarticle}
\usepackage[T1]{fontenc}
\usepackage[utf8]{inputenc}
\usepackage[letterpaper]{geometry}
\usepackage{babel}
\usepackage{array}
\usepackage{verbatim}
\usepackage{float}
\usepackage{booktabs}
\usepackage{calc}
\usepackage{framed}
\usepackage{amsmath}
\usepackage{amssymb}
\usepackage{cancel}
\usepackage{graphicx}
\usepackage{setspace}
\usepackage{wasysym}
\makeatletter

\providecommand{\tabularnewline}{\\}
\newcommand{\lyxdot}{.}

\numberwithin{equation}{section}
\numberwithin{figure}{section}
\numberwithin{table}{section}

\usepackage{bbm}
\usepackage{tensind}
\tensordelimiter{?}
\usepackage{multicol}
\usepackage[labelfont={sc,small},textfont=small]{caption}

\makeatother

\begin{document}
\title{Riemannian Geometry: Definitions, Pictures, and Results\\
\quad{}}
\author{\doublespacing{}\textsc{Adam Marsh}}
\date{\textit{\normalsize{}August 17, 2022}}
\maketitle
\begin{abstract}
A pedagogical but concise overview of Riemannian geometry is provided,
in the context of usage in physics. The emphasis is on defining and
visualizing concepts and relationships between them, as well as listing
common confusions, alternative notations and jargon, and relevant
facts and theorems. Special attention is given to detailed figures
and geometric viewpoints, some of which would seem to be novel to
the literature. Topics are avoided which are well covered in textbooks,
such as historical motivations, proofs and derivations, and tools
for practical calculations. As much material as possible is developed
for manifolds with connection (omitting a metric) to make clear which
aspects can be readily generalized to gauge theories. The presentation
in most cases does not assume a coordinate frame or zero torsion,
and the coordinate-free, tensor, and Cartan formalisms are developed
in parallel. 
\end{abstract}
\tableofcontents{}

\section{Introduction}

Riemannian geometry is fundamental to general relativity, and is also
the foundational inspiration for gauge theories. This bifurcation
has led to many presentations tending towards either the specific
(e.g. presented in tensor notation assuming a coordinate frame and
zero torsion) or the abstract (e.g. using the language of fiber bundles).
Here we attempt to cover the material in a way that makes clear the
relationships between different approaches and notations, while emphasizing
intuitive geometric meanings.

In the presentation we try to take an approach which is useful both
as a learning tool complementary to other resources, and as a reference
which concisely covers the relevant topics. This ends up consisting
mainly of clear definitions along with related results. We also attempt
to ``take pictures seriously,'' by making explicit the assumptions
being made and the quantities being depicted. Thus the three main
components are definitions, pictures, and results.

A series of appendices are included which cover relevant material
referred to in the presentation. These appendices can either be read
before the main presentation or referred to as necessary.

Throughout the paper, warnings concerning a common confusion or easily
misunderstood concept are separated from the core material by boxes,
as are intuitive interpretations or heuristic views that help in understanding
a particular concept. Quantities are written in \textbf{bold} when
first mentioned or defined.

\section{\label{sec:Parallel-transport}Parallel transport}

\subsection{The parallel transporter}

By definition, for a vector $w$ at a point $p$ of an $n$-dimensional
manifold $M$, \textbf{parallel transport}\index{parallel transport}
assigns a vector $\parallel_{C}\left(w\right)$ at another point $q$
that is dependent upon a specific path $C$ in $M$ from $p$ to $q$.

To see that this dependence upon the path matches our intuition, we
can consider a vector transported in what we might consider to be
a ``parallel'' fashion along the edges of an eighth of a sphere.
In this example, the sphere is embedded in $\mathbb{R}^{3}$ and the
concept of ``parallel'' corresponds to incremental vectors along
the path having a projection onto the original tangent plane that
is parallel to the original vector.

\begin{figure}[H]
\noindent \begin{centering}
\includegraphics[width=0.55\columnwidth]{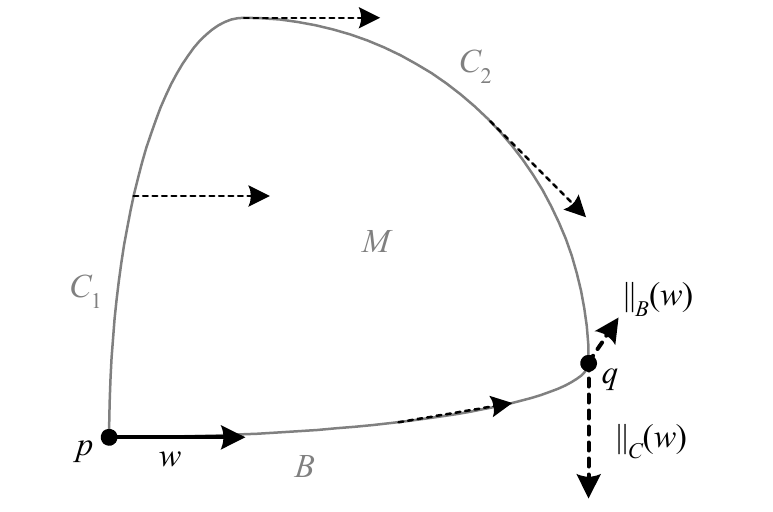}
\par\end{centering}
\caption{A vector $w$ transported in what we intuitively consider to be a
\textquotedblleft parallel\textquotedblright{} way along two different
paths ($B$ and $C=C_{1}+C_{2}$) on a surface results in two different
vectors.}
\end{figure}

The \textbf{parallel transporter} is therefore a map 
\begin{equation}
\parallel_{C}\colon T_{p}M\to T_{q}M,
\end{equation}
where $C$ is a curve in $M$ from $p$ to $q$ and $T_{p}M$ is the
tangent space at $p$ (see Section \ref{subsec:Tangent-vectors-and-differential-forms}).
To match our intuition we also require that this map be linear (i.e.
parallel transport is assumed to preserve the vector space structure
of the tangent space); that it be the identity for vanishing $C$;
that if $C=C_{1}+C_{2}$ then $\parallel_{C}=\parallel_{C_{2}}\parallel_{C_{1}}$;
and that the dependence on $C$ be smooth (this is most easily defined
in the context of fiber bundles, which we will not cover here). If
we then choose a frame on $U\subset M$, we have bases for each tangent
space that provide isomorphisms $T_{p}U\cong\mathbb{R}^{n}$, $T_{q}U\cong\mathbb{R}^{n}$.
Thus the parallel transporter can be viewed as a map 
\begin{equation}
\parallel^{\lambda}{}_{\mu}\colon\left\{ C\right\} \to GL\left(n,\mathbb{R}\right)
\end{equation}
from the set of curves on $U$ to the Lie group $GL\left(n,\mathbb{R}\right)$
of general linear transformations on $\mathbb{R}^{n}$; however, it
is important to note that the values of $\parallel^{\lambda}{}_{\mu}$
depend upon the choice of frame. 

\subsection{\label{subsec:The-covariant-derivative}The covariant derivative}

Having defined the parallel transporter, we can now consider the \textbf{covariant
derivative}

\begin{equation}
\begin{aligned}\nabla_{v}w & \equiv\underset{\varepsilon\rightarrow0}{\textrm{lim}}\frac{1}{\varepsilon}\left(w\left|_{p+\varepsilon v}\right.-\parallel_{C}\left(w\left|_{p}\right.\right)\right)\\
 & =\underset{\varepsilon\rightarrow0}{\textrm{lim}}\frac{1}{\varepsilon}\left(\parallel_{-C}\left(w\left|_{p+\varepsilon v}\right.\right)-w\left|_{p}\right.\right),
\end{aligned}
\end{equation}
where $C$ is an infinitesimal curve starting at $p$ with tangent
$v$. At a point $p$, $\nabla_{v}w$ compares the value of $w$ at
$p+\varepsilon v$ to its value at $p$ after being parallel transported
to $p+\varepsilon v$, or equivalently in the limit $\varepsilon\rightarrow0$,
the value of $w$ at $p$ to its value at $p+\varepsilon v$ after
being parallel transported back to $p$. (see Section \ref{subsec:Tangent-vectors-and-differential-forms}
on how $p+\varepsilon v$ is well-defined in the limit $\varepsilon\rightarrow0$).

\begin{figure}[H]
\noindent \begin{centering}
\includegraphics[width=0.8\columnwidth]{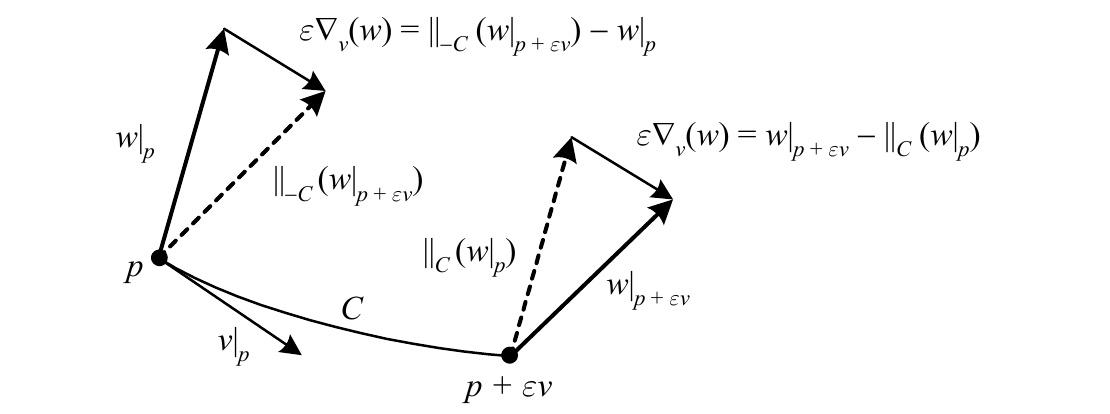}
\par\end{centering}
\caption{The covariant derivative $\nabla_{v}w$ is the difference between
a vector field $w$ and its parallel transport in the direction $v$. }
\end{figure}

\begin{framed}%
\noindent \sun{} In this and future depictions of vector derivatives,
the situation is simplified by focusing on the change in the vector
field $w$ while showing the ``transport'' of $w$ as a parallel
displacement. This has the advantage of highlighting the equivalency
of defining the derivative at either 0 or $\varepsilon$ in the limit
$\varepsilon\rightarrow0$. Depicting $\parallel_{C}\left(w\left|_{p}\right.\right)$
as a non-parallel vector at $p+\varepsilon v$ would be more accurate,
but would obscure this fact. We also will follow the picture here
in using words to characterize derivatives: namely, ``the difference''
is short for ``the difference per unit $\varepsilon$ to order $\varepsilon$
in the limit $\varepsilon\rightarrow0$.''\end{framed}Two properties of $\nabla_{v}w$ that are easy to verify are that
is is linear in $v$, and that for a function $f$ on $M$ it obeys
the rule 
\begin{equation}
\begin{aligned}\nabla_{v}\left(fw\right) & =v\left(f\right)w+f\nabla_{v}\left(w\right)\\
 & =\mathrm{d}f\left(v\right)w+f\nabla_{v}\left(w\right).
\end{aligned}
\end{equation}
As we will see in Section \ref{subsec:The-covariant-derivative-on-the-tensor-algebra},
this is the Leibniz rule (see Appendix \ref{subsec:Derivations})
for the covariant derivative generalized to the tensor algebra. See
Section \ref{subsec:The-differential-and-pullback} for a review of
the differential $\mathrm{d}$ and the relation $v(f)=\mathrm{d}f(v)$.
Note that $\nabla_{v}w$ is a directional derivative, i.e. it depends
only upon the value of $v$ at $p$; $v$ is in effect used only to
choose a direction. In contrast, the Lie derivative $L_{v}w$ (see
Section \ref{subsec:The-Lie-derivative-of-a-vector-field}) requires
$v$ to be a vector field, since $w$ is in this case compared to
its value after being ``transported'' by the local flow of $v$,
and so depends on the derivative of $v$ at $p$.

\noindent %
\begin{framed}%
\noindent $\triangle$ It is important to remember that there is no
way to ``transport'' a vector on a manifold without introducing
some extra structure. \end{framed}

Instead of parallel transport, one can consider the covariant derivative
as the fundamental structure being added to the manifold. In this
case it is useful to define the covariant derivative along a smooth
parametrized curve $C(t)$ by using the tangent to the curve as the
direction, i.e. 
\begin{equation}
\frac{\mathrm{D}}{\mathrm{d}t}w\equiv\mathrm{D}_{t}w\equiv\nabla_{\dot{C}(t)}w,
\end{equation}
where $\dot{C}(t)$ is the tangent to $C$ at $t$. $\mathrm{D}_{t}w$
is sometimes called the \textbf{absolute derivative}\index{absolute derivative}
(AKA intrinsic derivative\index{intrinsic derivative}) and its definition
only requires that $w$ be defined along the curve $C(t)$. We can
then define the parallel transport of $w\left|_{p}\right.$ along
$C(t)$ as the vector field $w$ that satisfies $\mathrm{D}_{t}w=0$. 

\noindent %
\begin{framed}%
\noindent $\triangle$ The notation for the absolute derivative is
potentially confusing since the implicitly referenced curve $C(t)$
does not appear in the expression $\mathrm{D}_{t}w$.\end{framed}

\subsection{The connection }

If we view $\nabla$ as a map from two vector fields $v$ and $w$
to a third vector field $\nabla_{v}w$, it is called an \textbf{affine
connection}\index{connection}. Note that since no use has been made
of coordinates or frames in the definition of $\nabla$, it is a frame-independent
quantity (see Appendix \ref{sec:Differentiable-manifolds} for a review
of coordinates and frames).

Since $\nabla_{v}$ is linear in $v$, and depends only on its local
value, we can regard $\nabla$ as a 1-form on $M$. If we choose a
frame $e_{\mu}$ on $M$ with corresponding dual frame $\beta^{\mu}$,
we can define the \textbf{connection 1-form}\index{connection 1-form}
\begin{equation}
\Gamma^{\lambda}{}_{\mu}\left(v\right)\equiv\beta^{\lambda}\left(\nabla_{v}e_{\mu}\right).
\end{equation}
$\Gamma^{\lambda}{}_{\mu}\left(v\right)$ is the $\lambda^{\textrm{th}}$
component of the difference between the frame $e_{\mu}$ and its parallel
transport in the direction $v$.

From its definition, it is clear that $\Gamma^{\lambda}{}_{\mu}$
is a frame-dependent object, and additionally it is not local since
it is formed from the derivative of the frame; therefore it cannot
be viewed as the components of a tensor (see Appendix \ref{sec:Tensors-and-forms}
for a review of tensors and forms).

At a point $p$, the value of $\Gamma^{\lambda}{}_{\mu}\left(v\right)$
is an infinitesimal linear transformation on $T_{p}M$, i.e. $\Gamma^{\lambda}{}_{\mu}$
is a frame-dependent 1-form whose values sit in the Lie algebra $gl\left(n,\mathbb{R}\right)$.
Using the notation for algebra- and vector-valued forms defined in
Section \ref{subsec:Algebra-valued-exterior-forms}, we can then write
\begin{equation}
\check{\Gamma}\left(v\right)\vec{w}\equiv\Gamma^{\lambda}{}_{\mu}\left(v\right)w^{\mu}e_{\lambda}=\left(\nabla_{v}e_{\mu}\right)w^{\mu},
\end{equation}
where we view $\vec{w}$ as a $\mathbb{R}^{n}$-valued 0-form. The
vector $\check{\Gamma}\left(v\right)\vec{w}$ measures the difference
between the frame and its parallel transport in the direction $v$,
weighted by the components of $w$.

\noindent %
\begin{framed}%
\noindent $\triangle$ It is important to remember that $\check{\Gamma}\left(v\right)\vec{w}$
is related to the difference between the frame and its parallel transport,
while $\nabla_{v}w$ measures the difference between $w$ and its
parallel transport; thus unlike $\nabla_{v}w$, $\check{\Gamma}\left(v\right)\vec{w}$
depends only upon the local value of $w$, but takes values that are
frame-dependent.\end{framed}%
\begin{framed}%
\noindent $\triangle$ Since we have used the frame to view $\check{\Gamma}$
as a $gl\left(n,\mathbb{R}\right)$-valued 1-form, i.e. a matrix-valued
1-form, $\vec{w}$ must be viewed as a frame-dependent column vector
of components. We could instead view $\check{\Gamma}$ as a $gl\left(\mathbb{R}^{n}\right)$-valued
1-form and $\vec{w}$ as a frame-independent intrinsic vector. In
this case the action of $\check{\Gamma}$ on $\vec{w}$ would be frame-independent,
but the value of $\check{\Gamma}$ itself would remain frame-dependent.
We choose to use matrix-valued forms due to the need below to take
the exterior derivative of component functions, but the abstract viewpoint
is important to keep in mind when generalizing to fiber bundles.\end{framed}

\subsection{The covariant derivative in terms of the connection }

$\nabla_{v}w$ can be written in terms of $\check{\Gamma}$ by using
the Leibniz rule from Section \ref{subsec:The-covariant-derivative}
with $w^{\mu}$ as frame-dependent functions:

\begin{equation}
\begin{aligned}\nabla_{v}w & =\nabla_{v}\left(w^{\mu}e_{\mu}\right)\\
 & =v\left(w^{\mu}\right)e_{\mu}+w^{\mu}\nabla_{v}\left(e_{\mu}\right)\\
 & =\mathrm{d}w^{\mu}\left(v\right)e_{\mu}+\check{\Gamma}\left(v\right)\vec{w}\\
 & \equiv\mathrm{d}\vec{w}\left(v\right)+\check{\Gamma}\left(v\right)\vec{w}
\end{aligned}
\end{equation}
Here we again view $\vec{w}$ as a $\mathbb{R}^{n}$-valued 0-form,
so that $\mathrm{d}\vec{w}\left(v\right)\equiv\mathrm{d}w^{\mu}\left(v\right)e_{\mu}$.
Thus $\mathrm{d}\vec{w}\left(v\right)$ is the change in the components
of $w$ in the direction $v$, making it frame-dependent even though
$w$ is not. Note that although $\nabla_{v}w$ is a frame-independent
quantity, both terms on the right hand side are frame-dependent. This
is depicted in the following figure.

\begin{figure}[H]
\noindent \begin{centering}
\includegraphics[width=0.75\columnwidth]{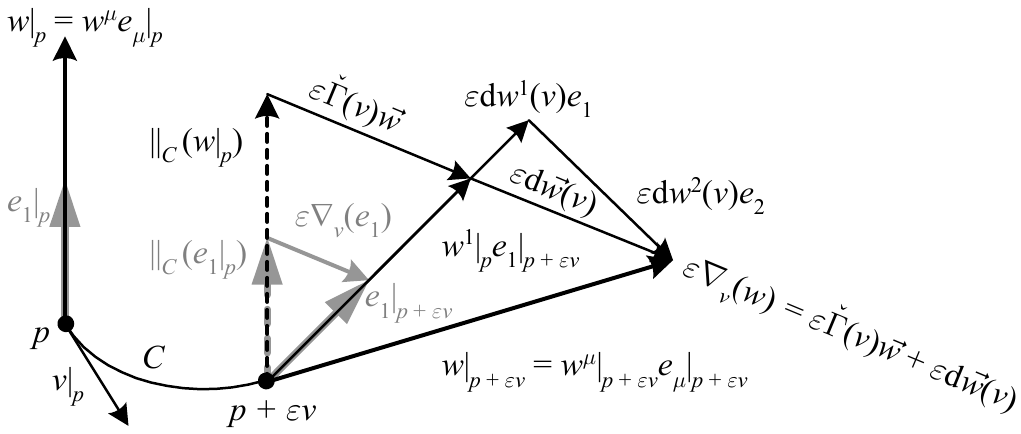}
\par\end{centering}
\caption{\label{fig:The-connection}Relationships between the frame, parallel
transport, covariant derivative, and connection for a vector $w$
parallel to $e_{1}$ at a point $p$.}
\end{figure}

\noindent %
\begin{framed}%
\noindent $\sun$ The relation $\nabla_{v}w=\check{\Gamma}\left(v\right)\vec{w}+\mathrm{d}\vec{w}\left(v\right)$
can be viewed as roughly saying that the change in $w$ under parallel
transport is equal to the change in the frame relative to its parallel
transport plus the change in the components of $w$ in that frame. \end{framed}

If the 1-form $\Gamma^{\lambda}{}_{\mu}\left(v\right)$ itself is
written using component notation, we arrive at the \textbf{connection
coefficients}\index{connection coefficients}

\begin{equation}
\Gamma^{\lambda}{}_{\mu\sigma}\equiv\Gamma^{\lambda}{}_{\mu}\left(e_{\sigma}\right)=\beta^{\lambda}\left(\nabla_{e_{\sigma}}e_{\mu}\right).
\end{equation}
$\Gamma^{\lambda}{}_{\mu\sigma}$ thus measures the $\lambda^{\mathrm{th}}$
component of the difference between $e_{\mu}$ and its parallel transport
in the direction $e_{\sigma}$.

\noindent %
\begin{framed}%
\noindent $\triangle$ This notation is potentially confusing, as
it makes $\Gamma^{\lambda}{}_{\mu\sigma}$ look like the components
of a tensor, which it is not: it is a derivative of the component
of the frame indexed by $\mu$, and therefore is not only locally
frame-dependent but also depends upon values of the frame at other
points, so that it is not a multilinear mapping on its local arguments.
Similarly, $\mathrm{d}\vec{w}$ looks like a frame-independent exterior
derivative, but it is not: it is the exterior derivative of the frame-dependent
components of $w$. \end{framed}

\noindent %
\begin{framed}%
\noindent $\triangle$ The ordering of the lower indices of $\Gamma^{\lambda}{}_{\mu\sigma}$
is not consistent across the literature (e.g. \cite{Wald} vs \cite{MTW}).
This is sometimes not remarked upon, possibly due to the fact that
in typical circumstances in general relativity (a coordinate frame
and zero torsion, to be defined in Section \ref{subsec:Torsion}),
the connection coefficients are symmetric in their lower indices.
\begin{comment}
See e.g. Wald Eq. 3.1.15 vs MTW
\end{comment}
\end{framed}

\noindent It is common to extend abstract index notation (see Section
\ref{subsec:Abstract-index-notation}) to be able to express the covariant
derivative in terms of the connection coefficients as follows:

\begin{equation}
\begin{aligned}\nabla_{e_{\mu}}w & =\mathrm{d}w^{\lambda}\left(e_{\mu}\right)e_{\lambda}+\Gamma^{\lambda}{}_{\sigma}\left(e_{\mu}\right)w^{\sigma}e_{\lambda}\\
\Rightarrow\nabla_{a}w^{b}\equiv\left(\nabla_{e_{a}}w\right)^{b} & =e_{a}\left(w^{b}\right)+\Gamma^{b}{}_{ca}w^{c}\\
\Rightarrow\nabla_{a}w^{b} & =\partial_{a}w^{b}+\Gamma^{b}{}_{ca}w^{c}
\end{aligned}
\end{equation}
Here we have also defined $\partial_{a}f\equiv\partial_{e_{a}}f=\mathrm{d}f(e_{a})=e_{a}(f)$,
which is then extended to $\partial_{v}f\equiv v^{a}\partial_{a}f$.
This notation is also sometimes supplemented to use a comma to indicate
partial differentiation and a semicolon to indicate covariant differentiation,
so that the above becomes
\begin{equation}
w^{b}{}_{;a}=w^{b}{}_{,a}+\Gamma^{b}{}_{ca}w^{c}.
\end{equation}

\noindent The extension of index notation to derivatives has several
potentially confusing aspects: 
\begin{itemize}
\item $\nabla_{a}$ and $\partial_{a}$ written alone are not 1-forms
\item Greek indices indicate only that a specific basis (frame) has been
chosen (\cite{Wald} pp. 23-26), but do not distinguish between a
general frame, where $\partial_{\mu}f\equiv\mathrm{d}f(e_{\mu})$,
and a coordinate frame, where $\partial_{\mu}f\equiv\partial f/\partial x^{\mu}$
\item $\nabla_{a}w^{b}\equiv(\nabla_{e_{a}}w)^{b}$, so since $\nabla_{v}w$
is linear in $v$, $\nabla_{a}w^{b}$ is in fact a tensor of type
$\left(1,1\right)$; a more accurate notation might be $(\nabla w)^{b}{}_{a}$
\item $w^{b}$ in the expression $\partial_{a}w^{b}\equiv\mathrm{d}w^{b}(e_{a})$
is not a vector, it is a set of frame-dependent component functions
labeled by $b$ whose change in the direction $e_{a}$ is being measured
\item The above means that, consistent with the definition of the connection
coefficients, we have $\nabla_{a}e_{b}=0+e_{c}\Gamma^{c}{}_{ba}$,
since the components of the frame itself by definition do not change
\item When using a coordinate frame based on curvilinear coordinates in
Euclidean space, parallel transport is implicit in taking partial
derivatives of vectors, resulting in the above being expressed as
$\partial_{\mu}e_{\lambda}=e_{\sigma}\Gamma^{\sigma}{}_{\lambda\mu}$
\item As previously noted, neither $\Gamma^{b}{}_{ca}$ nor $\Gamma^{b}{}_{ca}w^{c}$
are tensors 
\end{itemize}
We will nevertheless use this notation for many expressions going
forward, as it is frequently used in general relativity.

\noindent %
\begin{framed}%
\noindent $\triangle$ It is important to remember that expressions
involving $\nabla_{a}$, $\partial_{a}$, and $\Gamma^{c}{}_{ba}$
must be handled carefully, as none of these are consistent with the
original concept of indices denoting tensor components.\end{framed}%
\begin{framed}%
\noindent $\triangle$ Some texts will distinguish between the labels
of basis vectors and abstract index notation by using expressions
such as $(e_{i})^{a}$. We will not follow this practice, as it makes
difficult the convenient method of matching indexes in expressions
such as $\partial_{a}w^{b}\equiv\mathrm{d}w^{b}(e_{a})$.\end{framed}%
\begin{framed}%
\noindent $\triangle$ If we choose coordinates $x^{\mu}$ and use
a coordinate frame so that $\partial_{\mu}\equiv\partial/\partial x^{\mu}$,
we have the usual relation $\partial_{\mu}\partial_{\nu}f=\partial_{\nu}\partial_{\mu}f$.
However, this is not necessarily implied by the Greek indices alone,
which only indicate that a particular frame has been chosen. For index
notation in general, mixed partials do not commute, since $\partial_{a}\partial_{b}f-\partial_{b}\partial_{a}f=e_{a}(e_{b}(f))-e_{b}(e_{a}(f))=[e_{a},e_{b}](f)=[e_{a},e_{b}]^{c}\partial_{c}f$,
which only vanishes in a holonomic frame. \end{framed}

\subsection{The parallel transporter in terms of the connection }

We can also consider the parallel transport of a vector $w$ along
an infinitesimal curve $C$ with tangent $v$. Referring to Fig. \ref{fig:The-connection},
we see that to order $\varepsilon$ the components $w^{\mu}$ transform
according to

\begin{equation}
\parallel^{\lambda}{}_{\mu}\left(C\right)w^{\mu}=w^{\lambda}-\varepsilon\Gamma^{\lambda}{}_{\mu}\left(v\right)w^{\mu},
\end{equation}
where $v$ is tangent to the curve $C$, and these components are
with respect to the frame at the new point after infinitesimal parallel
transport. Using this relation, we can build up a frame-dependent
expression for the parallel transporter for finite $C$ by multiplying
terms $\left(1-\varepsilon\Gamma\left|_{p}\right.\right)$ where $\Gamma\left|_{p}\right.$
is used to denote the matrix $\Gamma^{\lambda}{}_{\mu}\left(v\left|_{p}\right.\right)$
evaluated on the tangent $v\left|_{p}\right.$ at successive points
$p$ along $C$. The limit of this process is the \textbf{path-ordered
exponential}\index{path-ordered exponential}

\begin{equation}
\begin{aligned}\parallel^{\lambda}{}_{\mu}\left(C\right) & =\underset{\varepsilon\rightarrow0}{\textrm{lim}}\left(1-\varepsilon\Gamma\left|_{q-\varepsilon}\right.\right)\left(1-\varepsilon\Gamma\left|_{q-2\varepsilon}\right.\right)\dotsm\left(1-\varepsilon\Gamma\left|_{p+\varepsilon}\right.\right)\left(1-\varepsilon\Gamma\left|_{p}\right.\right)\\
 & \equiv P\textrm{exp}\left(-\underset{C}{\int}\Gamma^{\lambda}{}_{\mu}\right),
\end{aligned}
\end{equation}
whose definition is based on the expression for the exponential

\begin{equation}
e^{x}=\underset{n\rightarrow\infty}{\textrm{lim}}\left(1+\frac{x}{n}\right)^{n}=\underset{\varepsilon\rightarrow0}{\textrm{lim}}\left(1+\varepsilon x\right)^{1/\varepsilon}.
\end{equation}

\noindent Note that the above expression for $\parallel^{\lambda}{}_{\mu}\left(C\right)$
exponentiates frame-dependent values in $gl\left(n,\mathbb{R}\right)$
to yield a frame-dependent value in $GL\left(n,\mathbb{R}\right)$. 

\subsection{\label{subsec:Geodesics-and-normal-coordinates}Geodesics and normal
coordinates}

Following the example of the Lie derivative (see Section \ref{subsec:The-Lie-derivative-of-a-vector-field}),
we can consider parallel transport of a vector $v$ in the direction
$v$ as generating a local flow. More precisely, for any vector $v$
at a point $p\in M$, there is a curve $\phi_{v}(t)$, unique for
some $-\varepsilon<t<\varepsilon$, such that $\phi_{v}(0)=p$ and
$\dot{\phi}_{v}\left(t\right)=\parallel_{\phi}(v)$, the last expression
indicating that the tangent to $\phi_{v}$ at $t$ is equal to the
parallel transport of $v$ along $\phi_{v}$ from $\phi_{v}(0)$ to
$\phi_{v}(t)$. This curve is called a \textbf{geodesic}\index{geodesic},
and its tangent vectors are all parallel transports of each other.
This means that for all tangent vectors $v$ to the curve, $\nabla_{v}v=0$,
so that geodesics are ``the closest thing to straight lines'' on
a manifold with parallel transport. 

Expressing a geodesic as a parametrized curve $C^{\mu}(t)$ with tangent
$v^{\mu}\left(t\right)\equiv\dot{C}^{\mu}\left(t\right)$ in given
coordinates, we can write 
\begin{equation}
\begin{aligned}\nabla_{v}v & =v^{\lambda}\left(\partial_{\lambda}v^{\mu}+\Gamma^{\mu}{}_{\sigma\lambda}v^{\sigma}\right)\\
 & =\partial_{v}\left(v^{\mu}\right)+\Gamma^{\mu}{}_{\sigma\lambda}v^{\sigma}v^{\lambda}\\
 & =\frac{\mathrm{d}}{\mathrm{d}t}\left(\frac{\mathrm{d}C^{\mu}}{\mathrm{d}t}\right)+\Gamma^{\mu}{}_{\sigma\lambda}\frac{\mathrm{d}C^{\sigma}}{\mathrm{d}t}\frac{\mathrm{d}C^{\lambda}}{\mathrm{d}t}\\
 & =\frac{\mathrm{d}^{2}C^{\mu}}{\mathrm{d}t^{2}}+\Gamma^{\mu}{}_{\sigma\lambda}\frac{\mathrm{d}C^{\sigma}}{\mathrm{d}t}\frac{\mathrm{d}C^{\lambda}}{\mathrm{d}t}=0,
\end{aligned}
\end{equation}
where the last line is called the \textbf{geodesic equation}\index{geodesic equation},
and in the third line we use the fact that the change of the 0-form
$v^{\mu}$ in the $v$ direction is equal to the derivative of the
function $v^{\mu}\left(t\right)$ with respect to $t$.

Now we can define the \index{exponential map}\textbf{exponential
map} at $p$ to be $\mathrm{exp}(v)\equiv\phi_{v}(1)$, which will
be well-defined for values of $v$ around the origin that map to some
$U\subset M$ containing $p$. Finally, choosing a basis for $T_{p}U$
provides an isomorphism $T_{p}U\cong\mathbb{R}^{n}$, allowing us
to define \textbf{geodesic normal coordinates}\index{geodesic normal coordinates}
(AKA normal coordinates) $\mathrm{exp}^{-1}\colon U\to\mathbb{R}^{n}$.
It can be shown (see \cite{Kobayashi} Vol. 1 pp148-149) that in a
coordinate frame at the origin $p$ of geodesic normal coordinates,
we have $\Gamma^{\lambda}{}_{\mu\sigma}=-\Gamma^{\lambda}{}_{\sigma\mu}$;
this implies that for zero torsion (to be defined in Section \ref{subsec:Torsion}),
the connection coefficients vanish at $p$.

\begin{figure}[H]
\begin{centering}
\includegraphics[width=0.5\columnwidth]{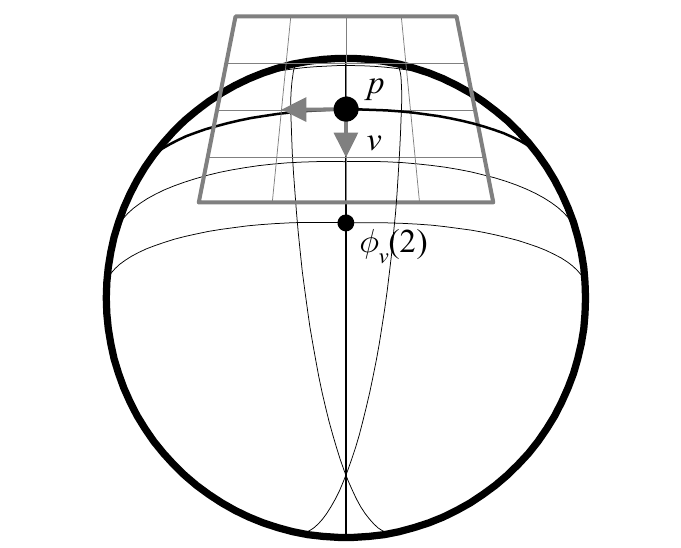}
\par\end{centering}
\caption{Geodesic normal coordinates at $p$ map points on a manifold to vectors
at $p$ tangent to the geodesic passing through both points. In the
figure $\mathrm{exp}(2v)=\phi_{v}(2)$, so the coordinate of the point
$\phi_{v}(2)\in M$ is $2v\in T_{p}M$.}
\end{figure}

\subsection{Summary }

In general, a ``manifold with connection\index{manifold with connection}''
is one with an additional structure that ``connects'' the different
tangent spaces of the manifold to one another in a linear fashion.
Specifying any one of the above connection quantities, the covariant
derivative, or the parallel transporter equivalently determines this
structure. The following tables summarize the situation.

\begin{table}[H]
\begin{tabular*}{1\columnwidth}{@{\extracolsep{\fill}}|l|>{\raggedright}p{0.22\columnwidth}|>{\raggedright}p{0.25\columnwidth}|>{\raggedright}p{0.3\columnwidth}|}
\hline 
Construct & Argument(s) & Value & Dependencies\tabularnewline
\hline 
\hline 
$\parallel_{C}$ & $v\in T_{p}M$ & $\parallel_{C}\left(v\right)\in T_{q}M$ & Path $C$ from $p$ to $q$\tabularnewline
\hline 
$\parallel^{\lambda}{}_{\mu}$ & Path $C$ & $\parallel^{\lambda}{}_{\mu}\left(C\right)\in GL$ & Frame on $M$\tabularnewline
\hline 
$\nabla_{v}$ & $w\in TM$ & $\nabla_{v}w\in T_{p}M$ & $v\in T_{p}M$\tabularnewline
\hline 
$\nabla$ & $v\in T_{p}M$, $w\in TM$ & $\nabla_{v}w\in T_{p}M$ & None\tabularnewline
\hline 
$\Gamma^{\lambda}{}_{\mu}$ & $v\in T_{p}M$ & $\Gamma^{\lambda}{}_{\mu}\left(v\right)\in gl$ & Frame on $M$\tabularnewline
\hline 
$\check{\Gamma}\left(v\right)$ & $\vec{w}\in T_{p}M$ & $\check{\Gamma}\left(v\right)\vec{w}\in T_{p}M$ & Frame on $M$, $v\in T_{p}M$\tabularnewline
\hline 
$\Gamma^{\lambda}{}_{\mu\sigma}$ & None & Connection coefficient & Frame on $M$\tabularnewline
\hline 
\end{tabular*}

\caption{Constructions related to the connection. Each construct above is considered
at a point $p$; to determine a manifold with connection it must be
defined for every point in $M$.}
\end{table}
Below we review the intuitive meanings of the various vector derivatives.

\begin{table}[H]
\begin{tabular*}{1\columnwidth}{@{\extracolsep{\fill}}|l|>{\raggedright}m{0.56\columnwidth}|}
\hline 
Vector derivative & Meaning\tabularnewline
\hline 
\hline 
$L_{v}w\equiv\underset{\varepsilon\rightarrow0}{\textrm{lim}}\left(w\left|_{p+\varepsilon v}\right.-\mathrm{d}\Phi_{\varepsilon}\left(w\left|_{p}\right.\right)\right)/\varepsilon$ & \noindent The difference between $w$ and its transport by the local
flow of $v$.\tabularnewline
\hline 
$\nabla_{v}w\equiv\underset{\varepsilon\rightarrow0}{\textrm{lim}}\left(w\left|_{p+\varepsilon v}\right.-\parallel_{C}\left(w\left|_{p}\right.\right)\right)/\varepsilon$ & \noindent The difference between $w$ and its parallel transport in
the direction $v$.\tabularnewline
\hline 
$\frac{\mathrm{D}}{\mathrm{d}t}w\equiv\mathrm{D}_{t}w\equiv\nabla_{\dot{C}(t)}w$ & \noindent The difference between $w$ and its parallel transport in
the direction tangent to $C(t)$.\tabularnewline
\hline 
$\Gamma^{\lambda}{}_{\mu}\left(v\right)\equiv\beta^{\lambda}\left(\nabla_{v}e_{\mu}\right)$ & \noindent The $\lambda^{\textrm{th}}$ component of the difference
between $e_{\mu}$ and its parallel transport in the direction $v$.\tabularnewline
\hline 
$\check{\Gamma}\left(v\right)\equiv\nabla_{v}\left(T_{p}M\right)$ & \noindent The infinitesimal linear transformation on the tangent space
that takes the parallel transported frame to the frame in the direction
$v$.\tabularnewline
\hline 
$\check{\Gamma}\left(v\right)\vec{w}\equiv\Gamma^{\lambda}{}_{\mu}\left(v\right)w^{\mu}e_{\lambda}=\left(\nabla_{v}e_{\mu}\right)w^{\mu}$ & \noindent The difference between the frame and its parallel transport
in the direction $v$, weighted by the components of $w$.\tabularnewline
\hline 
$\Gamma^{\lambda}{}_{\mu\sigma}\equiv\Gamma^{\lambda}{}_{\mu}\left(e_{\sigma}\right)=\beta^{\lambda}\left(\nabla_{\sigma}e_{\mu}\right)$ & \noindent The $\lambda^{\textrm{th}}$ component of the difference
between $e_{\mu}$ and its parallel transport in the direction $e_{\sigma}$.\tabularnewline
\hline 
$\mathrm{d}\vec{w}\left(v\right)\equiv\mathrm{d}w^{\mu}\left(v\right)e_{\mu}$ & \noindent The change in the frame-dependent components of $w$ in
the direction $v$.\tabularnewline
\hline 
$\partial_{a}w^{b}\equiv\mathrm{d}w^{b}(e_{a})$ & \noindent The change in the $b^{\mathrm{th}}$ frame-dependent component
of $w$ in the direction $e_{a}$.\tabularnewline
\hline 
$\nabla_{a}w^{b}\equiv(\nabla_{e_{a}}w)^{b}$ & \noindent The $b^{\mathrm{th}}$ component of the difference between
$w$ and its parallel transport in the direction $e_{a}$.\tabularnewline
\hline 
\end{tabular*}

\caption{Definitions and meanings of vector derivatives.}
\end{table}
Other quantities in terms of the connection:
\begin{itemize}
\item $\nabla_{v}w=\mathrm{d}\vec{w}\left(v\right)+\check{\Gamma}\left(v\right)\vec{w}$
\item $\nabla_{a}w^{b}=\partial_{a}w^{b}+\Gamma^{b}{}_{ca}w^{c}$
\item $\parallel^{\lambda}{}_{\mu}\left(C\right)w^{\mu}=w^{\mu}-\varepsilon\Gamma^{\lambda}{}_{\mu}\left(v\right)w^{\mu}$
\qquad{}(for infinitesimal $C$ with tangent $v$)
\item $\parallel^{\lambda}{}_{\mu}\left(C\right)w^{\mu}=P\textrm{exp}\left(-\int_{C}\Gamma^{\lambda}{}_{\mu}\right)w^{\mu}$
\end{itemize}

\section{\label{sec:Manifolds-with-connection}Manifolds with connection }

All of the above constructs used to define a manifold with connection
manipulate vectors, which means they can be naturally extended to
operate on arbitrary tensor fields on $M$. This is the usual approach
taken in general relativity; however, one can alternatively focus
on $k$-forms on $M$, an approach that generalizes more directly
to gauge theories in physics. This viewpoint is sometimes called the
\textbf{Cartan formalism}\index{Cartan formalism}. We will cover
both approaches. 

\noindent %
\begin{framed}%
\noindent $\triangle$ Note that a manifold with connection includes
no concept of length or distance (a metric). It is important to remember
that unless noted, nothing in this section depends upon this extra
structure.\end{framed}

\subsection{\label{subsec:The-covariant-derivative-on-the-tensor-algebra}The
covariant derivative on the tensor algebra}

If we define the covariant derivative of a function to coincide with
the normal derivative, i.e. $\nabla_{a}f\equiv\partial_{a}f$, then
we can use the Leibniz rule to define the covariant derivative of
a 1-form. This is sometimes described as making the covariant derivative
``commute with contractions,'' where for a 1-form $\varphi$ and
a vector $v$ we require
\begin{equation}
\begin{aligned}\nabla_{a}\left(\varphi_{b}v^{b}\right) & \equiv\left(\nabla_{a}\varphi_{b}\right)v^{b}+\varphi_{b}\left(\nabla_{a}v^{b}\right)\\
 & =\left(\nabla_{a}\varphi_{b}\right)v^{b}+\varphi_{b}\left(\partial_{a}v^{b}+\Gamma^{b}{}_{ca}v^{c}\right).
\end{aligned}
\end{equation}
At the same time, choosing a frame and treating $\varphi_{b}$ and
$v^{b}$ as frame-dependent functions on $M$, we have
\begin{equation}
\begin{aligned}\nabla_{a}\left(\varphi_{b}v^{b}\right) & \equiv\partial_{a}\left(\varphi_{b}v^{b}\right)\\
 & =\left(\partial_{a}\varphi_{b}\right)v^{b}+\varphi_{b}\left(\partial_{a}v^{b}\right),
\end{aligned}
\end{equation}
so that equating the two we arrive at
\begin{equation}
\nabla_{a}\varphi_{b}\equiv\partial_{a}\varphi_{b}-\Gamma^{c}{}_{ba}\varphi_{c}.
\end{equation}
As with vectors, the partial derivative $\partial_{a}\varphi_{b}$
acts upon the frame-dependent components of the 1-form. 

We can then extend the covariant derivative to be a derivation on
the tensor algebra (see Section \ref{subsec:Derivations}) by following
the above logic for each covariant and contravariant component:

\begin{equation}
\begin{aligned}\nabla_{a}T^{b_{1}\ldots b_{m}}{}_{c_{1}\ldots c_{n}} & \equiv\partial_{a}T^{b_{1}\ldots b_{m}}{}_{c_{1}\ldots c_{n}}\\
 & \phantom{{}=}+\underset{j=1}{\overset{m}{\sum}}\Gamma^{b_{j}}{}_{da}T^{b_{1}\ldots b_{j-1}db_{j+1}\ldots b_{m}}{}_{c_{1}\ldots c_{n}}\\
 & \phantom{{}=}-\underset{j=1}{\overset{n}{\sum}}\Gamma^{d}{}_{c_{j}a}T^{b_{1}\ldots b_{m}}{}_{c_{1}\ldots c_{j-1}dc_{j+1}\ldots c_{n}}
\end{aligned}
\end{equation}
Note that since the covariant derivative of a 0-form is $\nabla_{a}f=\partial_{a}f=\partial_{e_{a}}f=e_{a}(f)$,
we then have $\nabla_{v}f=v^{a}\nabla_{a}f=v^{a}e_{a}(f)=v(f)$. 

The concept of parallel transport along a curve $C$ can be extended
to the tensor algebra as well, by parallel transporting all vector
arguments backwards to the starting point of $C$, applying the tensor,
then parallel transporting the resulting vectors forward to the endpoint
of $C$. So for example the parallel transport of a tensor $T^{a}{}_{b}$
is defined as

\begin{equation}
\begin{aligned}\parallel_{C}\left(T^{a}{}_{b}\right) & \equiv?\parallel^{a}{}_{c}?\left(C\right)?T^{c}{}_{d}??\parallel^{d}{}_{b}?\left(-C\right)\\
 & =\left(1-\varepsilon?\Gamma^{a}{}_{c}?\left(v\right)\right)?T^{c}{}_{d}?\left(1+\varepsilon?\Gamma^{d}{}_{b}?\left(v\right)\right),
\end{aligned}
\end{equation}
where for infinitesimal $C$ with tangent $v$ we have $\parallel_{C}^{-1}=\parallel_{-C}=1+\varepsilon\check{\Gamma}\left(v\right)$
since $\parallel_{C}=1-\varepsilon\check{\Gamma}\left(v\right)$. 

\begin{figure}[H]
\noindent \begin{centering}
\includegraphics[width=0.62\columnwidth]{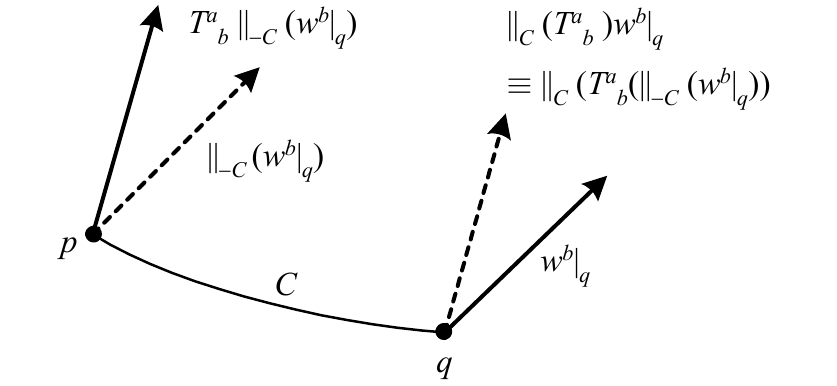}
\par\end{centering}
\caption{The parallel transport of a tensor can be defined by parallel transporting
all vector arguments backwards to the starting point, applying the
tensor, then parallel transporting the resulting vectors forward to
the endpoint.}
\end{figure}
With this definition, the covariant derivative $\nabla_{a}T$ can
be viewed as ``the difference between $T$ and its parallel transport
in the direction $e_{a}$.''

\noindent %
\begin{framed}%
\noindent $\triangle$ It can sometimes be confusing when using the
extended covariant derivative as to what type of tensor it is being
applied to. For example, $w^{b}$ in the expression $\partial_{a}w^{b}$
is not a vector, it is a set of frame-dependent functions labeled
by $b$; yet this expression can in theory also be written $\nabla_{a}w^{b}$,
in which case there is no indication that the covariant derivative
is acting on these functions instead of the vector $w^{b}$.\end{framed}

\noindent %
\begin{framed}%
\noindent $\triangle$ When the covariant derivative is used as a
derivation on the tensor algebra, care must be taken with relations,
since their forms can change considerably based upon what arguments
are applied and whether index notation is used. In particular, $(\nabla_{a}\nabla_{b}-\nabla_{b}\nabla_{a})f=\nabla_{a}(\partial_{b}f)-\nabla_{b}(\partial_{a}f)$
is not a ``mixed partials'' expression, since $(\partial_{a}f)$
is a 1-form. And as we will see, $(\nabla_{a}\nabla_{b}-\nabla_{b}\nabla_{a})f$
is a different construction than $(\nabla_{a}\nabla_{b}-\nabla_{b}\nabla_{a})w^{c}$,
which is different from $(\nabla_{u}\nabla_{v}-\nabla_{v}\nabla_{u})w$.
It is important to realize that an expression such as $\nabla_{a}\nabla_{b}-\nabla_{b}\nabla_{a}$
without context has no unambiguous meaning.\end{framed}

\noindent %
\begin{framed}%
\noindent $\triangle$ It is important to remember that since expressions
like $\partial_{a}w^{b}$ and $\Gamma^{c}{}_{ba}$ are not tensors,
applying $\nabla_{d}$ to them is not well-defined (unless we consider
them as arrays of functions and are applying $\nabla_{d}=\partial_{d}$).\end{framed}

\subsection{\label{subsec:The-exterior-covariant-derivative-of-vector-valued-forms}The
exterior covariant derivative of vector-valued forms}

A vector field $w$ on $M$ can be viewed as a vector-valued 0-form.
As noted previously, the covariant derivative $\nabla_{v}w$ is linear
in $v$ and depends only on its local value, and so can be viewed
as a vector-valued 1-form $\mathrm{D}\vec{w}(v)\equiv\nabla_{v}w$.
$\mathrm{D}\vec{w}$ is called the \textbf{exterior covariant derivative}\index{exterior covariant derivative}
of the vector-valued 0-form $\vec{w}$. This definition is then extended
to vector-valued $k$-forms $\vec{\varphi}$ by following the example
of the exterior derivative $\mathrm{d}$ (see Section \ref{subsec:The-exterior-derivative-of-a-k-form}):

\begin{equation}
\begin{aligned} & \mathrm{D}\vec{\varphi}\left(v_{0},\dotsc,v_{k}\right)\\
 & \equiv\underset{j=0}{\overset{k}{\sum}}\left(-1\right)^{j}\nabla_{v_{j}}\left(\vec{\varphi}\left(v_{0},\dotsc,v_{j-1},v_{j+1},\dotsc,v_{k}\right)\right)\\
 & \phantom{{}=}+\underset{i<j}{\sum}\left(-1\right)^{i+j}\vec{\varphi}\left(\left[v_{i},v_{j}\right],v_{0},\dotsc,v_{i-1},v_{i+1},\dotsc,v_{j-1},v_{j+1},\dotsc,v_{k}\right)
\end{aligned}
\end{equation}
For example, if $\vec{\varphi}$ is a vector-valued 1-form, 
\begin{equation}
\mathrm{D}\vec{\varphi}\left(v,w\right)\equiv\nabla_{v}\vec{\varphi}\left(w\right)-\nabla_{w}\vec{\varphi}\left(v\right)-\vec{\varphi}\left(\left[v,w\right]\right).
\end{equation}
So while the first term of $\mathrm{d}\varphi$ takes the difference
between the scalar values of $\varphi(w)$ along $v$, the first term
of $\mathrm{D}\vec{\varphi}$ takes the difference between the vector
values of $\vec{\varphi}(w)$ along $v$ after parallel transporting
them to the same point (which is required to compare them). At a point
$p$, $\mathrm{D}\vec{\varphi}\left(v,w\right)$ can thus be viewed
as the “sum of $\vec{\varphi}$ on the boundary of the surface defined
by its arguments after being parallel transported back to $p$,” and
if we use $\Vert_{\varepsilon v}$ to denote parallel transport along
an infinitesimal curve with tangent $v$, we can write 
\begin{equation}
\begin{aligned}\varepsilon^{2}\mathrm{D}\vec{\varphi}\left(v,w\right) & =\Vert_{-\varepsilon v}\vec{\varphi}\left(\varepsilon w\left|_{p+\varepsilon v}\right.\right)-\vec{\varphi}\left(\varepsilon w\left|_{p}\right.\right)\\
 & -\Vert_{-\varepsilon w}\vec{\varphi}\left(\varepsilon v\left|_{p+\varepsilon w}\right.\right)+\vec{\varphi}\left(\varepsilon v\left|_{p}\right.\right)\\
 & -\vec{\varphi}\left(\varepsilon^{2}\left[v,w\right]\right).
\end{aligned}
\end{equation}

\noindent 
\begin{figure}[H]
\begin{centering}
\includegraphics[width=0.7\columnwidth]{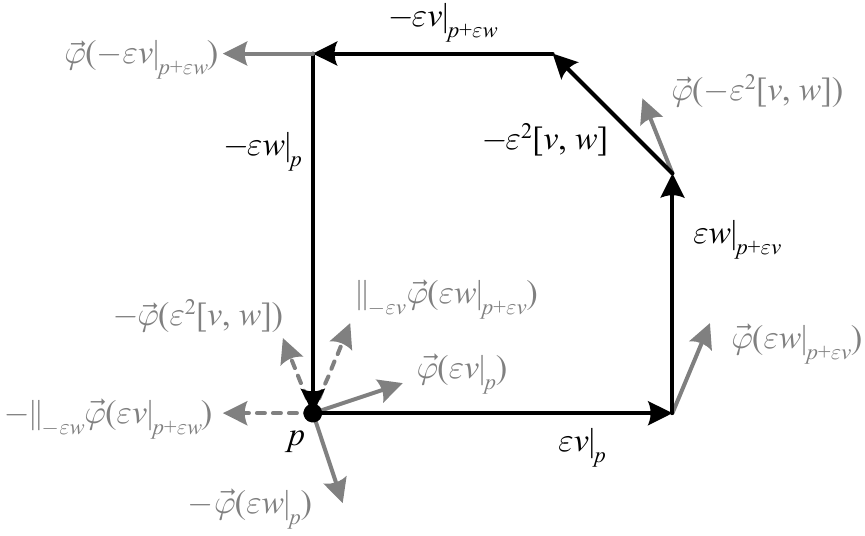}
\par\end{centering}
\caption{The exterior covariant derivative $\mathrm{D}\vec{\varphi}\left(v,w\right)$
sums the vectors $\vec{\varphi}$ along the boundary of the surface
defined by $v$ and $w$ by parallel transporting them to the same
point. Note that the \textquotedblleft completion of the parallelogram\textquotedblright{}
$[v,w]$ is already of order $\varepsilon^{2}$, so its parallel transport
has no effect to this order. }
\end{figure}

From its definition, it is clear that $\mathrm{D}\vec{\varphi}$ is
a frame-independent quantity. In terms of the connection, we must
consider $\vec{w}$ as a frame-dependent $\mathbb{R}^{n}$-valued
0-form, so that

\begin{equation}
\mathrm{D}\vec{w}\left(v\right)=\nabla_{v}w=\mathrm{d}\vec{w}\left(v\right)+\check{\Gamma}\left(v\right)\vec{w}.
\end{equation}
For a $\mathbb{R}^{n}$-valued $k$-form $\vec{\varphi}$ we find
that

\begin{equation}
\mathrm{D}\vec{\varphi}=\mathrm{d}\vec{\varphi}+\check{\Gamma}\wedge\vec{\varphi},
\end{equation}
where the exterior derivative is defined to apply to the frame-dependent
components, i.e. $\mathrm{d}\vec{\varphi}(v_{0}\ldots v_{k})\equiv\mathrm{d}\varphi^{\mu}(v_{0}\ldots v_{k})e_{\mu}$.
Recall that $\check{\Gamma}$ is a $gl(n,\mathbb{R})$-valued 1-form,
so that for example if $\vec{\varphi}$ is a $\mathbb{R}^{n}$-valued
1-form then 
\begin{equation}
\begin{aligned}(\check{\Gamma}\wedge\vec{\varphi})\left(v,w\right) & \equiv\check{\Gamma}\left(v\right)\vec{\varphi}\left(w\right)-\check{\Gamma}\left(w\right)\vec{\varphi}\left(v\right)\\
 & =?\Gamma^{\lambda}{}_{\mu}?\left(v\right)\varphi^{\mu}\left(w\right)-?\Gamma^{\lambda}{}_{\mu}?\left(w\right)\varphi^{\mu}\left(v\right).
\end{aligned}
\end{equation}

\noindent %
\begin{framed}%
\noindent $\triangle$ As with the covariant derivative, it is important
to remember that $\mathrm{D}\vec{\varphi}$ is frame-independent while
$\mathrm{d}\vec{\varphi}$ and $\check{\Gamma}$ are not. \end{framed}

The set of vector-valued forms can be viewed as an infinite-dimensional
algebra by defining multiplication via the vector field commutator;
it turns out that $\mathrm{D}$ does not satisfy the Leibniz rule
in this algebra and so is not a derivation (see Appendix \ref{subsec:Derivations}).
However, following the above reasoning one can extend the definition
of $\mathrm{D}$ to the algebra of tensor-valued forms, or the subset
of anti-symmetric tensor-valued forms; $\mathrm{D}$ then is a derivation
with respect to the tensor product in the former case and a graded
derivation with respect to the exterior product in the latter case.
We will not pursue either of these two generalizations.

\subsection{The exterior covariant derivative of algebra-valued forms }

Recalling from Section \ref{subsec:The-covariant-derivative-on-the-tensor-algebra}
the definition of parallel transport of a tensor, we can view a $gl(n,\mathbb{R})$-valued
0-form $\check{\Theta}$ as a tensor of type $\left(1,1\right)$,
so that the infinitesimal parallel transport of $\check{\Theta}$
along $C$ with tangent $v$ is
\begin{equation}
\parallel_{C}(\check{\Theta})=\left(1-\varepsilon\check{\Gamma}\left(v\right)\right)\check{\Theta}\left(1+\varepsilon\check{\Gamma}\left(v\right)\right).
\end{equation}
We can now follow the reasoning used to define the covariant derivative
of a vector in terms of the connection

\begin{equation}
\begin{aligned}\nabla_{v}w & \equiv\underset{\varepsilon\rightarrow0}{\textrm{lim}}\frac{1}{\varepsilon}\left(w\left|_{p+\varepsilon v}\right.-\parallel_{C}w\left|_{p}\right.\right)\\
 & =\underset{\varepsilon\rightarrow0}{\textrm{lim}}\frac{1}{\varepsilon}\left(\vec{w}\left|_{p+\varepsilon v}\right.-\left(1-\varepsilon\check{\Gamma}\left(v\right)\right)\vec{w}\left|_{p}\right.\right)\\
 & =\underset{\varepsilon\rightarrow0}{\textrm{lim}}\frac{1}{\varepsilon}\left(w^{\mu}\left|_{p+\varepsilon v}\right.-w^{\mu}\left|_{p}\right.+\varepsilon?\Gamma^{\mu}{}_{\lambda}?\left(v\right)w^{\lambda}\left|_{p}\right.\right)e_{\mu}\left|_{p+\varepsilon v}\right.\\
 & =\mathrm{d}\vec{w}\left(v\right)+\check{\Gamma}\left(v\right)\vec{w}
\end{aligned}
\end{equation}
to give the covariant derivative of a $gl(n,\mathbb{R})$-valued 0-form 

\begin{equation}
\begin{aligned}\nabla_{v}\check{\Theta} & \equiv\underset{\varepsilon\rightarrow0}{\textrm{lim}}\frac{1}{\varepsilon}\left(\check{\Theta}\left|_{p+\varepsilon v}\right.-\parallel_{C}\left(\check{\Theta}\left|_{p}\right.\right)\right)\\
 & =\underset{\varepsilon\rightarrow0}{\textrm{lim}}\frac{1}{\varepsilon}\left(\check{\Theta}\left|_{p+\varepsilon v}\right.-\left(1-\varepsilon\check{\Gamma}\left(v\right)\right)\check{\Theta}\left|_{p}\right.\left(1+\varepsilon\check{\Gamma}\left(v\right)\right)\right)\\
 & =\mathrm{d}\check{\Theta}\left(v\right)+\check{\Gamma}\left(v\right)\check{\Theta}-\check{\Theta}\check{\Gamma}\left(v\right)\\
 & =\mathrm{d}\check{\Theta}\left(v\right)+\left[\check{\Gamma},\check{\Theta}\right]\left(v\right)\\
 & =\mathrm{d}\check{\Theta}\left(v\right)+\left(\check{\Gamma}[\wedge]\check{\Theta}\right)\left(v\right).
\end{aligned}
\end{equation}
Here we have only kept terms to order $\varepsilon$, followed previous
convention to define $\mathrm{d}\check{\Theta}\left(v\right)\equiv\mathrm{d}?\Theta^{\mu}{}_{\lambda}?\beta^{\lambda}e_{\mu}$,
and defined the Lie commutator $[\check{\Gamma},\check{\Theta}]$
in terms of the multiplication of the $gl(n,\mathbb{R})$-valued forms
$\check{\Gamma}$ and $\check{\Theta}$, which (see Section \ref{subsec:Algebra-valued-exterior-forms}
for notation) as a 1-form is equivalent to $\check{\Gamma}[\wedge]\check{\Theta}$.
$\nabla_{v}\check{\Theta}$ is then ``the difference between the
linear transformation $\check{\Theta}$ and its parallel transport
in the direction $v$.''

The above definition of the covariant derivative can then be extended
to arbitrary $gl(n,\mathbb{R})$-valued $k$-forms by defining

\begin{equation}
\mathrm{D}\check{\Theta}\equiv\mathrm{d}\check{\Theta}+\check{\Gamma}[\wedge]\check{\Theta},
\end{equation}
which can be shown to be equivalent to the construction used for $\mathbb{R}^{n}$-valued
$k$-forms in Section \ref{subsec:The-exterior-covariant-derivative-of-vector-valued-forms}.
For example, for a $gl(n,\mathbb{R})$-valued 1-form $\check{\Theta}$,
we have 
\begin{equation}
\mathrm{D}\check{\Theta}\left(v,w\right)\equiv\nabla_{v}\check{\Theta}\left(w\right)-\nabla_{w}\check{\Theta}\left(v\right)-\check{\Theta}\left(\left[v,w\right]\right),
\end{equation}
with the covariant derivatives acting on the value of $\check{\Theta}$
as a tensor of type $\left(1,1\right)$. So at a point $p$, $\mathrm{D}\check{\Theta}\left(v,w\right)$
can be viewed as the “sum of $\check{\Theta}$ on the boundary of
the surface defined by its arguments after being parallel transported
back to $p$.” With respect to the set of $gl(n,\mathbb{R})$-valued
forms under the exterior product using the Lie commutator $[\wedge]$,
$\mathrm{D}$ is a graded derivation and for a $gl(n,\mathbb{R})$-valued
$k$-form $\check{\Theta}$ satisfies the Leibniz rule 
\begin{equation}
\mathrm{D}(\check{\Theta}[\wedge]\check{\Psi})=\mathrm{D}\check{\Theta}[\wedge]\check{\Psi}+\left(-1\right)^{k}\check{\Theta}[\wedge]\mathrm{D}\check{\Psi}.
\end{equation}

\subsection{\label{subsec:Torsion}Torsion }

Given a frame $e_{\mu}$, we can view the dual frame $\beta^{\mu}$
as a vector-valued 1-form that simply returns its vector argument:
\begin{equation}
\vec{\beta}\left(v\right)\equiv\beta^{\mu}\left(v\right)e_{\mu}=v.
\end{equation}
Clearly this is a frame-independent object. The \textbf{torsion}\index{torsion}
is then defined to be the exterior covariant derivative 
\begin{equation}
\vec{T}\equiv\mathrm{D}\vec{\beta}.
\end{equation}
In terms of the connection, we must consider $\vec{\beta}$ as a frame-dependent
$\mathbb{R}^{n}$-valued 1-form, which gives us the torsion as a $\mathbb{R}^{n}$-valued
2-form 
\begin{equation}
\vec{T}=\mathrm{d}\vec{\beta}+\check{\Gamma}\wedge\vec{\beta}.
\end{equation}
This definition of $\vec{T}$ is sometimes called \textbf{Cartan's
first structure equation}\index{Cartan's first structure equation}.

In terms of the covariant derivative, the torsion 2-form is

\begin{equation}
\begin{aligned}\vec{T}\left(v,w\right) & \equiv\nabla_{v}\left(\vec{\beta}\left(w\right)\right)-\nabla_{w}\left(\vec{\beta}\left(v\right)\right)-\vec{\beta}\left(\left[v,w\right]\right)\\
 & =\nabla_{v}w-\nabla_{w}v-\left[v,w\right].
\end{aligned}
\end{equation}
For a torsion-free connection\index{torsion-free connection} in a
holonomic frame, we then have $\nabla_{\sigma}e_{\mu}=\nabla_{\mu}e_{\sigma}$,
which means that the connection coefficients are symmetric in their
lower indices, i.e. 
\begin{equation}
?\Gamma^{\lambda}{}_{\mu\sigma}?\equiv\beta^{\lambda}\left(\nabla_{\sigma}e_{\mu}\right)=\beta^{\lambda}\left(\nabla_{\mu}e_{\sigma}\right)=?\Gamma^{\lambda}{}_{\sigma\mu}?.
\end{equation}
For this reason, a torsion-free connection is also called a \textbf{symmetric
connection}\index{symmetric connection}.

From the definition in terms of the exterior covariant derivative,
we can view the torsion as the “sum of the boundary vectors of the
surface defined by its arguments after being parallel transported
back to $p$,” i.e. the torsion measures the amount by which the boundary
of a loop fails to close after being parallel transported. From the
definition in terms of the covariant derivative, we arrive in the
figure below at another interpretation where, like the Lie derivative
$L_{v}w$ (see Section \ref{subsec:The-Lie-derivative-of-a-vector-field}),
$\vec{T}(v,w)$ ``completes the parallelogram'' formed by its vector
arguments, but this parallelogram is formed by parallel transport
instead of local flow. Note however that the torsion vector has the
opposite sign as the Lie derivative. 

\begin{figure}[H]
\noindent \begin{centering}
\includegraphics[width=0.55\columnwidth]{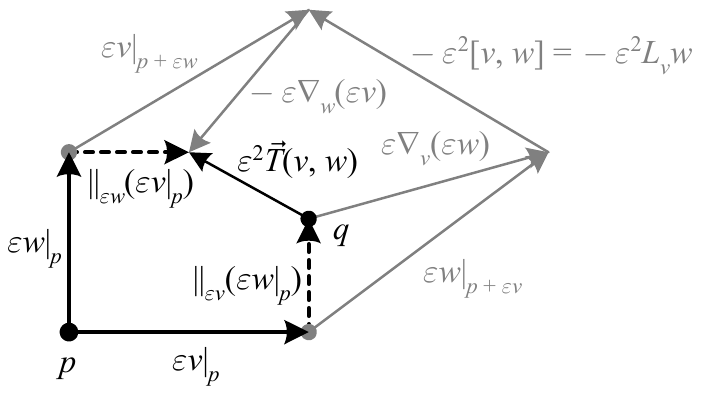}
\par\end{centering}
\caption{The torsion vector $\vec{T}\left(v,w\right)$, constructed above starting
at the point $q$, \textquotedblleft completes the parallelogram\textquotedblright{}
formed by parallel transport. $\Vert_{\varepsilon v}$ denotes parallel
transport along an infinitesimal curve with tangent $v$.}
\end{figure}

Zero torsion then means that moving infinitesimally along $v$ followed
by the parallel transport of $w$ is the same as moving infinitesimally
along $w$ followed by the parallel transport of $v$. Non-zero torsion
signifies that ``a loop made of parallel transported vectors is not
closed.'' 

As this geometric interpretation suggests, and as is evident from
the expression $\vec{T}\equiv\mathrm{D}\vec{\beta}$, one can verify
algebraically that despite being defined in terms of derivatives $\vec{T}(v,w)$
in fact only depends on the local values of $v$ and $w$, and thus
can be viewed as a tensor of type $\left(1,2\right)$:

\begin{equation}
?T^{c}{}_{ab}?v^{a}w^{b}\equiv v^{a}\nabla_{a}w^{c}-w^{a}\nabla_{a}v^{c}-[v,w]^{c}
\end{equation}
Another relation can be obtained for the torsion tensor by applying
its vector value to a function $f$ before moving into index notation:

\begin{equation}
\begin{aligned}\vec{T}\left(v,w\right)(f) & \equiv\left(\nabla_{v}w\right)(f)-\left(\nabla_{w}v\right)(f)-\left[v,w\right](f)\\
\Rightarrow?T^{c}{}_{ab}?v^{a}w^{b}\nabla_{c}f & =\left(v^{a}\nabla_{a}w^{b}\right)\nabla_{b}f-\left(w^{b}\nabla_{b}v^{a}\right)\nabla_{a}f\\
 & \phantom{{}=}-\left[v^{a}\nabla_{a}\left(w^{b}\nabla_{b}f\right)-w^{b}\nabla_{b}\left(v^{a}\nabla_{a}f\right)\right]\\
\Rightarrow?T^{c}{}_{ab}?\nabla_{c}f & =\nabla_{b}\nabla_{a}f-\nabla_{a}\nabla_{b}f
\end{aligned}
\end{equation}
Here we have used the Leibniz rule and recalled that $v(f)=\nabla_{v}f=v^{a}\nabla_{a}f$
and $[v,w](f)=v(w(f))-w(v(f))$ (see Section \ref{subsec:Tangent-vectors-and-differential-forms}).
In terms of the connection coefficients $\Gamma^{c}{}_{ab}=\beta^{c}\nabla_{b}e_{a}$
we have

\begin{equation}
\begin{aligned}?T^{c}{}_{ab}? & =\beta^{c}\vec{T}\left(e_{a},e_{b}\right)\\
 & =\beta^{c}\nabla_{a}e_{b}-\beta^{c}\nabla_{b}e_{a}-\beta^{c}[e_{a},e_{b}]\\
 & =\Gamma^{c}{}_{ba}-\Gamma^{c}{}_{ab}-[e_{a},e_{b}]^{c}.
\end{aligned}
\end{equation}

\noindent %
\begin{framed}%
\noindent $\triangle$ Note that zero torsion thus always means that
$\nabla_{a}\nabla_{b}f=\nabla_{b}\nabla_{a}f$ (and $[v,w]=L_{v}w=\nabla_{v}w-\nabla_{w}v$),
but it only means $?\Gamma^{\lambda}{}_{\mu\sigma}?=?\Gamma^{\lambda}{}_{\sigma\mu}?$
in a holonomic frame.\end{framed}

In the above figure, the failure of the parallel transported vectors
to meet can be viewed as either due to their lengths changing or due
to their being rotated out of the plane of the figure. As we will
see, the latter interpretation is more relevant for Riemannian manifolds,
where parallel transport leaves lengths invariant. In Einstein-Cartan
theory in physics, non-zero torsion is associated with spin in matter.
A suggestive example along these lines that highlights the rotation
aspect of torsion is Euclidean $\mathbb{R}^{3}$ with parallel transport
defined by translation, except in the $x$ direction where parallel
transport rotates a vector clockwise by an angle proportional to the
distance transported. As we will see in the next section, this parallel
transport has torsion but no curvature.

\begin{figure}[H]
\begin{centering}
\includegraphics[width=0.5\columnwidth]{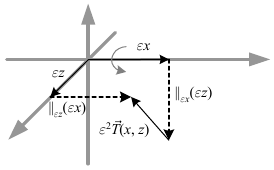}
\par\end{centering}
\caption{An example of non-zero torsion suggestive of spin.}
\end{figure}

The zero torsion expression $[v,w]=\nabla_{v}w-\nabla_{w}v$ means
that we can replace partial with covariant derivatives in the usual
expression for the Lie derivative of a vector field: 
\begin{equation}
\begin{aligned}\left(L_{v}w\right)^{a} & =\left[v,w\right]^{a}\\
 & =v^{b}\partial_{b}w^{a}-w^{b}\partial_{b}v^{a}\\
 & \overset{\cancel{T}}{=}v^{b}\nabla_{b}w^{a}-w^{b}\nabla_{b}v^{a}
\end{aligned}
\end{equation}
This can be extended to the Lie derivative of a general tensor, so
that in the case of zero torsion we have
\begin{equation}
\begin{aligned}L_{v}?T^{a_{1}\dots a_{m}}{}_{b_{1}\dots b_{n}}? & =v^{c}\nabla_{c}?T^{a_{1}\dots a_{m}}{}_{b_{1}\dots b_{n}}?\\
 & -\sum_{j=1}^{m}\left(\nabla_{c}v^{a_{j}}\right)?T^{a_{1}\dots a_{j-1}ca_{j+1}\dots a_{m}}{}_{b_{1}\dots b_{n}}?\\
 & +\sum_{j=1}^{n}\left(\nabla_{b_{j}}v^{c}\right)?T^{a_{1}\dots a_{m}}{}_{b_{1}\dots b_{j-1}cb_{j+1}\dots b_{n}}?.
\end{aligned}
\label{eq:Lie covariant}
\end{equation}

\subsection{\label{subsec:Curvature}Curvature }

The exterior covariant derivative $\mathrm{D}$ parallel transports
its values on the boundary before summing them, and therefore we do
not expect it to mimic the property $\mathrm{d}^{2}=0$ (see Section
\ref{subsec:The-exterior-derivative-of-a-1-form}). Indeed it does
not; instead, for a vector field $w$ viewed as a vector-valued 0-form
$\vec{w}$, we have

\begin{equation}
\left(\mathrm{D}^{2}\vec{w}\right)(u,v)\equiv\check{R}\left(u,v\right)\vec{w}=\nabla_{u}\nabla_{v}w-\nabla_{v}\nabla_{u}w-\nabla_{\left[u,v\right]}w,
\end{equation}
which defines the \textbf{curvature 2-form}\index{curvature 2-form}
$\check{R}$, which is $gl(\mathbb{R}^{n})$-valued. From its definition,
$\check{R}\vec{w}$ is a frame-independent quantity, and thus if $\vec{w}$
is considered as a vector-valued 0-form, $\check{R}$ is frame-independent
as well. In the (more common) case that we view $\vec{w}$ as a frame-dependent
$\mathbb{R}^{n}$-valued 0-form, $\check{R}$ must be considered to
be $gl(n,\mathbb{R})$-valued, and is thus a frame-dependent matrix.
A connection with zero curvature is called \textbf{flat}\index{connection!flat}\index{flat connection},
as is any region of $M$ with a flat connection.

For a general $\mathbb{R}^{n}$-valued form $\vec{\varphi}$ it is
not hard to arrive at an expression for $\check{R}$ in terms of the
connection: 

\begin{equation}
\mathrm{D}^{2}\vec{\varphi}=\left(\mathrm{d}\check{\Gamma}+\check{\Gamma}\wedge\check{\Gamma}\right)\wedge\vec{\varphi}\equiv\check{R}\wedge\vec{\varphi}
\end{equation}
Note that $\mathrm{D}\check{\Gamma}=\mathrm{d}\check{\Gamma}+\check{\Gamma}[\wedge]\check{\Gamma}$
is a similar but distinct construction, since e.g. 
\begin{equation}
(\check{\Gamma}\wedge\check{\Gamma})\left(v,w\right)=\check{\Gamma}\left(v\right)\check{\Gamma}\left(w\right)-\check{\Gamma}\left(w\right)\check{\Gamma}\left(v\right),
\end{equation}
while 
\begin{equation}
\begin{aligned}(\check{\Gamma}[\wedge]\check{\Gamma})\left(v,w\right) & =[\check{\Gamma}\left(v\right),\check{\Gamma}\left(w\right)]-[\check{\Gamma}\left(w\right),\check{\Gamma}\left(v\right)]\\
 & =2(\check{\Gamma}\wedge\check{\Gamma})\left(v,w\right).
\end{aligned}
\end{equation}
Thus we have 
\begin{equation}
\begin{aligned}\check{R} & \equiv\mathrm{d}\check{\Gamma}+\check{\Gamma}\wedge\check{\Gamma}\\
 & =\mathrm{d}\check{\Gamma}+\frac{1}{2}\check{\Gamma}[\wedge]\check{\Gamma}.
\end{aligned}
\end{equation}
The definition of $\check{R}$ in terms of $\check{\Gamma}$ is sometimes
called \textbf{Cartan's second structure equation}\index{Cartan's second structure equation}.
An immediate property from the definition of $\check{R}$ is 
\begin{equation}
\check{R}(u,v)=-\check{R}(v,u),
\end{equation}
which allows us to write e.g. for a vector-valued 1-form $\vec{\varphi}$
\begin{equation}
\begin{aligned}\left(\mathrm{D}^{2}\vec{\varphi}\right)(u,v,w) & \equiv\left(\check{R}\wedge\vec{\varphi}\right)\left(u,v,w\right)\\
 & =\check{R}\left(u,v\right)\vec{\varphi}(w)+\check{R}\left(v,w\right)\vec{\varphi}(u)+\check{R}\left(w,u\right)\vec{\varphi}(v).
\end{aligned}
\end{equation}

Constructing the same picture as can be done for the double exterior
derivative (see Section \ref{subsec:The-exterior-derivative-of-a-1-form}),
we put 
\[
\mathrm{D}^{2}\vec{w}\equiv\mathrm{D}\vec{\varphi},
\]
where 
\[
\vec{\varphi}(v)\equiv\mathrm{D}\vec{w}(v)=\nabla_{v}w.
\]
Expanding both derivatives in terms of parallel transport, we find
in the following figure that as we sum values around the boundary
of the surface defined by its arguments, $\mathrm{D}^{2}$ fails to
cancel the endpoint and starting point at the far corner. Examining
the values of these non-canceling points, we can view the curvature
as ``the difference between $w$ when parallel transported around
the two opposite edges of the boundary of the surface defined by its
arguments.'' 

\begin{figure}[H]
\begin{centering}
\includegraphics[width=0.7\columnwidth]{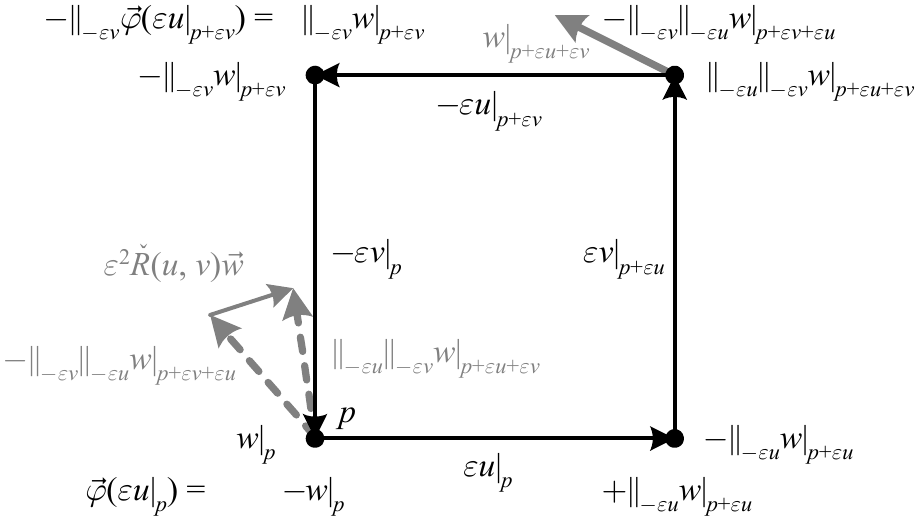}
\par\end{centering}
\caption{$\check{R}\left(u,v\right)\vec{w}=\left(\mathrm{D}^{2}\vec{w}\right)(u,v)$
is \textquotedblleft the difference between $w$ when parallel transported
around the two opposite edges of the boundary of the surface defined
by its arguments.\textquotedblright{} In the figure we assume vanishing
Lie bracket for simplicity, so that $v\left|_{p+\varepsilon u+\varepsilon v}\right.=v\left|_{p+\varepsilon v+\varepsilon u}\right.$. }
\end{figure}

In terms of the connection, we can use the path integral formulation
to examine the parallel transporter around the closed path $L\equiv\partial S$
defined by the surface $S\equiv\left(\varepsilon u\wedge\varepsilon v\right)$
to order $\varepsilon^{2}$. This calculation after some work (see
\cite{GockelerSchucker} pp. 51-53) yields

\begin{equation}
\begin{aligned}\parallel_{L}(w) & =P\textrm{exp}\left(-\int_{L}\check{\Gamma}\right)\vec{w}\\
 & =w-\int_{S}\left(\mathrm{d}\check{\Gamma}+\check{\Gamma}\wedge\check{\Gamma}\right)\vec{w}\\
 & =w-\varepsilon^{2}\check{R}\left(u,v\right)\vec{w},
\end{aligned}
\end{equation}
where we have dropped the indices since $L$ is a closed path and
thus $\parallel_{L}$ is basis-independent. Thus the curvature can
be viewed as ``the difference between $w$ and its parallel transport
around the boundary of the surface defined by its arguments.'' 

\begin{figure}[H]
\noindent \begin{centering}
\includegraphics[width=0.72\columnwidth]{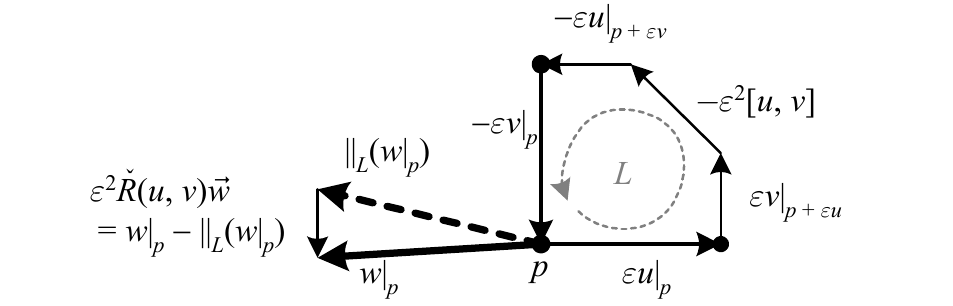}
\par\end{centering}
\caption{$\check{R}\left(u,v\right)\vec{w}$ is \textquotedblleft the difference
between $w$ and its parallel transport around the boundary of the
surface defined by its arguments.\textquotedblright}
\end{figure}

As this picture suggests, one can verify algebraically that the value
of $\check{R}\left(u,v\right)\vec{w}$ at a point $p$ only depends
upon the value of $w$ at $p$, even though it can be defined in terms
of $\nabla w$, which depends upon nearby values of $w$. Similarly,
$\check{R}\left(u,v\right)\vec{w}$ at a point $p$ only depends upon
the values of $u$ and $v$ at $p$, even though it can be defined
in terms of $[u,v]$, which depends upon their vector field values
(note that $\nabla_{u}\nabla_{v}w$ depends upon the vector field
values of both $v$ and $w$). Finally, $\check{R}$ (as a $gl(\mathbb{R}^{n})$-valued
2-form) is frame-independent, even though it can be defined in terms
of $\check{\Gamma}$, which is not. Thus the curvature can be viewed
as a tensor of type $\left(1,3\right)$, called the \textbf{Riemann
curvature tensor}\index{Riemann curvature tensor} (AKA Riemann tensor,
curvature tensor\index{curvature tensor}, Riemann–Christoffel tensor):

\begin{equation}
\begin{aligned}?R^{c}{}_{dab}?u^{a}v^{b}w^{d} & \equiv u^{a}\nabla_{a}\left(v^{b}\nabla_{b}w^{c}\right)-v^{b}\nabla_{b}\left(u^{a}\nabla_{a}w^{c}\right)-[u,v]^{d}\nabla_{d}w^{c}\\
 & =u^{a}v^{b}\nabla_{a}\nabla_{b}w^{c}-u^{a}v^{b}\nabla_{b}\nabla_{a}w^{c}+?T^{d}{}_{ab}?u^{a}v^{b}\nabla_{d}w^{c}\\
\Rightarrow?R^{c}{}_{dab}?w^{d} & =\left(\nabla_{a}\nabla_{b}-\nabla_{b}\nabla_{a}+?T^{d}{}_{ab}?\nabla_{d}\right)w^{c}
\end{aligned}
\end{equation}
Here we have used the Leibniz rule and recalled that $[u,v]^{d}=u^{a}\nabla_{a}v^{d}-v^{b}\nabla_{b}u^{d}-?T^{d}{}_{ab}?u^{a}v^{b}$. 

To obtain an expression in terms of the connection coefficients, we
first examine the double covariant derivative, recalling that $\nabla_{b}w^{c}$
is a tensor:
\begin{equation}
\begin{aligned}\nabla_{a}\left(\nabla_{b}w^{c}\right) & =\partial_{a}\nabla_{b}w^{c}+\Gamma^{c}{}_{fa}\nabla_{b}w^{f}-\Gamma^{f}{}_{ba}\nabla_{f}w^{c}\\
 & =\partial_{a}\partial_{b}w^{c}+\partial_{a}(\Gamma^{c}{}_{fb}w^{f})\\
 & \phantom{{}=}+\Gamma^{c}{}_{fa}\partial_{b}w^{f}+\Gamma^{c}{}_{fa}\Gamma^{f}{}_{gb}w^{g}-\Gamma^{f}{}_{ba}\nabla_{f}w^{c}\\
 & =\partial_{a}\partial_{b}w^{c}+\partial_{a}\Gamma^{c}{}_{fb}w^{f}\\
 & \phantom{{}=}+\Gamma^{c}{}_{fb}\partial_{a}w^{f}+\Gamma^{c}{}_{fa}\partial_{b}w^{f}\\
 & \phantom{{}=}+\Gamma^{c}{}_{fa}\Gamma^{f}{}_{gb}w^{g}-\Gamma^{f}{}_{ba}\nabla_{f}w^{c}.
\end{aligned}
\end{equation}
When we subtract the same expression with $a$ and $b$ reversed,
we recognize that for the functions $w^{c}$ we have $\partial_{a}\partial_{b}w^{c}-\partial_{b}\partial_{a}w^{c}=[e_{a},e_{b}]^{d}\partial_{d}w^{c}$,
that the second line $\Gamma^{c}{}_{fb}\partial_{a}w^{f}+\Gamma^{c}{}_{fa}\partial_{b}w^{f}$
vanishes, and that $\Gamma^{f}{}_{ba}-\Gamma^{f}{}_{ab}=[e_{a},e_{b}]^{f}+?T^{f}{}_{ab}?$,
so that
\begin{equation}
\begin{aligned}\left(\nabla_{a}\nabla_{b}-\nabla_{b}\nabla_{a}\right)w^{c} & =[e_{a},e_{b}]^{d}\partial_{d}w^{c}+\partial_{a}\Gamma^{c}{}_{fb}w^{f}-\partial_{b}\Gamma^{c}{}_{fa}w^{f}\\
 & \phantom{{}=}+\Gamma^{c}{}_{fa}\Gamma^{f}{}_{gb}w^{g}-\Gamma^{c}{}_{fb}\Gamma^{f}{}_{ga}w^{g}\\
 & \phantom{{}=}-\left([e_{a},e_{b}]^{f}+?T^{f}{}_{ab}?\right)\nabla_{f}w^{c},
\end{aligned}
\end{equation}
and thus relabeling dummy indices to obtain an expression in terms
of $w^{d}$, we arrive at

\begin{equation}
\begin{aligned}?R^{c}{}_{dab}?w^{d} & =\left(\nabla_{a}\nabla_{b}-\nabla_{b}\nabla_{a}+?T^{d}{}_{ab}?\nabla_{d}\right)w^{c}\\
 & =\left(\partial_{a}\Gamma^{c}{}_{db}-\partial_{b}\Gamma^{c}{}_{da}+\Gamma^{c}{}_{fa}\Gamma^{f}{}_{db}-\Gamma^{c}{}_{fb}\Gamma^{f}{}_{da}-[e_{a},e_{b}]^{f}\Gamma^{c}{}_{df}\right)w^{d}.
\end{aligned}
\end{equation}
This expression follows much more directly from the expression $\check{R}\equiv\mathrm{d}\check{\Gamma}+\check{\Gamma}\wedge\check{\Gamma}$,
but the above derivation from the covariant derivative expression
is included here to clarify other presentations which are sometimes
obscured by the quirks of index notation for covariant derivatives. 

\noindent %
\begin{framed}%
\noindent $\triangle$ The derivation above makes clear how the expression
for the curvature in terms of the covariant derivative simplifies
to $?R^{c}{}_{dab}?w^{d}=\left(\nabla_{a}\nabla_{b}-\nabla_{b}\nabla_{a}\right)w^{c}$
for zero torsion but is unchanged in a holonomic frame, while in contrast
the expression in terms of the connection coefficients is unchanged
for zero torsion but in a holonomic frame simplifies to omit the term
$[e_{a},e_{b}]^{f}\Gamma^{c}{}_{df}w^{d}$. \end{framed}%
\begin{framed}%
\noindent $\triangle$ Note that the sign and the order of indices
of $R$ as a tensor are not at all consistent across the literature.\end{framed}

\subsection{First Bianchi identity }

If we take the exterior covariant derivative of the torsion, we get

\begin{equation}
\mathrm{D}\vec{T}=\mathrm{DD}\vec{\beta}=\check{R}\wedge\vec{\beta}.
\end{equation}
This is called the \textbf{first} (AKA algebraic) \textbf{Bianchi
identity}\index{first Bianchi identity}\index{Bianchi identity!first}.
Using the antisymmetry of $\check{R}$, we can write the first Bianchi
identity explicitly as

\begin{equation}
\mathrm{D}\vec{T}(u,v,w)=\check{R}(u,v)\vec{w}+\check{R}(v,w)\vec{u}+\check{R}(w,u)\vec{v}.
\end{equation}
In the case of zero torsion, this identity becomes $\check{R}\wedge\vec{\beta}=0$,
which in index notation can be written $?R^{c}{}_{[dab]}?=0$. 

We can find a geometric interpretation for this identity by first
constructing a variant of our picture of $\check{R}(u,v)\vec{w}$
as the change in $\vec{w}$ after being parallel transported in opposite
directions around a loop. Taking advantage of our previous result
that $\check{R}(u,v)\vec{w}$ only depends upon the local values of
$u$ and $v$, we are free to construct their vector field values
such that $[u,v]=0$. We then examine the difference between $\vec{w}$
being parallel transported in each direction halfway around the loop.
For infinitesimal parallel transport from a point $p$ along a curve
$C$ with tangent $v$ we have $\parallel_{\varepsilon v}(w\left|_{p}\right.)\equiv\parallel_{C}(w\left|_{p}\right.)=w\left|_{p+\varepsilon v}\right.-\varepsilon\nabla_{v}w\left|_{p}\right.$.
Therefore we find that 
\begin{equation}
\begin{aligned}\parallel_{\varepsilon u}\parallel_{\varepsilon v}(w\left|_{p}\right.) & =\parallel_{u}\left(w\left|_{p+\varepsilon v}\right.-\varepsilon\nabla_{v}w\left|_{p}\right.\right)\\
 & =w\left|_{p+\varepsilon v+\varepsilon u}\right.-\varepsilon\nabla_{v}w\left|_{p+\varepsilon u}-\varepsilon\nabla_{u}w\left|_{p+\varepsilon v}\right.+\varepsilon^{2}\nabla_{u}\nabla_{v}w\left|_{p},\right.\right.
\end{aligned}
\end{equation}
so that

\begin{equation}
\begin{aligned}\parallel_{\varepsilon u}\parallel_{\varepsilon v}(w\left|_{p}\right.)-\parallel_{\varepsilon v}\parallel_{\varepsilon u}(w\left|_{p}\right.) & =\varepsilon^{2}\nabla_{u}\nabla_{v}w\left|_{p}\right.-\varepsilon^{2}\nabla_{v}\nabla_{u}w\left|_{p}\right.\\
 & =\varepsilon^{2}\check{R}(u,v)\vec{w},
\end{aligned}
\end{equation}
since $[u,v]=0$ means that $w\left|_{p+\varepsilon v+\varepsilon u}\right.=w\left|_{p+\varepsilon u+\varepsilon v}\right.$.
In the case of zero torsion, we can further take advantage of our
freedom in choosing the vector field values of $u$ and $v$ by requiring
them to equal their parallel transports, i.e. $v\left|_{p+\varepsilon u}\right.\equiv\parallel_{\varepsilon u}(v\left|_{p}\right.)$
and $u\left|_{p+\varepsilon v}\right.\equiv\parallel_{\varepsilon v}(u\left|_{p}\right.)$,
preserving the property $[u,v]=0$ due to the vanishing torsion. 

\begin{figure}[H]
\noindent \begin{centering}
\includegraphics[width=0.66\columnwidth]{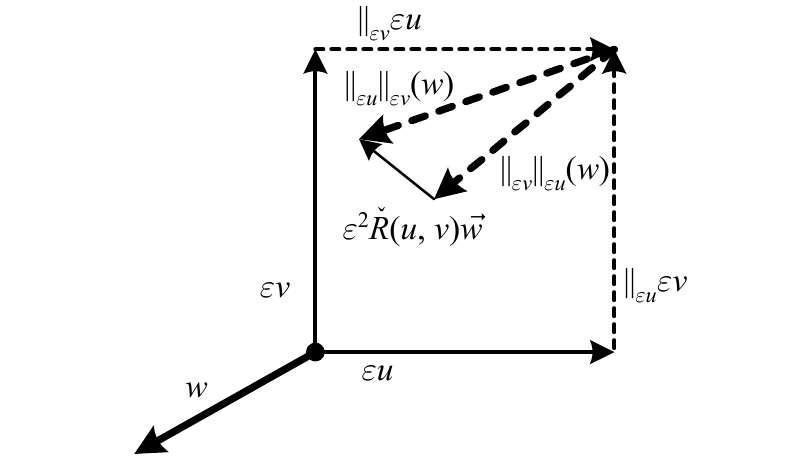}
\par\end{centering}
\caption{A slight variant of $\check{R}\left(u,v\right)\vec{w}$ viewed as
\textquotedblleft the difference between $w$ when parallel transported
around the two opposite edges of the boundary of the surface defined
by its arguments.\textquotedblright{} In the case of zero torsion,
the boundary can be built from parallel transports instead of vector
field values.}
\end{figure}

Thus, still assuming zero torsion, we can construct a cube from the
parallel transports of $u$, $v$, and $w$. This construction reveals
that the first Bianchi identity corresponds to the fact that the three
curvature vectors form a triangle, i.e. their sum is zero. 

\begin{figure}[H]
\noindent \begin{centering}
\includegraphics[width=0.7\columnwidth]{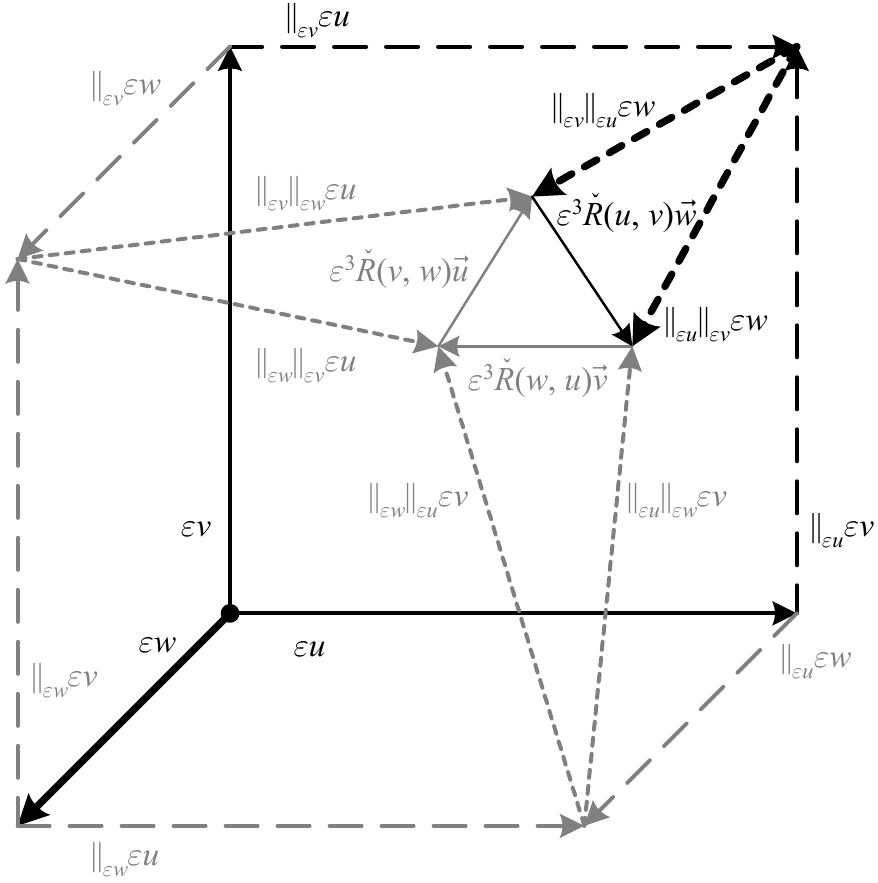}
\par\end{centering}
\caption{The first Bianchi identity reflects the fact that for zero torsion,
the far corners of a cube made of parallel transported vectors do
not meet, and their separation is made up of the differences in parallel
transport via opposite edges of each face. Note that the corners of
the triangle are points since vanishing torsion means that e.g. $\varepsilon u+\parallel_{\varepsilon u}(\varepsilon w)=\varepsilon w+\parallel_{\varepsilon w}(\varepsilon u)$,
so that the top point of the triangle reflects this equality parallel
transported by $\varepsilon v$.}
\end{figure}

\subsection{Second Bianchi identity }

If we take the exterior covariant derivative of the curvature, we
get

\begin{equation}
\mathrm{D}\check{R}=0.
\end{equation}
This is called the \textbf{second Bianchi identity}\index{second Bianchi identity}\index{Bianchi identity!second},
and can be verified algebraically from the definition $\check{R}\equiv\mathrm{d}\check{\Gamma}+\check{\Gamma}\wedge\check{\Gamma}$.
We can write this identity more explicitly as

\begin{equation}
\begin{aligned}0 & =\mathrm{D}\check{R}(u,v,w)\vec{a}\\
 & =\nabla_{u}\check{R}(v,w)\vec{a}+\nabla_{v}\check{R}(w,u)\vec{a}+\nabla_{w}\check{R}(u,v)\vec{a}\\
 & \phantom{{}=}-\check{R}([u,v],w)\vec{a}-\check{R}([v,w],u)\vec{a}-\check{R}([w,u],v)\vec{a},
\end{aligned}
\end{equation}
where we have used the antisymmetry of $\check{R}$ and the covariant
derivative acts on the value of $\check{R}$ as a tensor of type $\left(1,1\right)$.
Working this expression into tensor notation and using the tensor
expression for the torsion in terms of the commutator, we find that

\begin{equation}
\begin{aligned}0 & =\nabla_{e}?R^{c}{}_{dab}?+\nabla_{a}?R^{c}{}_{dbe}?+\nabla_{b}?R^{c}{}_{dea}?\\
 & \phantom{{}=}-?R^{c}{}_{dfe}??T^{f}{}_{ab}?-?R^{c}{}_{dfa}??T^{f}{}_{be}?-?R^{c}{}_{dfb}??T^{f}{}_{ea}?,
\end{aligned}
\end{equation}
or 
\begin{equation}
?R^{c}{}_{d[ab;e]}?=?R^{c}{}_{df[e}??T^{f}{}_{ab]}?,
\end{equation}
and in the case of zero torsion, $?R^{c}{}_{d[ab;e]}?=0$.

Geometrically, the second Bianchi identity can be seen as reflecting
the same ``boundary of a boundary'' idea as that of $\mathrm{d}^{2}=0$
in Fig. \ref{fig:The-3-form}, except that here we are parallel transporting
a vector $\vec{a}$ around each face that makes up the boundary of
the cube. As in the previous section, we can take advantage of the
fact that $\check{R}(v,w)\vec{a}$ only depends upon the local value
of $\vec{a}$, constructing its vector field values such that e.g.
$\vec{a}\left|_{p+\varepsilon u}\right.=\parallel_{\varepsilon u}(\vec{a}\left|_{p}\right.)$,
giving us 
\begin{equation}
\begin{aligned}\varepsilon\nabla_{u}\check{R}(v,w)\vec{a} & =\check{R}(v\left|_{p+\varepsilon u}\right.,w\left|_{p+\varepsilon u}\right.)\vec{a}\left|_{p+\varepsilon u}\right.-\parallel_{\varepsilon u}\check{R}(v,w)\parallel_{\varepsilon u}^{-1}\vec{a}\left|_{p+\varepsilon u}\right.\\
 & =\check{R}(v\left|_{p+\varepsilon u}\right.,w\left|_{p+\varepsilon u}\right.)\parallel_{\varepsilon u}\vec{a}-\parallel_{\varepsilon u}\check{R}(v,w)\vec{a}.
\end{aligned}
\end{equation}
The first term parallel transports $\vec{a}$ along $\varepsilon u$
and then around the parallelogram defined by $v$ and $w$ at $p+\varepsilon u$,
while the second parallel transports $\vec{a}$ around the parallelogram
defined by $v$ and $w$ at $p$, then along $\varepsilon u$. Thus
in the case of vanishing Lie commutators (e.g. a holonomic frame),
we construct a cube from the vector fields $u$, $v$, and $w$, and
find that the second Bianchi identity reflects the fact that $\mathrm{D}\check{R}(u,v,w)\vec{a}$
parallel transports $\vec{a}$ along each edge of the cube an equal
number of times in opposite directions, thus canceling out any changes.

\begin{figure}[H]
\noindent \begin{centering}
\includegraphics[width=0.85\columnwidth]{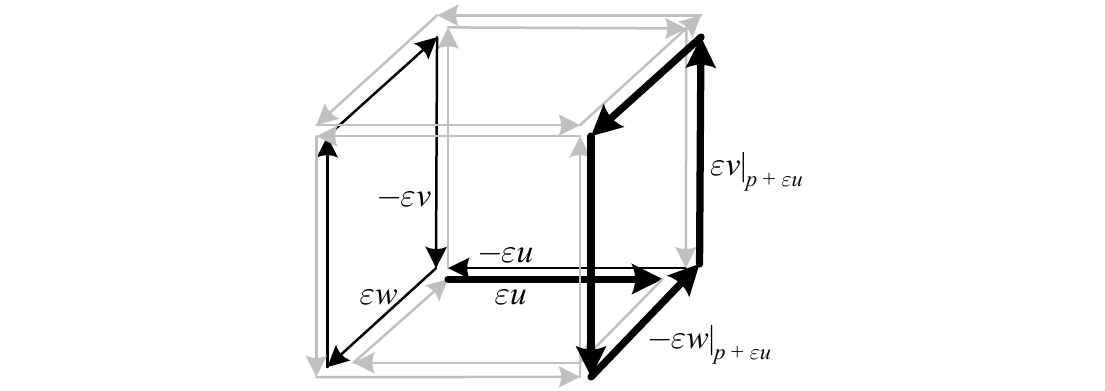}
\par\end{centering}
\caption{The second Bianchi identity reflects the fact that for vanishing Lie
commutators, $\mathrm{D}\check{R}(u,v,w)\vec{a}$ parallel transports
$\vec{a}$ along each edge of the cube made of the three vector field
arguments an equal number of times in opposite directions, thus canceling
out any changes. Above, $\varepsilon\nabla_{u}\check{R}(v,w)\vec{a}=\check{R}(v\left|_{p+\varepsilon u}\right.,w\left|_{p+\varepsilon u}\right.)\parallel_{\varepsilon u}\vec{a}-\parallel_{\varepsilon u}\check{R}(v,w)\vec{a}$
is highlighted by the bold arrows representing the path along which
$\vec{a}$ is parallel transported in the first term, and by the remaining
dark arrows representing the path along which $\vec{a}$ is parallel
transported in the second term.}
\end{figure}

In the case of non-vanishing torsion, where there is a non-vanishing
commutator $\vec{T}(u,v)=-[u,v]\neq0$, we find that the cube gains
a ``shaved edge,'' and that the extra non-vanishing term $-\check{R}([u,v],w)\vec{a}$
in $\mathrm{D}\check{R}$ maintains the ``boundary of a boundary''
logic by adding a loop of parallel transports of $\vec{a}$ in the
proper direction around the new ``face'' created.

\begin{figure}[H]
\noindent \begin{centering}
\includegraphics[width=0.85\columnwidth]{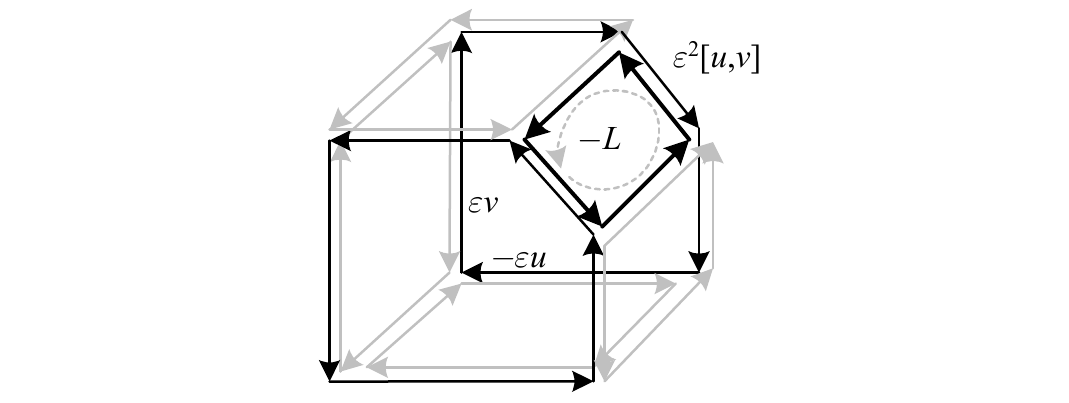}
\par\end{centering}
\caption{In the case of non-vanishing torsion and thus commutator, the extra
term $-\check{R}([u,v],w)\vec{a}$ in $\mathrm{D}\check{R}$ maintains
the cancellation of face boundaries by adding a loop $L$ around the
new \textquotedblleft shaved edge\textquotedblright{} created.}
\end{figure}

\section{Introducing the metric}

\subsection{The Riemannian metric}

A (pseudo) metric tensor (see Section \ref{subsec:Abstract-index-notation})
is a (pseudo) inner product $\left\langle v,w\right\rangle $ on a
vector space $V$ that can be represented by a symmetric tensor $g_{ab}$,
and thus can be used to lower and raise indices on tensors. A \textbf{(pseudo)
Riemannian metric}\index{Riemannian metric} (AKA metric) is a (pseudo)
metric tensor field on a manifold $M$, making $M$ a \textbf{(pseudo)
Riemannian manifold}\index{Riemannian manifold}. 

A metric defines the length (norm) of tangent vectors, and can thus
be used to define the length $L$ of a curve $C$ via parametrization
and integration:
\begin{equation}
\begin{aligned}L(C) & \equiv\int\left\Vert \dot{C}(t)\right\Vert \mathrm{d}t\\
 & =\int\sqrt{\left\langle \dot{C}(t),\dot{C}(t)\right\rangle }\mathrm{d}t
\end{aligned}
\end{equation}
This also turns any (non-pseudo) Riemannian manifold into a metric
space, with distance function $d(x,y)$ defined to be the minimum
length curve connecting the two points $x$ and $y$; this curve is
called a (Riemannian) geodesic, and it locally minimizes the distance
between any of its points. With a pseudo-Riemannian metric the distance
may be instead maximized, and this extremal distance is only locally
valid since e.g. the curve may eventually self-intersect as the equator
on a sphere does. It can be shown that for any tangent vector $v$
on a Riemannian manifold there is a unique geodesic $C_{v}(d)$ parametrized
by distance whose tangent is $v$; one can then define the exponential
map by $\mathrm{exp}(v)\equiv C_{v}(1)$. 

\noindent %
\begin{framed}%
\noindent \sun{} With a metric, our intuitive picture of a manifold
loses its ``stretchiness'' via the introduction of length and angles;
but having only intrinsically defined properties, the manifold can
still be e.g. rolled up like a piece of paper if imagined as flat
and embedded in a larger space. \end{framed}

If the coordinate frame of $x^{\mu}$ is orthonormal at a point $p\in M^{n}$
in a Riemannian manifold, for arbitrary coordinates $y^{\mu}$ we
can consider the components of the metric tensor in the two coordinate
frames to find that 
\begin{equation}
\begin{aligned}g_{\mu\nu}\mathrm{d}y^{\mu}\mathrm{d}y^{\nu} & =\delta_{\lambda\sigma}\mathrm{d}x^{\lambda}\mathrm{d}x^{\sigma}\\
 & =\delta_{\lambda\sigma}\frac{\partial x^{\lambda}}{\partial y^{\mu}}\mathrm{d}y^{\mu}\frac{\partial x^{\sigma}}{\partial y^{\nu}}\mathrm{d}y^{\nu}\\
 & =\left[J_{x}(y)\right]^{T}\left[J_{x}(y)\right]\mathrm{d}y^{\mu}\mathrm{d}y^{\nu}\\
\Rightarrow\mathrm{det}\left(g_{\mu\nu}\right) & =\left[\mathrm{det}\left(J_{x}\left(y\right)\right)\right]^{2},
\end{aligned}
\end{equation}
where $J_{x}(y)$ is the Jacobian matrix (see Section \ref{subsec:The-differential-and-pullback})
and we have used the fact that $\mathrm{det}(A^{T}A)=[\mathrm{det}(A)]^{2}$.
Thus the volume of an region $U\in M^{n}$ corresponding to $R\in\mathbb{R}^{n}$
in the coordinates $x^{\mu}$ is 
\begin{equation}
V(U)=\int_{R}\sqrt{\mathrm{det}(g)}\mathrm{d}x^{1}\ldots\mathrm{d}x^{n},
\end{equation}
where $\mathrm{det}(g)$ is the determinant of the metric tensor as
a matrix in the coordinate frame $\partial/\partial x^{\mu}$. In
the context of a pseudo-Riemannian manifold $\mathrm{det}(g)$ can
be negative, and the integrand 
\begin{equation}
\mathrm{d}V\equiv\sqrt{\left|\mathrm{det}(g)\right|}\mathrm{d}x^{1}\ldots\mathrm{d}x^{n}
\end{equation}
is called the \textbf{volume element}\index{volume element}, or when
written as a form $\mathrm{d}V\equiv\sqrt{\left|\mathrm{det}(g)\right|}\mathrm{d}x^{1}\wedge\cdots\wedge\mathrm{d}x^{n}$
it is called the \textbf{volume form}\index{volume form}. In physical
applications $\mathrm{d}V$ usually denotes the \textbf{volume pseudo-form}\index{volume pseudo-form},
which gives a positive value regardless of orientation. Note that
if the coordinate basis is orthonormal then $\left|\mathrm{det}(g)\right|=1$;
thus these definitions are consistent with those typically defined
on $\mathbb{R}^{n}$. Sometimes one defines a volume form on a manifold
without defining a metric; in this case the metric (and connection)
is not uniquely determined.

\noindent %
\begin{framed}%
\noindent $\triangle$ The symbol $g$ is frequently used to denote
$\mathrm{det}(g)$, and sometimes $\sqrt{\left|\mathrm{det}(g)\right|}$,
in addition to denoting the metric tensor itself.\end{framed}

We can use the inner product to define an \textbf{orthonormal frame}\index{orthonormal frame}\index{frame!orthonormal}
on $M$. In four dimensions an orthonormal frame is also called a
\textbf{tetrad}\index{tetrad} (AKA vierbein\index{vierbein}). Any
frame on a manifold can be defined to be an orthonormal frame, which
is equivalent to defining the metric (which in the orthonormal frame
is $g_{ab}=\eta_{ab}$). An orthonormal holonomic frame exists on
a region of $M$ if and only if that region is flat. Thus in general,
given a set of coordinates on $M$, we have to choose between using
either a non-coordinate orthonormal frame or a non-orthonormal coordinate
frame.

The \textbf{Hopf-Rinow theorem}\index{Hopf-Rinow theorem} says that
a connected Riemannian manifold $M$ is complete as a metric space
(or equivalently, all closed and bounded subsets are compact) if and
only if it is \textbf{geodesically complete}\index{geodesically complete},
meaning that the exponential map is defined for all vectors at some
$p\in M$. If $M$ is geodesically complete at $p$, then it is at
all points on the manifold, so this property can also be used to state
the theorem. This theorem is not valid for pseudo-Riemannian manifolds;
any (pseudo) Riemannian manifold that is geodesically complete is
called a \textbf{geodesic manifold}\index{geodesic manifold}. 

As noted previously, a Riemannian metric can be defined on any differentiable
manifold. In general, not every manifold admits a pseudo-Riemannian
metric, and in particular not every 4-manifold admits a Minkowski
metric, but 4-manifolds that are noncompact, parallelizable, or compact,
connected and of Euler characteristic 0 all do.

In the same way that differentiable manifolds are equivalent if they
are related by a diffeomorphism, Riemannian manifolds are equivalent
if they are related by an \textbf{isometry}\index{isometry}, a diffeomorphism
$\Phi\colon M\to N$ that preserves the metric, i.e. $\forall v,w\in TM$,
$\left\langle v,w\right\rangle \left|_{p}\right.=\left\langle \mathrm{d}\Phi_{p}(v),\mathrm{d}\Phi_{p}(w)\right\rangle \left|_{\Phi(p)}\right.$.
Also like diffeomorphisms, the isometries of a manifold form a group;
for example, the group of isometries of Minkowski space is the Poincaré
group. A vector field whose one-parameter diffeomorphisms are isometries
is called a \textbf{Killing field}\index{Killing field}, also called
a \textbf{Killing vector}\index{Killing vector} since it can be shown
(\cite{Petersen} pp. 188-189) that a Killing field is determined
by a vector at a single point along with its covariant derivatives.
A Killing field thus satisfies $L_{v}g_{ab}=0$, which using eq. (\ref{eq:Lie covariant})
for a Levi-Civita connection (see next section) is equivalent to 
\begin{equation}
\nabla_{a}v_{b}+\nabla_{b}v_{a}=0,
\end{equation}
called the \textbf{Killing equation}\index{Killing equation} (AKA
Killing condition\index{Killing condition}).

We can then consider isometric immersions and embeddings, and ask
whether every Riemannian manifold can be embedded in some $\mathbb{R}^{n}$.
The \textbf{Nash embedding theorem}\index{Nash embedding theorem}
provides an affirmative answer, and it can also be shown that every
pseudo-Riemannian manifold can be isometrically embedded in some $\mathbb{R}^{n}$
with some signature while maintaining arbitrary differentiability
of the metric.

\subsection{\label{subsec:The-Levi-Civita-connection}The Levi-Civita connection }

A connection on a Riemannian manifold $M$ is called a \textbf{metric
connection}\index{metric connection}\index{connection!metric} (AKA
metric compatible connection, isometric connection\index{isometric connection}\index{connection!isometric})
if its associated parallel transport respects the metric, i.e. it
preserves lengths and angles. More precisely, $\forall v,w\in TM$,
we require that 
\begin{equation}
\left\langle \parallel_{C}(v),\parallel_{C}(w)\right\rangle =\left\langle v,w\right\rangle 
\end{equation}
for any curve $C$ in $M$. 

In terms of the metric, this can be written $g_{ab}\parallel_{C}v^{a}\parallel_{C}w^{b}=g_{ab}v^{a}w^{b}$.
But recalling that the parallel transport of tensors just transports
the arguments, we also have $\left(\parallel_{-C}g_{ab}\right)v^{a}w^{b}=g_{ab}\parallel_{C}v^{a}\parallel_{C}w^{b}$,
so that we must have $\parallel_{-C}g_{ab}=g_{ab}$, or $\nabla_{c}g_{ab}=0$.
Using the Leibniz rule for the covariant derivative over the tensor
product, we can derive a Leibniz rule over the inner product: 
\begin{equation}
\begin{aligned}\nabla_{c}\left(g_{ab}v^{a}w^{b}\right) & =0+g_{ab}\nabla_{c}v^{a}w^{b}+g_{ab}v^{a}\nabla_{c}w^{b}\\
\Rightarrow\nabla_{u}\left\langle v,w\right\rangle  & =\left\langle \nabla_{u}v,w\right\rangle +\left\langle v,\nabla_{u}w\right\rangle 
\end{aligned}
\end{equation}
Requiring this relationship to hold is an equivalent way to define
a metric connection. In terms of the connection coefficients, a metric
connection then satisfies
\begin{equation}
\begin{aligned}\nabla_{c}g_{ab} & =\partial_{c}g_{ab}-\Gamma^{d}{}_{ac}g_{db}-\Gamma^{d}{}_{bc}g_{ad}=0.\\
\Rightarrow\partial_{c}g_{ab} & =\Gamma{}_{abc}+\Gamma{}_{bac},
\end{aligned}
\end{equation}
where we write $\Gamma{}_{abc}\equiv\Gamma^{d}{}_{bc}g_{ad}$, which
again it is important to note is not tensor. By considering $\partial_{c}\left(g^{ad}g_{df}\right)=\partial_{c}\left(\delta^{a}{}_{f}\right)=0$,
we arrive at the complementary expression
\begin{equation}
\begin{aligned}\partial_{c}g^{ab} & =-g^{ad}g^{bf}\partial_{c}g_{df}\\
 & =-\left(\Gamma^{ab}{}_{c}+\Gamma^{ba}{}_{c}\right).
\end{aligned}
\end{equation}

The \textbf{Levi-Civita connection}\index{Levi-Civita connection}\index{connection!Levi-Civita}
(AKA Riemannian connection\index{Riemannianconnection}\index{connection!Riemannian},
Christoffel connection\index{Christoffel connection}\index{connection!Christoffel})
is then the torsion-free metric connection on a (pseudo) Riemannian
manifold $M$. The \textbf{fundamental theorem of Riemannian geometry}\index{fundamental theorem of Riemannian geometry}
states that for any (pseudo) Riemannian manifold the Levi-Civita connection
exists and is unique. On the other hand, an arbitrary connection can
only be the Levi-Civita connection for some metric if it is torsion-free
and preserves lengths; moreover, this metric is unique only up to
a scaling factor (excepting special cases, e.g. if the manifold is
a product space there can be a scaling factor for each factor space;
in physics, this corresponds to a choice of units).

For a metric connection, the curvature then must take values that
are infinitesimal rotations, i.e. $\check{R}$ is $o(r,s)$-valued.
Thus if we eliminate the influence of the signature by lowering the
first index, the first two indices of the curvature tensor are anti-symmetric:
\begin{equation}
R_{cdab}=-R{}_{dcab}
\end{equation}
Using the anti-symmetry of the other indices and the first Bianchi
identity, this leads to another commonly noted symmetry 
\begin{equation}
R_{cdab}=R{}_{abcd}.
\end{equation}

The Leibniz rule for the covariant derivative over the inner product
along with the zero torsion relation $\nabla_{v}w=\nabla_{w}v+\left[v,w\right]$
can be used to derive an expression called the \textbf{Koszul formula}\index{Koszul formula}:
\begin{equation}
\begin{aligned}2\left\langle \nabla_{u}v,w\right\rangle = & \nabla_{u}\left\langle v,w\right\rangle +\nabla_{v}\left\langle w,u\right\rangle -\nabla_{w}\left\langle u,v\right\rangle \\
 & -\left\langle u,[v,w]\right\rangle +\left\langle v,[w,u]\right\rangle +\left\langle w,[u,v]\right\rangle 
\end{aligned}
\end{equation}
Substituting in the frame vector fields and eliminating the metric
tensor from the left hand side, we arrive at an expression for the
connection in terms of the metric:

\begin{equation}
\begin{aligned}2\Gamma^{c}{}_{ba}=g^{cd}( & \partial_{a}g_{bd}+\partial_{b}g_{da}-\partial_{d}g_{ab}\\
 & -g_{af}[e_{b},e_{d}]^{f}+g_{bf}[e_{d},e_{a}]^{f}+g_{df}[e_{a},e_{b}]^{f})
\end{aligned}
\end{equation}

On a (pseudo) Riemannian manifold, the connection coefficients for
the Levi-Civita connection in a coordinate basis $\Gamma^{\lambda}{}_{\mu\sigma}$
are called the \textbf{Christoffel symbols}\index{Christoffel symbols},
and are sometimes denoted $\{\substack{\lambda\\
\mu\sigma
}
\}$ or $\{\substack{\mu\sigma\\
\lambda
}
\}$. At a point $p\in U\subset M$, an orthonormal basis for $T_{p}U$
can be used to form geodesic normal coordinates, which are then called
\textbf{Riemann normal coordinates}\index{Riemann normal coordinates}.
Recalling from Section \ref{subsec:Geodesics-and-normal-coordinates}
that with zero torsion the connection coefficients vanish at $p$,
we can apply the covariant derivative to the metric tensor to conclude
that the partial derivatives of the metric $g_{\mu\nu}=\eta_{\mu\nu}$
all also vanish at $p$. %
\begin{framed}%
\noindent \sun{} The vanishing of the Christoffel symbols at the origin
of Riemann normal coordinates is frequently used to simplify the derivation
of tensor relations which are then, being frame-independent, seen
to be true in any coordinate system or frame (and if the origin was
chosen arbitrarily, at any point). In particular, the covariant and
partial derivatives are equivalent at the origin of Riemann normal
coordinates.\end{framed}

\subsection{Independent quantities and dependencies}

From their definitions, the parallel transport and connection in general
determine each other. It can be shown that every manifold admits a
connection, and every other connection can be obtained by adding a
frame-independent $gl\left(\mathbb{R}^{n}\right)$-valued 1-form (tensor
field of type $(1,2)$) to it. If the curvature is given over $M$,
there is at most one metric (apart from special cases, up to a scaling
factor, and for $n>2$) whose Levi-Civita connection yields this curvature.

If we choose coordinate charts and use coordinate frames on $M^{n}$,
we can calculate the number of independent functions and equations
associated with the various quantities and relations we have covered,
and use them to verify the associated dependencies. 

\begin{table}[H]
\begin{tabular*}{1\columnwidth}{@{\extracolsep{\fill}}|>{\raggedright}p{0.4\columnwidth}|l|>{\raggedright}p{0.15\columnwidth}|}
\hline 
Quantity / relation & Viewpoint & Count\tabularnewline
\hline 
\hline 
Metric & Symmetric matrix of functions & $n(n+1)/2$\tabularnewline
\hline 
Coordinate frame & Fixed & 0\tabularnewline
\hline 
Connection & $gl$-valued (matrix-valued) 1-form & $n^{3}$\tabularnewline
\hline 
Metric condition & Derivative of metric & $n^{2}(n+1)/2$\tabularnewline
\hline 
Torsion-free condition & Vector-valued 2-form & $n^{2}(n-1)/2$\tabularnewline
\hline 
\end{tabular*}

\caption{Independent function and equation counts in a coordinate frame.}
\end{table}
The choice of coordinates determines the frame, leaving the geometry
of the Riemannian manifold defined by the $n(n+1)/2$ functions of
the metric. A torsion-free connection consists of $n^{3}-n^{2}(n-1)/2=n^{2}(n+1)/2$
functions. The metric condition is exactly this number of equations,
allowing us in general to solve for the connection if the metric is
known, or vice-versa (up to a constant scaling factor). 

Alternatively, we can look at things in a orthonormal frame:

\begin{table}[H]
\begin{tabular*}{1\columnwidth}{@{\extracolsep{\fill}}|>{\raggedright}p{0.5\columnwidth}|l|>{\raggedright}p{0.15\columnwidth}|}
\hline 
Quantity / relation & Viewpoint & Count\tabularnewline
\hline 
\hline 
Metric & Fixed & 0\tabularnewline
\hline 
Orthonormal frame & $n$ vector fields & $n^{2}$\tabularnewline
\hline 
Change of orthonormal frame & $SO$-valued 0-form & $n(n-1)/2$\tabularnewline
\hline 
Connection & $so$-valued 1-form  & $n^{2}(n-1)/2$\tabularnewline
\hline 
Metric condition & Automatically satisfied & 0\tabularnewline
\hline 
Torsion-free condition & Vector-valued 2-form & $n^{2}(n-1)/2$\tabularnewline
\hline 
\end{tabular*}

\caption{Independent function and equation counts in an orthonormal frame.}
\end{table}
Here the metric is fixed, defined by the frame, which consists of
$n^{2}$ functions, but is determined only up to a change of orthonormal
frame (rotation); this yields $n^{2}-n(n-1)/2=n(n+1)/2$ functions,
consistent with the metric function count above. The torsion-free
condition is the same number of equations as the connection has functions,
so that in general the torsion-free connection can be determined by
the orthonormal frame.

\subsection{\label{subsec:The-divergence-and-conserved-quantities}The divergence
and conserved quantities}

The divergence of a vector field $u$ (see Section \ref{subsec:The-exterior-derivative-of-a-k-form})
can be generalized to a pseudo-Riemannian manifold of signature $\left(r,s\right)$
by defining 
\begin{equation}
\mathrm{div}(u)\equiv(-1)^{s}*\mathrm{d}(*(u^{\flat})).
\end{equation}
Using the relations $i_{u}\Omega=*(u^{\flat})$ (see Section \ref{subsec:Relationships-between-derivations})
and $(-1)^{s}A=(*A)\Omega$ for $A\in\Lambda^{n}M^{n}$ (see Section
\ref{subsec:The-Hodge-star}), we have 
\begin{equation}
\begin{aligned}\mathrm{d}(i_{u}\Omega) & =\mathrm{d}(*(u^{\flat}))\\
 & =(-1)^{s}*\mathrm{d}(*(u^{\flat}))\Omega\\
 & =\mathrm{div}(u)\Omega.
\end{aligned}
\label{eq:divergence}
\end{equation}
Using $i_{u}\mathrm{d}+\mathrm{d}i_{u}=L_{u}$ we then arrive at $\mathrm{div}(u)\Omega=L_{u}\Omega$,
or as it is more commonly written 
\begin{equation}
\mathrm{div}(u)\mathrm{d}V=L_{u}\mathrm{d}V.
\end{equation}
Thus we can say that $\mathrm{div}(u)$ is ``the fraction by which
a unit volume changes when transported by the flow of $u$,'' and
if $\mathrm{div}(u)=0$ then we can say that ``the flow of $u$ leaves
volumes unchanged.'' Expanding the volume element in coordinates
$x^{\lambda}$ we can obtain an expression for the divergence in terms
of these coordinates,
\begin{equation}
\mathrm{div}(u)=\frac{1}{\sqrt{\left|\mathrm{det}(g)\right|}}\partial_{\lambda}\left(u^{\lambda}\sqrt{\left|\mathrm{det}(g)\right|}\right).
\end{equation}
Note that both this metric-dependent expression and the expression
$\nabla_{a}u^{a}$ (sometimes called the \textbf{covariant divergence}\index{covariant divergence})
in terms of the Levi-Civita connection are coordinate-independent
and equal to $\partial_{a}u^{a}$ in Riemann normal coordinates, confirming
our expectation that for zero torsion we have 
\begin{equation}
\mathrm{div}(u)=\nabla_{a}u^{a}.
\end{equation}
\begin{figure}[H]
\noindent \begin{centering}
\includegraphics[width=0.5\columnwidth]{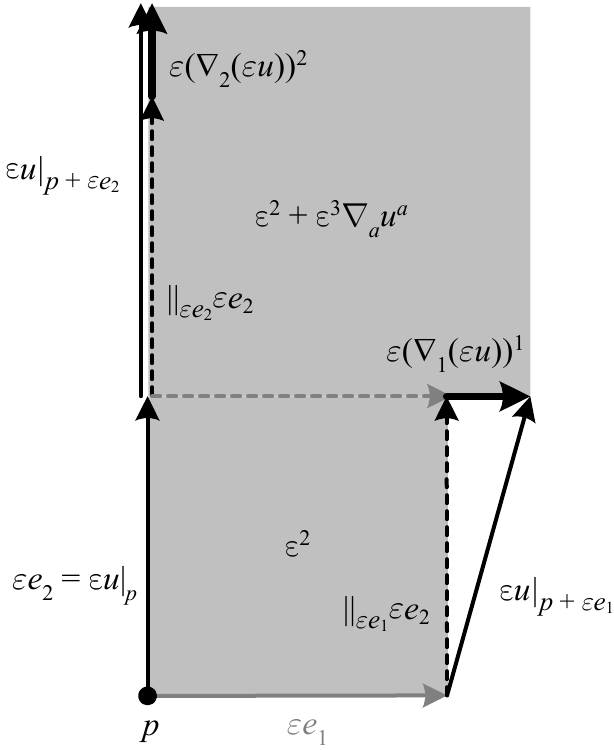}
\par\end{centering}
\caption{The divergence measures the change in volume due to the flow. Here
we assume zero torsion, and that the vector field $u$ has unit length
at point $p$, and choose an orthonormal frame which aligns $e_{2}$
with $u$. Each covariant derivative extends a face of the volume,
with their sum being proportional to the total change in volume. Note
that the upper right corner is of order $\varepsilon^{4}$ and so
can be neglected, and e.g. any component of $\nabla_{1}u$ orthogonal
to $e_{1}$ leaves the volume unchanged, since a more accurate depiction
would include the volume with edge $-\varepsilon e_{1}$, where by
linearity this component would be in the opposite direction and thus
cancel the volume change. Also note that non-zero torsion would reduce
the top edge $\parallel_{\varepsilon e_{2}}\varepsilon e_{1}$ by
$\varepsilon^{2}?T^{1}{}_{1b}?u^{b}$, which must be added back by
subtracting this component, matching the algebraic result.}
\end{figure}

Using the relation $\mathrm{div}(u)\Omega=\mathrm{d}(i_{u}\Omega)$
from eq. (\ref{eq:divergence}), along with Stokes' theorem, we recover
the classical divergence theorem\index{divergence theorem}
\begin{equation}
\begin{aligned}\int_{V}\mathrm{div}(u)\mathrm{d}V & =\int_{\partial V}i_{u}\mathrm{d}V\\
 & =\int_{\partial V}\left\langle u,n\right\rangle \mathrm{d}S,
\end{aligned}
\end{equation}
where $V$ is an $n$-dimensional compact submanifold of $M^{n}$,
$n$ is the unit normal vector to $\partial V$, and $\mathrm{d}S\equiv i_{n}\mathrm{d}V$
is the induced volume element (``surface element\index{surface element}'')
for $\partial V$. In the case of a Riemannian metric, this can be
thought of as reflecting the intuitive fact that ``the change in
a volume due to the flow of $u$ is equal to the net flow across that
volume's boundary.'' If $\mathrm{div}(u)=0$ then we can say that
``the net flow of $u$ across the boundary of a volume is zero.''
We can also consider an infinitesimal $V$, so that the divergence
at a point measures ``the net flow of $u$ across the boundary of
an infinitesimal volume.'' 

In physics, one considers the divergence of the \textbf{current vector}
\index{current vector}(AKA current density\index{current density},
flux\index{flux}, \index{flux density}flux density) $j\equiv\rho u$
of a physical flow in space at a moment in time, where $\rho$ is
the density of the physical quantity $Q$ and $u$ is thus a velocity
field; e.g. in $\mathbb{R}^{3}$, $j$ has units $Q/(\mathrm{length})^{2}(\mathrm{time})$.
For a flat Riemannian metric on the manifold representing space, the
\textbf{continuity equation}\index{continuity equation} (AKA equation
of continuity\index{equation of continuity}) is
\begin{equation}
\frac{\mathrm{d}q}{\mathrm{d}t}=\Sigma-\int_{\partial V}\left\langle j,\hat{n}\right\rangle \mathrm{d}S,
\end{equation}
where $q$ is the amount of $Q$ contained in $V$, $t$ is time,
and $\Sigma$ is the rate of $Q$ being created within $V$. The continuity
equation thus states the intuitive fact that the change of $Q$ within
$V$ equals the amount generated less the amount which passes through
$\partial V$. 

Using the divergence theorem, we can then obtain the differential
form of the continuity equation
\begin{equation}
\frac{\partial\rho}{\partial t}=\sigma-\mathrm{div}(j),
\end{equation}
where $\sigma$ is the amount of $Q$ generated per unit volume per
unit time. This equation then states the intuitive fact that at a
point, the change in density of $Q$ equals the amount generated less
the amount that moves away. Positive $\sigma$ is referred to as a
\textbf{source}\index{source} of $Q$, and negative $\sigma$ a \textbf{sink}\index{sink}.
If $\sigma=0$ then we say that $Q$ is a \textbf{conserved quantity}\index{conserved quantity}
and refer to the continuity equation as a (local) \textbf{conservation
law}\index{conservation law}. 

Under a flat Lorentzian metric, we can combine $\rho$ and $j$ into
the \textbf{four-current}\index{four-current} 
\begin{equation}
J\equiv(\rho,j^{\mu}),
\end{equation}
and express the continuity equation with $\sigma=0$ as 
\begin{equation}
\mathrm{div}(J)=0,
\end{equation}
whereupon $J$ is called a \textbf{conserved current}\index{conserved current}.
Note that if any curvature is present (but no torsion), when we split
out the time component we recover a Riemannian divergence but introduce
a source due to the non-zero Christoffel symbols
\begin{equation}
\begin{aligned}\nabla_{\mu}J^{\mu} & =\partial_{\mu}J^{\mu}+?\Gamma^{\mu}{}_{\nu\mu}?J^{\nu}\\
 & =\partial_{t}\rho+\nabla_{i}j^{i}+\left(?\Gamma^{\mu}{}_{t\mu}?\rho+?\Gamma^{t}{}_{it}?j^{i}\right),
\end{aligned}
\end{equation}
where $t$ is the negative signature component and the index $i$
goes over the remaining positive signature components. Thus, since
the Christoffel symbols are coordinate-dependent, in the presence
of curvature there is in general no coordinate-independent conserved
quantity associated with a vanishing Lorentzian divergence. A conserved
current nevertheless means that the quantity is conserved in finite
volumes of spacetime, in the sense that $\int_{\partial V}\left\langle J,\hat{n}\right\rangle \mathrm{d}S=0$
over any spacetime volume $V$, and the continuity equation holds
for the components of the coordinate-dependent quantity $\mathfrak{J}\equiv J\sqrt{\left|\mathrm{det}(g)\right|}$,
since
\begin{equation}
\begin{aligned}\partial_{\mu}\mathfrak{J}^{\mu} & =\partial_{t}\mathfrak{J}^{t}+\partial_{i}\mathfrak{J}^{i}\\
 & =\partial_{t}\mathfrak{J}^{t}+\nabla_{i}\mathfrak{J}^{i}=0.
\end{aligned}
\end{equation}
\begin{framed}%
\noindent \sun{} \textbf{Noether's theorem}\index{Noether's theorem}
derives conserved currents from transformations (``symmetries'')
on the variables of an expression called the \textbf{action}\index{action}
that leave it unchanged.\end{framed}

\subsection{Ricci and sectional curvature}

The \textbf{Ricci curvature tensor}\index{Ricci curvature tensor}
(AKA Ricci tensor\index{Ricci tensor}) is formed by contracting two
indices in the Riemann curvature tensor:

\begin{equation}
\begin{aligned}?R_{ab}? & \equiv?R^{c}{}_{acb}?\\
\mathrm{Ric}(v,w) & \equiv?R{}_{ab}?v^{a}w^{b}
\end{aligned}
\end{equation}
Using the symmetries of the Riemann tensor for a metric connection
along with the first Bianchi identity for zero torsion, it is easily
shown that the Ricci tensor for the Levi-Civita connection is symmetric.
A pseudo-Riemannian manifold is said to have constant Ricci curvature,
or to be an \textbf{Einstein manifold}\index{Einstein manifold},
if the Ricci tensor is a constant multiple of the metric tensor.

Since the Ricci tensor is symmetric for zero torsion, by the spectral
theorem it can be diagonalized on a Riemannian manifold and thus is
determined by 
\begin{equation}
\mathrm{Ric}(v)\equiv\mathrm{Ric}(v,v),
\end{equation}
which is called the \textbf{Ricci curvature function}\index{Ricci curvature function}
(AKA Ricci function\index{Ricci function}). Note that the Ricci function
is not a 1-form since it is not linear in $v$. Choosing a basis that
diagonalizes $R_{ab}$ is equivalent to choosing our basis vectors
to line up with the directions that yield extremal values of the Ricci
function on the unit vectors $\mathrm{Ric}(\hat{v},\hat{v})$ (or
equivalently, the principal axes of the ellipsoid / hyperboloid $\mathrm{Ric}(v,v)=1$). 

Finally, if we raise one of the indices of the Ricci tensor and contract
we arrive at the \textbf{Ricci scalar}\index{Ricci scalar} (AKA scalar
curvature\index{scalar curvature}): 
\begin{equation}
\begin{aligned}R & \equiv g^{ab}R_{ab}\end{aligned}
\end{equation}
For a Riemannian manifold $M^{n}$, the Ricci scalar can thus be viewed
as $n$ times the average of the Ricci function on the set of unit
tangent vectors. 

\noindent %
\begin{framed}%
\noindent $\triangle$ The Ricci function and Ricci scalar are sometimes
defined as averages instead of contractions (sums), introducing extra
factors in terms of the dimension $n$ to the above definitions.\end{framed}

The Ricci function in terms of the curvature 2-form in an orthonormal
frame $e_{\mu}$ (dropping the hats to avoid clutter) on a pseudo-Riemannian
manifold $M^{n}$ naturally splits into terms which each also measure
curvature:
\begin{equation}
\mathrm{Ric}(e_{\mu})=\sum_{i\neq\mu}g_{ii}\left\langle \check{R}(e_{i},e_{\mu})\vec{e}_{\mu},e_{i}\right\rangle 
\end{equation}
The term $i=\mu$ vanishes due to the anti-symmetry of $\check{R}$.
The $(n-1)$ non-zero terms are each called a \textbf{sectional curvature}\index{sectional curvature},
which in general is defined as%
\begin{comment}
See Gravitational Curvature by Frankel
\end{comment}
\begin{equation}
\begin{aligned}K(v,w) & \equiv\frac{\left\langle \check{R}(v,w)\vec{w},v\right\rangle }{\left\langle v,v\right\rangle \left\langle w,w\right\rangle -\left\langle v,w\right\rangle ^{2}}\\
\Rightarrow K(e_{i},e_{j}) & =g_{ii}g_{jj}\left\langle \check{R}(e_{i},e_{j})\vec{e}_{j},e_{i}\right\rangle \\
\Rightarrow\mathrm{Ric}(e_{\mu}) & =\sum_{i\neq\mu}g_{\mu\mu}K(e_{i},e_{\mu})\\
\Rightarrow R & =\sum_{j}g_{jj}\mathrm{Ric}(e_{j})\\
 & =\sum_{i\neq j}K(e_{i},e_{j})\\
 & =2\sum_{i<j}K(e_{i},e_{j}).
\end{aligned}
\end{equation}
Note that the sectional curvature is not a 2-form since it is not
linear in its arguments; in fact it is constructed to only depend
on the plane defined by them, and therefore is symmetric and defined
to vanish for equal arguments. Thus for a Riemannian manifold, the
Ricci function of a unit vector $\mathrm{Ric}(\hat{v})$ can be viewed
as $(n-1)$ times the average of the sectional curvatures of the planes
that include $\hat{v}$, and the Ricci scalar can be viewed as $n$
times the average of all the Ricci functions. For a pseudo-Riemannian
manifold, the Ricci scalar is twice the sum of all sectional curvatures,
or $n(n-1)$ times the average of all sectional curvatures, whose
count is the binomial coefficient $n$ choose 2 or $n(n-1)/2$.

The sectional curvatures completely determine the Riemann tensor,
but in general the Ricci tensor alone does not for manifolds of dimension
greater than 3. However, the Riemann tensor is determined by the Ricci
tensor together with the \textbf{Weyl curvature tensor}\index{Weyl curvature tensor}
(AKA Weyl tensor\index{Weyl tensor}, conformal tensor\index{conformal tensor}),
whose definition (not reproduced here) removes all contractions of
the Riemann tensor, so that it is the ``trace-free part of the curvature''
(i.e. all of its contractions vanish). The Weyl tensor is only defined
and non-zero for dimensions $n>3$. 

The \textbf{Einstein tensor}\index{Einstein tensor} is defined as
\begin{equation}
\begin{aligned}G(v,w) & \equiv\mathrm{Ric}(v,w)-\frac{R}{2}g(v,w)\\
G_{ab} & =R_{ab}-\frac{R}{2}g_{ab}.
\end{aligned}
\end{equation}
If we define $G\equiv g^{ab}G_{ab}$ then we find that $R_{ab}=G_{ab}-Gg_{ab}/(n-2)$,
so that the Einstein tensor vanishes iff the Ricci tensor does. Now,
for zero torsion the Einstein tensor is symmetric, and by the spectral
theorem can be diagonalized at a given point in an orthonormal basis,
which also diagonalizes the Ricci tensor. In terms of the sectional
curvature, we have 
\begin{equation}
\begin{aligned}G(e_{\mu},e_{\mu}) & =-g_{\mu\mu}\sum_{\begin{subarray}{c}
i<j\\
i,j\neq\mu
\end{subarray}}K(e_{i},e_{j}).\end{aligned}
\end{equation}
Thus for a Riemannian manifold, the Einstein tensor $G(\hat{v},\hat{v})$
applied to a unit vector twice can be viewed as $-\left\langle \hat{v},\hat{v}\right\rangle (n-1)(n-2)/2$
times the average of the sectional curvatures of the planes orthogonal
to $\hat{v}$. Using the second Bianchi identity with zero torsion
it can be shown (\cite{FrankelGravCurv} pp. 80-81) that the Einstein
tensor is also ``divergenceless,'' i.e. 
\begin{equation}
\nabla_{a}G^{ab}=0.
\end{equation}
For each value of $b$ in an orthonormal frame, this relation expressed
in terms of the Riemann curvature tensor can be seen to be equivalent
to the second Bianchi identity. Recall that unless the metric is flat,
there is no conserved quantity which can be associated with this vanishing
``divergence'' for a Lorentzian metric.

\noindent %
\begin{framed}%
\noindent $\triangle$ Frequent references to the divergencelessness
of the Einstein tensor being related to a conserved quantity usually
refer to some kind of particular context; one simple one is that in
the limit of zero curvature or infinitesimal volume, there is a set
of conserved quantities due to the above equation.\end{framed}

\subsection{Curvature and geodesics}

Geometrically, the Ricci function $\mathrm{Ric}(v)$ at a point $p\in M^{n}$
can be seen to measure the extent to which the area defined by the
geodesics emanating from the $(n-1)$-surface perpendicular to $v$
changes in the direction of $v$. Considering the three dimensional
case in an orthonormal frame (and again dropping the hats in $\hat{e}_{i}$
to avoid clutter), we have
\begin{equation}
\begin{aligned}\mathrm{Ric}(e_{2}) & =\left\langle \check{R}(e_{1},e_{2})\vec{e}_{2},e_{1}\right\rangle +\left\langle \check{R}(e_{3},e_{2})\vec{e}_{2},e_{3}\right\rangle \\
 & =K(e_{1},e_{2})+K(e_{3},e_{2}).
\end{aligned}
\end{equation}
If we form a cube made from parallel transported vectors as we did
for the first Bianchi identity, we can see that each sectional curvature
term in $\mathrm{Ric}(e_{2})$ takes an edge of the cube and measures
the length of the difference between the cube-aligned component of
its parallel transport in the $e_{2}$ direction and the edge of the
cube at a point parallel transported in the $e_{2}$ direction. 

\begin{figure}[H]
\noindent \begin{centering}
\includegraphics[width=0.7\columnwidth]{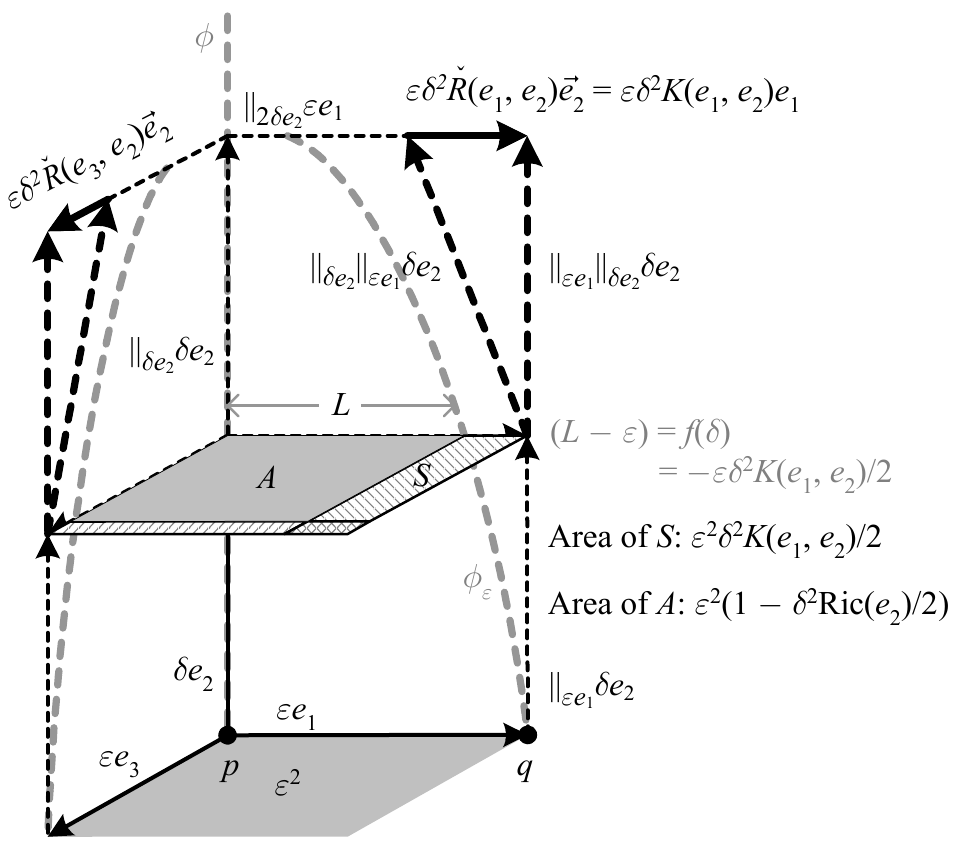}
\par\end{centering}
\caption{Each sectional curvature measures the convergence of geodesics, while
their sum forms the Ricci curvature function, which measures the change
in the area of the $(n-1)$-surface formed by geodesics perpendicular
to its argument. In the figure we assume without loss of generality
(see below) that $\check{R}(e_{1},e_{2})\vec{e}_{2}$ is parallel
to $e_{1}$.}
\end{figure}

The figure above details the sectional curvature $K(e_{1},e_{2})=\beta^{1}\check{R}(e_{1},e_{2})\vec{e}_{2}$
assuming that $\check{R}(e_{1},e_{2})\vec{e}_{2}$ is parallel to
$e_{1}$, so that $\left\langle \check{R}(e_{1},e_{2})\vec{e}_{2},e_{1}\right\rangle =\left\Vert \check{R}(e_{1},e_{2})\vec{e}_{2}\right\Vert $.
The parallel transport of $e_{2}$ along itself is depicted as parallel,
so that the geodesic parametrized by arclength $\phi(t)$ is a straight
line in the figure. The vector $\parallel_{\delta e_{2}}\parallel_{\varepsilon e_{1}}\delta e_{2}$
is the parallel transport of $\parallel_{\varepsilon e_{1}}\delta e_{2}$
by $\delta$ in the direction parallel to $e_{2}$, and therefore
the geodesic $\phi_{\varepsilon}(t)$ tangent to $\parallel_{\varepsilon e_{1}}\delta e_{2}$
at $q$ has tangent $\parallel_{\delta e_{2}}\parallel_{\varepsilon e_{1}}\delta e_{2}$
after moving a distance $\delta$. If we consider the function $f(t)$
whose value at $t=\delta$ is the quantity $(L-\varepsilon)$ in the
figure (i.e. $f(t)$ measures the offset of the geodesic from the
right edge of the stack of parallel cubes), its derivative is the
slope of the tangent, so that to lowest order in $t$ we have 
\begin{equation}
\begin{aligned}\dot{f}(t) & =-\varepsilon t^{2}K(e_{1},e_{2})/t\\
 & =-\varepsilon tK(e_{1},e_{2})\\
\Rightarrow f(t) & =-\varepsilon t^{2}K(e_{1},e_{2})/2.
\end{aligned}
\end{equation}
We can generalize this logic to arbitrary unit vectors $\hat{v}$
and $\hat{w}$ to conclude that $K(\hat{v},\hat{w})/2$ is the ``fraction
by which the geodesic parallel to $\hat{w}$ with separation direction
$\hat{v}$ bends towards $\hat{w}$.'' More precisely, in terms of
the distance function and the exponential map, to order $\varepsilon$
and $\delta^{2}$ we have 
\begin{equation}
d\left(\mathrm{exp}(\delta\hat{w}),\mathrm{exp}(\delta\parallel_{\varepsilon\hat{v}}\hat{w})\right)=\varepsilon\left(1-\frac{\delta^{2}}{2}K(\hat{v},\hat{w})\right).
\end{equation}
In the general case $L$ in the figure is the distance between two
geodesics infinitesimally separated in the $\hat{v}$ direction, so
if we define $L(t)$ as this distance at any point along the parametrized
geodesic tangent to $\hat{w}$, the above becomes 
\begin{equation}
\begin{aligned}L(t) & =L(0)\left(1-\frac{t^{2}}{2}K(\hat{v},\hat{w})\right)\\
\Rightarrow\left.\frac{\ddot{L}}{L}\right|_{t=0} & =-K(\hat{v},\hat{w}),
\end{aligned}
\end{equation}
where the double dots indicate the second derivative with respect
to $t$. Thus $K(\hat{v},\hat{w})$ is ``the acceleration of two
parallel geodesics in the $\hat{w}$ direction with initial separation
$\hat{v}$ towards each other as a fraction of the initial gap.''

Now, the distance $\left|\varepsilon-L\right|=\varepsilon\delta^{2}K(e_{1},e_{2})/2$
defines a strip $S$ bordering the surface orthogonal to $e_{2}$
a distance $\delta$ in the $e_{2}$ direction. This strip thus has
an area $\varepsilon^{2}\delta^{2}K(e_{1},e_{2})/2$. If we sum this
with the other strip of area $\varepsilon^{2}\delta^{2}K(e_{3},e_{2})/2$,
to order $\varepsilon^{2}$ and $\delta^{2}$ we measure the extent
to which the area $A$ defined by the geodesics emanating from the
surface perpendicular to $e_{2}$ changes in the direction of $e_{2}$.
But the sum of sectional curvatures is just the Ricci function, so
that in general $\mathrm{Ric}(v)/2$ is the ``fraction by which the
area defined by the geodesics emanating from the $(n-1)$-surface
perpendicular to $v$ changes in the direction of $v$.'' More precisely,
we can follow the same logic as above, defining the ``infinitesimal
geodesic $(n-1)$-area'' $A(t)$ along a parametrized geodesic tangent
to $v$, so that to order $\varepsilon^{2}$ and $t^{2}$ we have
\begin{equation}
\begin{aligned}A(t) & =\varepsilon^{2}\left(1-\frac{t^{2}}{2}\mathrm{Ric}(v)\right)\\
\Rightarrow\left.\frac{\ddot{A}}{A}\right|_{t=0} & =-\mathrm{Ric}(v).
\end{aligned}
\end{equation}
Thus $\mathrm{Ric}(v)$ is ``the acceleration of the parallel geodesics
emanating from the $(n-1)$-surface perpendicular to $v$ towards
each other as a fraction of the initial surface.'' Note that if our
previous assumption that $\check{R}(e_{1},e_{2})\vec{e}_{2}$ is parallel
to $e_{1}$ is dropped, the only impact is that of an $e_{3}$ component
on the area calculation; to address this, a more accurate picture
would be to extend the area to include all four quadrants defined
by both negative and positive values of $e_{1}$ and $e_{3}$, in
which case any change in area due to an $e_{3}$ component cancels.
In the case of a pseudo-Riemannian manifold, ``areas'' and ``volumes''
become less geometric concepts; however, we have a clear picture in
the case of a Lorentzian manifold that the Ricci function applied
to a time-like vector $v\equiv\partial/\partial x^{0}=\partial/\partial t$
tells us how the infinitesimal space-like volume $V$ of free-falling
particles (i.e. following geodesics) changes over time according to
\begin{equation}
\begin{aligned}\left.\frac{\ddot{V}}{V}\right|_{t=0} & =-\mathrm{Ric}(v)\\
 & =-R_{00}\\
 & =-R^{\mu}{}_{0\mu0}.
\end{aligned}
\end{equation}

\subsection{Jacobi fields and volumes}

Now let us consider a vector field $J(t)$ along the geodesic $\phi(t)$
such that $J(0)\equiv J\left|_{p}\right.=J\left|_{\phi(0)}\right.=e_{1}$
and $J(\delta)\equiv J\left|_{\phi(\delta)}\right.=(L/\varepsilon)\parallel_{\delta e_{2}}e_{1}$,
i.e. $J$ is the vector field ``between adjacent geodesics.'' 

\begin{figure}[H]
\noindent \begin{centering}
\includegraphics[width=0.65\columnwidth]{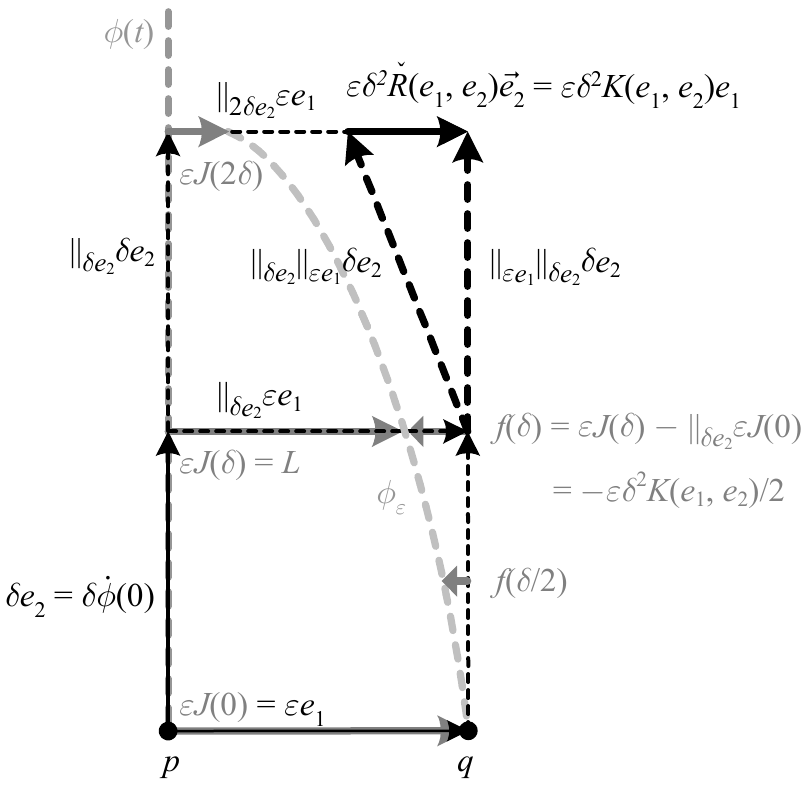}
\par\end{centering}
\caption{A Jacobi field is the vector field between adjacent geodesics, whose
construction creates a relationship between the covariant derivative
and the sectional curvature.}
\end{figure}
Then the function 
\begin{equation}
\begin{aligned}f(t) & =-\varepsilon t^{2}K(e_{1},e_{2})/2\\
 & =-\varepsilon t^{2}K(J,\dot{\phi})/2
\end{aligned}
\end{equation}
is the difference between $J$ and its parallel transport in the direction
tangent to $\phi$, i.e. it is the value of the covariant derivative
along $\phi$. Since this difference is of order $t^{2}$, at $t=0$
we have 
\begin{equation}
\mathrm{D}_{t}^{2}J=-K(J,\dot{\phi}),
\end{equation}
or dropping the assumption that $\check{R}(e_{1},e_{2})\vec{e}_{2}$
is parallel to $e_{1}$, 
\begin{equation}
\frac{\mathrm{D}^{2}J}{\mathrm{d}t^{2}}+\check{R}(J,\dot{\phi})\vec{\dot{\phi}}=0.
\end{equation}
Considered as an equation for all $J(t)$, this is called the \textbf{Jacobi
equation}\index{Jacobi equation}, with the vector field $J(t)$ that
satisfies it called a \textbf{Jacobi field}\index{Jacobi field}.
A more precise way to generalize our construction of $J$ is to define
a one-parameter family of geodesics $\phi_{s}(t)$, so that 
\begin{equation}
J(t)=\left.\frac{\partial\phi_{s}(t)}{\partial s}\right|_{s=0}.
\end{equation}
If $M$ is complete then every Jacobi field can be expressed in this
way for some family of geodesics. 

If we then consider the Jacobi fields $J_{v}(t)$ corresponding to
the geodesics $\phi_{v}(t)$ of tangent vectors $\left\Vert v\right\Vert =1$
parametrized by arclength and such that to order $t$ we have $\left\Vert J_{v}(1)\right\Vert =1$,
it can be shown (\cite{DoCarmo} pp. 114-115) that to order $t^{3}$
we have $\left\Vert J_{v}(t)\right\Vert =t(1-t^{2}K(J_{v},\dot{\phi}_{v})/6)$. 

\begin{figure}[H]
\noindent \begin{centering}
\includegraphics[width=0.7\columnwidth]{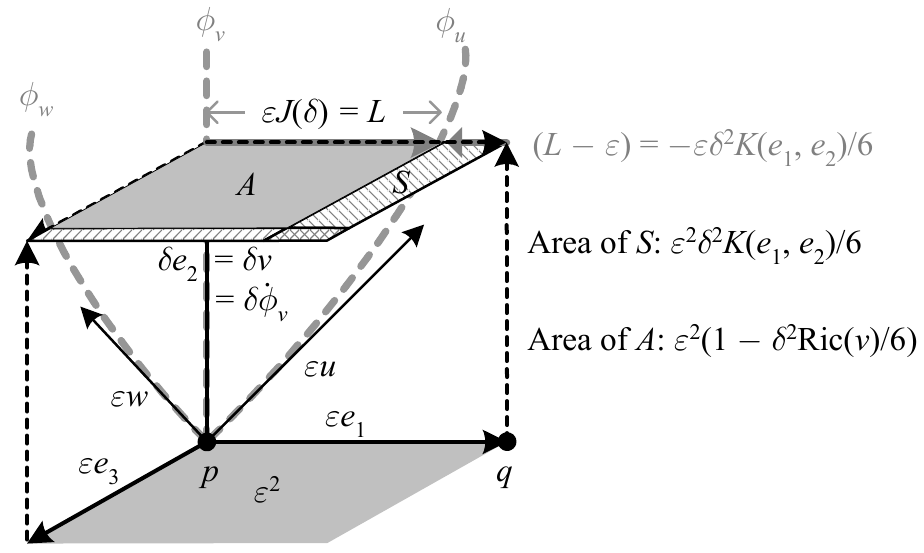}
\par\end{centering}
\caption{The infinitesimal geodesic area element derived from the Jacobi field
between radial geodesics.}
\end{figure}
This means that if we apply the previous reasoning for parallel geodesics
to these radial geodesics we have an ``infinitesimal geodesic $(n-1)$-area
element'' $A(t)=t^{2}(1-t^{2}\mathrm{Ric}(v)/6)$. Integrating this
over all values of $v$ gives for small $t=\varepsilon$ the surface
area of a geodesic $n$-ball of radius $\varepsilon$, which we denote
$\partial B_{\varepsilon}(M^{n})$. But this integral just averages
the values of the Ricci function, which is the Ricci scalar over the
dimension $n$, so that to order $\varepsilon^{2}$ we have 
\begin{equation}
\frac{\partial B_{\varepsilon}(M^{n})}{\partial B_{\varepsilon}(\mathbb{R}^{n})}=1-\frac{\varepsilon^{2}}{6n}R,
\end{equation}
and integrating over the radius we find (see \cite{GrayGeodesicBalls})
a similar relation for the volume of a geodesic sphere compared to
a Euclidean one of
\begin{equation}
\frac{B_{\varepsilon}(M^{n})}{B_{\varepsilon}(\mathbb{R}^{n})}=1-\frac{\varepsilon^{2}}{6(n+2)}R.
\end{equation}
Thus $\varepsilon^{2}R/6n$ is ``the fraction by which the surface
area of a geodesic $n$-ball of radius $\varepsilon$ is smaller than
it would be under a flat metric,'' and $\varepsilon^{2}R/6(n+2)$
is ``the fraction by which the volume of a geodesic $n$-ball of
radius $\varepsilon$ is smaller than it would be under a flat metric.''

Alternatively, we can use Riemann normal coordinates to express $v$
in our ``infinitesimal geodesic $(n-1)$-area element,'' whereupon
following similar logic to the above we find that, at points close
to the origin of our coordinates, to order $\left\Vert x\right\Vert ^{2}$
the volume element is 
\begin{equation}
\mathrm{d}V=\left(1-\frac{1}{6}R_{\mu\nu}x^{\mu}x^{\nu}\right)\mathrm{d}x^{1}\cdots\mathrm{d}x^{n},
\end{equation}
or using the expression of the volume element in terms the square
root of the determinant of the metric, again to order $\left\Vert x\right\Vert ^{2}$
we find that 
\begin{equation}
g_{\mu\nu}=\delta_{\mu\nu}-\frac{1}{3}R_{\mu\lambda\nu\sigma}x^{\lambda}x^{\sigma}.
\end{equation}

As is apparent from their definitions, the Ricci tensor and function
do not depend on the metric. We can attempt to find a metric-free
geometric interpretation by considering the concept of a \textbf{parallel
volume form}\index{parallel volume form}. This is defined as a volume
form which is invariant under parallel transport. We immediately see
that it is only possible to define such a form if parallel transport
around a loop does not alter volumes, i.e. that $\check{R}$ must
be $o(r,s)$-valued. This means that the connection is metric compatible,
so we can define one if we wish; but if we do not, and assume zero
torsion so that the Ricci tensor is symmetric, then our logic for
volumes remains valid and we can still take a metric-free view of
the expression for $\mathrm{d}V$ above as expressing the geodesic
volume as measured by the parallel volume form. Note that unlike the
Ricci tensor and function, the definitions here of the individual
sectional curvatures and scalar curvature do depend upon the metric.

\subsection{Summary}

Below, we review the intuitive meanings of the various relations we
have defined on a Riemannian manifold.

\begin{table}[H]
\begin{tabular*}{1\columnwidth}{@{\extracolsep{\fill}}|l|>{\raggedright}m{0.58\columnwidth}|}
\hline 
Relation & Meaning\tabularnewline
\hline 
\hline 
$\mathrm{div}(u)\mathrm{d}V=L_{u}\mathrm{d}V$ & $\mathrm{div}(u)$ is the fraction by which a unit volume changes
when transported by the flow of $u$.\tabularnewline
\hline 
$\begin{aligned}\int_{V}\mathrm{div}(u)\mathrm{d}V & =\int_{\partial V}i_{u}\mathrm{d}V\\
 & =\int_{\partial V}\left\langle u,n\right\rangle \mathrm{d}S
\end{aligned}
$ & The change in a volume due to transport by the flow of $u$ is equal
to the net flow of $u$ across that volume's boundary.\tabularnewline
\hline 
$\mathrm{div}(u)=0$ & $u$ having zero divergence means the flow of $u$ leaves volumes
unchanged, or the net flow of $u$ across the boundary of a volume
is zero.\tabularnewline
\hline 
$j\equiv\rho u$, $\rho$ is the density of $Q$ & The current vector $j$ is the vector whose length is the amount of
$Q$ per unit time crossing a unit area perpendicular to $j$\tabularnewline
\hline 
$\begin{aligned}\frac{\mathrm{d}q}{\mathrm{d}t} & =\Sigma-\int_{\partial V}\left\langle j,n\right\rangle \mathrm{d}S\end{aligned}
$ & The change in $q$ (the amount of $Q$ within $V$) equals the amount
generated less the amount which passes through $\partial V$.\tabularnewline
\hline 
$\begin{aligned}\frac{\partial\rho}{\partial t} & =\sigma-\mathrm{div}(j)\end{aligned}
$ & The change in the density of $Q$ at a point equals the amount generated
less the amount that moves away.\tabularnewline
\hline 
\end{tabular*}

\caption{Divergence and continuity relations and their intuitive meanings.}
\end{table}

\begin{table}[H]
\begin{tabular*}{1\columnwidth}{@{\extracolsep{\fill}}|l|>{\raggedright}m{0.56\columnwidth}|}
\hline 
Relation & Meaning\tabularnewline
\hline 
\hline 
$R\equiv g^{ab}R_{ab}$ & The Ricci scalar is $n$ times the average of the Ricci function on
the set of unit tangent vectors.\tabularnewline
\hline 
$\mathrm{Ric}(e_{\mu})=\underset{i\neq\mu}{\sum}g_{\mu\mu}K(e_{i},e_{\mu})$ & The Ricci function of a unit vector is $(n-1)$ times the average
of the sectional curvatures of the planes that include the vector.\tabularnewline
\hline 
$R=\underset{j}{\sum}g_{jj}\mathrm{Ric}(e_{j})$ & The Ricci scalar is $n$ times the average of all the Ricci functions.\tabularnewline
\hline 
$R=2\underset{i<j}{\sum}K(e_{i},e_{j})$ & The Ricci scalar is $n(n-1)$ times the average of all sectional curvatures.\tabularnewline
\hline 
$\begin{aligned}G(e_{\mu},e_{\mu}) & =-g_{\mu\mu}\sum_{\begin{subarray}{c}
i<j\\
i,j\neq\mu
\end{subarray}}K(e_{i},e_{j})\end{aligned}
$ & The Einstein tensor applied to a unit vector $\hat{v}$ twice is $-\left\langle \hat{v},\hat{v}\right\rangle (n-1)(n-2)/2$
times the average of the sectional curvatures of the planes orthogonal
to the vector.\tabularnewline
\hline 
$\begin{aligned}d\left(\mathrm{exp}(\delta\hat{w}),\mathrm{exp}(\delta\parallel_{\varepsilon\hat{v}}\hat{w})\right)\\
=\varepsilon\left(1-\frac{\delta^{2}}{2}K(\hat{v},\hat{w})\right)
\end{aligned}
$ & $K(\hat{v},\hat{w})/2$ is the fraction by which the geodesic parallel
to $\hat{w}$ starting $\hat{v}$ away bends towards $\hat{w}$. \tabularnewline
\hline 
$\left.\frac{\ddot{L}}{L}\right|_{t=0}=-K(\hat{v},\hat{w})$ & $K(\hat{v},\hat{w})$ is the acceleration of two parallel geodesics
in the $\hat{w}$ direction with initial separation direction $\hat{v}$
towards each other as a fraction of the initial gap.\tabularnewline
\hline 
$\left.\frac{\ddot{A}}{A}\right|_{t=0}=-\mathrm{Ric}(v)$ & $\mathrm{Ric}(v)/2$ is the fraction by which the area defined by
the geodesics emanating from the $(n-1)$-surface perpendicular to
$v$ changes in the direction of $v$.

$\mathrm{Ric}(v)$ is the acceleration of the parallel geodesics emanating
from the $(n-1)$-surface perpendicular to $v$ towards each other
as a fraction of the initial surface.\tabularnewline
\hline 
$\begin{aligned}\frac{\partial B_{\varepsilon}(M^{n})}{\partial B_{\varepsilon}(\mathbb{R}^{n})} & =1-\frac{\varepsilon^{2}}{6n}R\end{aligned}
$ & $\varepsilon^{2}R/6n$ is the fraction by which the surface area of
a geodesic $n$-ball of radius $\varepsilon$ is smaller than it would
be under a flat metric.\tabularnewline
\hline 
$\begin{aligned}\frac{B_{\varepsilon}(M^{n})}{B_{\varepsilon}(\mathbb{R}^{n})} & =1-\frac{\varepsilon^{2}}{6(n+2)}R\end{aligned}
$ & $\varepsilon^{2}R/6(n+2)$ is the fraction by which the volume of
a geodesic $n$-ball of radius $\varepsilon$ is smaller than it would
be under a flat metric.\tabularnewline
\hline 
\end{tabular*}

\caption{Relations defined on a Riemannian manifold $M^{n}$ and their intuitive
meanings.}
\end{table}

\appendix

\section*{Appendices \addcontentsline{toc}{section}{Appendices}}

\section{\label{sec:Tensors-and-forms}Tensors and forms}

It is assumed the reader is familiar with vector spaces and inner
products, as well as the tensor product and the exterior product (AKA
wedge product, Grassmann product). In the following, we will limit
our discussion to finite-dimensional real vector spaces $V=\mathbb{R}^{n}$;
generalization to complex scalars is straightforward.

\subsection{\label{subsec:The-structure-of-the-dual-space}The structure of the
dual space }

Given a finite-dimensional vector space $V$, the \textbf{dual space}\index{dual space}
$V^{*}$ is defined to be the set of linear mappings from $V$ to
the scalars (AKA the linear functionals\index{linear functional}
on $V$). The elements of $V^{*}$ can be added together and multiplied
by scalars, so $V^{*}$ is also a vector space, with the same dimension
as $V$.

\noindent %
\begin{framed}%
\noindent $\triangle$ Note that in general, the word ``dual\index{dual}''
is used for many concepts in mathematics; in particular, the dual
space has no relation to the Hodge dual (defined below). \end{framed}

An element $\varphi\colon V\to\mathbb{R}$ of $V^{*}$ is called a
\textbf{1-form}\index{1-form}. Given a pseudo inner product on $V$,
we can construct an isomorphism between $V$ and $V^{*}$ defined
by 
\begin{equation}
v\mapsto\left\langle v,\;\right\rangle ,
\end{equation}
i.e. $v\in V$ is mapped to the element of $V^{*}$ which maps any
vector $w\in V$ to $\left\langle v,w\right\rangle $. This isomorphism
then induces a corresponding pseudo inner product on $V^{*}$ defined
by 
\begin{equation}
\left\langle \left\langle v,\;\right\rangle ,\left\langle w,\;\right\rangle \right\rangle \equiv\left\langle v,w\right\rangle .
\end{equation}

An equivalent way to set up this isomorphism is to choose a basis
$e_{\mu}$ of $V$, and then form the \textbf{dual basis}\index{dual basis}
$\beta^{\nu}$ of $V^{*}$, defined to satisfy $\beta^{\lambda}(e_{\mu})=?\delta^{\lambda}{}_{\mu}?$.
The isomorphism between $V$ and $V^{*}$ is then defined by the correspondence
\begin{equation}
v=v^{\mu}e_{\mu}\mapsto(\eta_{\mu\lambda}v^{\mu})\beta^{\lambda}\equiv v_{\lambda}\beta^{\lambda},
\end{equation}
which is identical to the isomorphism induced by the pseudo inner
product on $V$ that makes $e_{\mu}$ orthonormal. Here we have used
the \textbf{Einstein summation convention}\index{summation convention},
i.e. a repeated index implies summation. Note that if $\left\langle e_{\mu},e_{\mu}\right\rangle =-1$
then $e_{\mu}\mapsto-\beta^{\mu}$. This isomorphism and its inverse
(usually in the context of Riemannian manifolds) are called the \textbf{musical
isomorphisms}\index{musical isomorphisms}, where if $v=v^{\mu}e_{\mu}$
and $\varphi=\varphi_{\mu}\beta^{\mu}$ we write
\begin{equation}
\begin{aligned}v^{\flat} & \equiv\left\langle v,\;\right\rangle \\
 & =\left(\eta_{\mu\lambda}v^{\lambda}\right)\beta^{\mu}\\
 & =v_{\mu}\beta^{\mu}\\
\varphi^{\sharp} & \equiv\left\langle \varphi,\;\right\rangle \\
 & =\left(\eta^{\mu\lambda}\varphi_{\lambda}\right)e_{\mu}\\
 & =\varphi^{\mu}e_{\mu}
\end{aligned}
\end{equation}
and call the $v^{\flat}$ the \textbf{flat}\index{flat} of the vector
$v$ and $\varphi^{\sharp}$ the \textbf{sharp}\index{sharp} of the
1-form $\varphi$.

\noindent %
\begin{framed}%
\noindent $\triangle$ It is important to remember that when the inner
product is not positive definite, the signs of components may change
under these isomorphisms. If the components are in terms of an arbitrary
(non-orthonormal) basis, then as we will see in Section \ref{subsec:Abstract-index-notation},
the components change their values as well, since $\eta_{\lambda\mu}$
is replaced by the metric tensor in the above analysis.\end{framed}

\noindent Note that since $\beta^{\lambda}(e_{\mu})=?\delta^{\lambda}{}_{\mu}?$
and $\left\langle e_{\mu},e_{\lambda}\right\rangle =\eta_{\mu\lambda}$
we have 
\begin{equation}
\begin{aligned}\varphi(v) & =\varphi_{\lambda}\beta^{\lambda}(v^{\mu}e_{\mu})\\
 & =\varphi_{\mu}v^{\mu}\\
 & =\eta_{\mu\lambda}\varphi^{\lambda}v^{\mu}\\
 & =\left\langle \varphi^{\sharp},v\right\rangle .
\end{aligned}
\end{equation}

\begin{figure}[H]
\noindent \begin{centering}
\includegraphics[width=0.5\columnwidth]{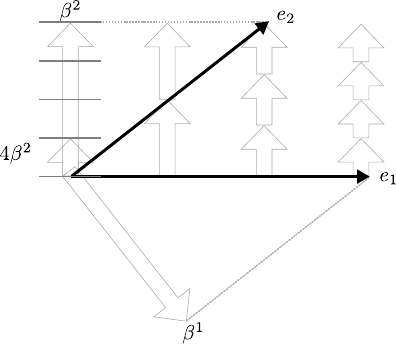}
\par\end{centering}
\caption{\label{fig:1-forms-vs-vectors}Depicting a 1-form $\varphi$ as the
associated vector $\varphi^{\Uparrow}$ or as a density of surfaces
has consequences that can be non-intuitive. When orthogonality corresponds
to right angles in a figure, an orthonormal basis and its dual basis
appear as identical arrows; in the figure, we see that for a non-orthonormal
basis, the dual basis does not appear to either be parallel to the
basis or to have identical lengths. We also see that quadrupling the
value of the 1-form means quartering its length in the figure, or
equivalently quadrupling the density of surfaces pierced by arrows.
This means that when depicting a linearly changing 1-form as above,
the length $L$ of the associated vector changes like $L\protect\mapsto L/(1+r\varepsilon)$
for some scaling factor $r$, which doesn't appear linear as a vector
representation would, whose length changes like $L\protect\mapsto L(1+r\varepsilon)$.}
\end{figure}

\noindent A 1-form acting on a vector can thus be viewed as yielding
a projection. Specifically, with a positive definite inner product,
$\varphi(v)/\left\Vert \varphi^{\sharp}\right\Vert $ is the length
of the projection of $v$ onto the ray defined by $\varphi^{\sharp}$.
If we then define 
\begin{equation}
\varphi^{\Uparrow}\equiv\varphi^{\sharp}/\left\Vert \varphi^{\sharp}\right\Vert ^{2},
\end{equation}
the length of this projection as a multiple of $\left\Vert \varphi^{\Uparrow}\right\Vert $
is 
\begin{equation}
\left\langle \varphi^{\Uparrow},v\right\rangle /\left\Vert \varphi^{\Uparrow}\right\Vert =\varphi(v).
\end{equation}
We can therefore represent a 1-form $\varphi$ as a ``receptacle''
$\varphi^{\Uparrow}$ which when applied to a vector ``arrow'' argument
$v$ yields the number of receptacles covered by the projection of
$v$ onto $\varphi^{\sharp}$, which is the value of $\varphi(v)$.
The advantage of this approach is that values can be calculated from
a figure absent a length scale. Another common graphical device is
to represent $\varphi$ as a density of ``surfaces'' where the value
of $\varphi(v)$ is the number of surfaces ``pierced'' by the arrow.
Figure \ref{fig:1-forms-vs-vectors} covers some non-intuitive aspects
of these visualizations.

\noindent %
\begin{framed}%
\noindent $\triangle$ It is important to remember that the practice
of depicting a 1-form $\varphi$ as the associated vector $\varphi^{\Uparrow}$
or as a density of surfaces has consequences that can be non-intuitive.\end{framed}

It is important to note that there is no \textbf{canonical isomorphism}\index{canonical isomorphism}
between $V$ and $V^{*}$, i.e. we cannot uniquely associate a 1-form
with a given vector without introducing extra structure, namely an
inner product or a preferred basis. Either structure will do: a choice
of basis is equivalent to the definition of the unique inner product
on $V$ that makes this basis orthonormal, which then induces the
same isomorphism as that induced by the dual basis.

In contrast, a canonical isomorphism $V\cong V^{**}$ can be made
via the association $v\in V\leftrightarrow\xi\in V^{**}$ with $\mathbb{\xi}\colon V^{*}\to\mathbb{R}$
defined by $\xi\left(\varphi\right)\equiv\varphi\left(v\right)$.
Thus $V$ and $V^{**}$ can be completely identified (for a finite-dimensional
vector space), and we can view $V$ as the dual of $V^{*}$, with
vectors regarded as linear mappings on 1-forms.

Vector components are often viewed as a column vector, which means
that 1-forms act on vector components as row vectors (which then are
acted on by matrices from the right). Under a change of basis we then
have the following relationships:

\begin{table}[H]
\begin{tabular*}{1\columnwidth}{@{\extracolsep{\fill}}|l|>{\raggedright}p{0.31\columnwidth}|>{\raggedright}p{0.35\columnwidth}|}
\hline 
 & Index notation & Matrix notation\tabularnewline
\hline 
\hline 
Basis & $e_{\mu}^{\prime}=A^{\lambda}{}_{\mu}e_{\lambda}$ & $e^{\prime}=eA$\tabularnewline
\hline 
Dual basis & $\beta^{\prime\mu}=(A^{-1})^{\mu}{}_{\lambda}\beta^{\lambda}$ & $\beta^{\prime}=A^{-1}\beta$\tabularnewline
\hline 
Vector components & $\left(v^{\mu}\right)^{\prime}=(A^{-1})^{\mu}{}_{\lambda}v^{\lambda}$ & $v^{\prime}=A^{-1}v$\tabularnewline
\hline 
1-form components & $\left(\varphi_{\mu}\right)^{\prime}=A^{\lambda}{}_{\mu}\varphi_{\lambda}$ & $\varphi^{\prime}=\varphi A$\tabularnewline
\hline 
\end{tabular*}

\caption{Transformations under a change of basis.}

Notes: We notationally distinguish between a changed vector $e_{\mu}^{\prime}$
and an unchanged vector with changed components $\left(v^{\mu}\right)^{\prime}$.
A 1-form will sometimes be viewed as a column vector, i.e. as the
transpose of the row vector (which is the sharp of the 1-form under
a Riemannian signature). Then we have $(\varphi^{\prime})^{T}=(\varphi A)^{T}=A^{T}\varphi^{T}$. 
\end{table}

\subsection{\label{subsec:Tensors}Tensors }

A \textbf{tensor}\index{tensor} of \textbf{type}\index{tensor type}
(AKA valence\index{tensor valence}) $\left(m,n\right)$ is defined
to be an element of the \textbf{tensor space}\index{tensor space}

\begin{equation}
V_{m,n}\equiv\left(V\otimes\dotsb\left(m\:\textrm{times}\right)\dotsb\otimes V\right)\otimes\left(V^{*}\otimes\dotsb\left(n\:\textrm{times}\right)\dotsb\otimes V^{*}\right).
\end{equation}
A \textbf{pure tensor}\index{pure tensor} (AKA simple or decomposable
tensor\index{decomposable tensor}\index{simple tensor}) of type\index{tensor type}
$\left(m,n\right)$ is one that can be written as the tensor product
of $m$ vectors and $n$ 1-forms; thus a general tensor is a sum of
pure tensors. The integer $\left(m+n\right)$ is called the \textbf{order}\index{tensor order}
(AKA degree\index{tensor degree}, rank) of the tensor, while the
tensor \textbf{dimension}\index{dimension} is that of $V$. Vectors
and 1-forms are then tensors of type $\left(1,0\right)$ and $\left(0,1\right)$.
The \textbf{rank}\index{tensor rank} (sometimes used to refer to
the order) of a tensor is the minimum number of pure tensors required
to express it as a sum. In ``tensor language\index{tensor language}''
vectors $v\in V$ are called \textbf{contravariant vectors}\index{contravariant vector}
and 1-forms $\varphi\in V^{*}$ are called \textbf{covariant vector}\index{covariant vector}\textbf{s}
(AKA covectors\index{covector}). A tensor of type $\left(k,0\right)$
is then called a \textbf{contravariant tensor}\index{contravariant tensor},
with \textbf{covariant tensors}\index{covariant tensor} being of
type $\left(0,k\right)$, and other tensor types being called \textbf{mixed
tensors}\index{mixed tensor}. Scalars can be considered tensors of
type $\left(0,0\right)$.

\noindent %
\begin{framed}%
\noindent $\triangle$ As noted above, the meanings of tensor rank
and order are often swapped in the literature. Another potential source
of confusion is that a mixed tensor is not the opposite of a pure
tensor.\end{framed}

The infinite direct sum of the tensor spaces of every type forms an
associative algebra. This algebra is also called the ``tensor algebra,”
and “tensor” sometimes refers to the general elements of this algebra,
in which case tensors as defined above are called \textbf{homogeneous
tensors}\index{homogeneous tensor}. In this book, we will always
use the term “tensor” to mean homogeneous tensor, while for “tensor
algebra” the inclusion of powers of the dual space will depend upon
context.

\subsection{Tensors as multilinear mappings }

There is an obvious multiplication of two 1-forms: the scalar multiplication
of their values. The resulting object $\varphi\psi\colon V\times V\to\mathbb{R}$
is a nondegenerate bilinear form on $V$. Viewed as an ``outer product''
on $V^{*}$, multiplication is trivially seen to be a bilinear operation,
i.e. $a\left(\varphi+\psi\right)\xi=a\varphi\xi+a\psi\xi$. Thus the
product of two 1-forms is isomorphic to their tensor product.

We can extend this to any tensor by viewing vectors as linear mappings
on 1-forms, and forming the isomorphism 
\begin{equation}
\bigotimes\varphi_{i}\mapsto\prod\varphi_{i}.
\end{equation}
Note that this isomorphism is not unique, since for example any real
multiple of the product would yield a multilinear form as well. However
it is canonical, since the choice does not impose any additional structure,
and is also consistent with considering scalars as tensors of type
$\left(0,0\right)$.

We can thus consider tensors to be multilinear mappings on $V^{*}$
and $V$. In fact, we can view a tensor of type $\left(m,n\right)$
as a mapping from $i<m$ 1-forms and $j<n$ vectors to the remaining
$\left(m-i\right)$ vectors and $\left(n-j\right)$ 1-forms. 

\begin{figure}[H]
\noindent \begin{centering}
\includegraphics[width=0.8\columnwidth]{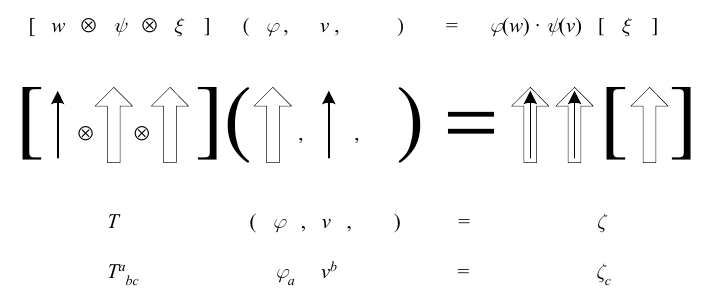}
\par\end{centering}
\caption{Different ways of depicting a pure tensor of type $\left(1,2\right)$.
The first line explicitly shows the tensor as a mapping from a 1-form
$\varphi$ and a vector $v$ to a 1-form $\xi$. The second line visualizes
vectors as arrows, and 1-forms as receptacles that when matched to
an arrow yield a scalar. The third line combines the constituent vectors
and 1-forms of the tensor into a single symbol $T$ while merging
the scalars into $\xi$ to define $\zeta$, and the last line adds
indices (covered in the next section).}
\end{figure}
A general tensor is a sum of pure tensors, so for example a tensor
of the form $\left(u\otimes\varphi\right)+\left(v\otimes\psi\right)$
can be viewed as a linear mapping that takes $\xi$ and $w$ to the
scalar $\xi\left(u\right)\cdot\varphi\left(w\right)+\xi\left(v\right)\cdot\psi\left(w\right)$.
Since the roles of mappings and arguments can be reversed, we can
simplify things further by viewing the arguments of a tensor as another
tensor: 
\begin{equation}
\begin{aligned}\left(u\otimes\varphi\right)\left(\xi\otimes w\right) & \equiv\left(u\otimes\varphi\right)\left(\xi,w\right)\\
 & =\left(\xi\otimes w\right)\left(u,\varphi\right)\\
 & =\xi\left(u\right)\cdot\varphi\left(w\right)
\end{aligned}
\end{equation}

\subsection{\label{subsec:Abstract-index-notation}Abstract index notation }

\textbf{Abstract index notation}\index{abstract index notation} uses
an upper Latin index to represent each contravariant vector component
of a tensor, and a lower index to represent each covariant vector
(1-form) component. We can see from the preceding figure that this
notation is quite compact and clearly indicates the type of each tensor
while re-using letters to indicate what ``slots” are to be used in
the mapping.

The tensor product of two tensors $?S^{a}{}_{b}?\otimes?T^{c}{}_{d}?$
is simply denoted $?S^{a}{}_{b}??T^{c}{}_{d}?$, and in this form
the operation is sometimes called the \textbf{tensor direct product}\index{tensor direct product}.
We may also consider a \textbf{contraction}\index{contraction} 
\begin{equation}
?T^{ab}{}_{bc}?=?T^{a}{}_{c}?,
\end{equation}
where two of the components of a tensor operate on each other to create
a new tensor with a reduced number of indices. For example, if $?T^{ab}{}_{c}?=v^{a}\otimes w^{b}\otimes\varphi_{c}$,
then $?T^{ab}{}_{b}?=\varphi(w)\cdot v^{a}$. Taking the tensor direct
product of two tensors and then contracting all opposite indices is
also called the contraction of the two tensors, i.e. the contraction
of $?S^{ab}{}_{c}?$ and $?T_{def}?$ is $?C_{cf}?=?S^{ab}{}_{c}??T{}_{abf}?$.
The contraction of any two symmetric indices with any two anti-symmetric
indices vanishes, e.g. if the (first) second tensor is (anti) symmetric
in the first two indices then 
\begin{equation}
S^{abc}T_{abd}=-S^{bac}T_{bad}=-S^{abc}T_{abd},
\end{equation}
where in the last step we relabel ``dummy'' indices summed over.
Similarly, any tensor with overlapping anti-symmetric and symmetric
indices vanishes, e.g. if the (first) second two indices are (anti)
symmetric then
\begin{equation}
T^{abc}=-T^{bac}=-T^{bca}=T^{cba}=T^{cab}=-T^{acb}=-T^{abc}.
\end{equation}

A (pseudo) inner product on $V$ is a symmetric bilinear mapping,
and thus corresponds to a symmetric tensor $g_{ab}$ called the \textbf{(pseudo)
metric tensor}\index{metric tensor}. The isomorphism $v\in V\mapsto v^{\flat}\in V^{*}$
induced by this pseudo inner product is then defined by 
\begin{equation}
v^{a}\mapsto v_{a}\equiv g_{ab}v^{b},
\end{equation}
and is called \textbf{index lowering}\index{index lowering}. The
\textbf{dual metric tensor}\index{dual metric tensor} (AKA conjugate
metric tensor\index{conjugate metric tensor}) is the corresponding
pseudo inner product on $V^{*}$ and is denoted $g^{ab}$, which provides
a consistent \textbf{index raising}\index{index raising} operation
since we immediately obtain $g^{ab}g_{ac}g_{bd}=g_{cd}$. We also
have the relation $v^{a}=g^{ab}v_{b}=g^{ab}g_{bc}v^{c}\Rightarrow g^{ab}g_{bc}=?g^{a}{}_{c}?=?\delta^{a}{}_{c}?$,
the identity mapping; thus $g^{ab}g_{ab}$ is equal to the dimension
of $V$. The inner product of two tensors of the same type is then
the contraction of their tensor direct product after index lowering/raising,
e.g. $\left\langle T^{ab},S^{cd}\right\rangle =T^{ab}S_{ab}=T^{ab}g_{ac}g_{bd}S^{cd}$.

\noindent %
\begin{framed}%
\noindent $\triangle$ It is important to remember that if $v$ is
a vector, the operation $v_{a}v^{a}$ implies index lowering, which
requires an inner product. In contrast, if $\varphi$ is a 1-form,
the operation $\varphi_{a}v^{a}$ is always valid regardless of the
presence of an inner product. \end{framed}

A symmetric or anti-symmetric tensor can be formed from a general
tensor by adding or subtracting versions with permuted indices. For
example, the combination $\left(T_{ab}+T_{ba}\right)/2$ is the \textbf{symmetrized
tensor}\index{symmetrized tensor} of $T$, i.e. exchanging any two
indices leaves it invariant. The \textbf{anti-symmetrized tensor}\index{anti-symmetrized tensor}
$\left(T_{ab}-T_{ba}\right)/2$ changes sign upon the exchange of
any two indices, and (only for tensors of order 2) yields the original
tensor $T_{ab}$ when added to the symmetrized tensor. The following
notation is common for tensors with $n$ indices, with the sums over
all permutations of indices:

\begin{flushleft}
\begin{equation}
\textrm{Symmetrization:}\;T_{\left(a_{1}\dots a_{n}\right)}\equiv\frac{1}{n!}\underset{\pi}{\sum}T_{a_{\pi\left(1\right)}\dots a_{\pi\left(n\right)}}
\end{equation}
\par\end{flushleft}

\begin{equation}
\textrm{Anti-symmetrization:}\;T_{\left[a_{1}\dots a_{n}\right]}\equiv\frac{1}{n!}\underset{\pi}{\sum}\textrm{sign}\left(\pi\right)T_{a_{\pi\left(1\right)}\dots a_{\pi\left(n\right)}}
\end{equation}
This operation can be performed on any subset of indices in a tensor,
as long as they are all covariant or all contravariant. Skipping indices
is denoted with vertical bars, as in $T_{\left(a|b|c\right)}=\left(T_{abc}+T_{cba}\right)/2$;
however, note that vertical bars alone are sometimes used to denote
a sum of ordered permutations, as in $T_{\left|abc\right|}=T_{abc}+T_{bca}+T_{cab}$.

\subsection{Tensors as multi-dimensional arrays }

In a given basis, a pure tensor of type $(m,n)$ can be written using
\textbf{component notation}\index{component notation} in the form

\begin{equation}
v^{1}\otimes\dotsb\otimes v^{m}\otimes\varphi_{1}\otimes\dotsb\otimes\varphi_{n}\equiv?T^{\mu_{1}\dots\mu_{m}}{}_{\lambda_{1}\dots\lambda_{n}}?e_{\mu_{1}}\otimes\dotsb\otimes e_{\mu_{m}}\otimes\beta^{\lambda_{1}}\otimes\dotsb\otimes\beta^{\lambda_{n}},
\end{equation}
where the Einstein summation convention is used in the second expression.
Note that the collection of terms into $T$ is only possible due to
the defining property of the tensor product being linear over addition.
The tensor product between basis elements is often dropped in such
expressions. Also note that this means that in terms of the tensor
as a multilinear mapping we have

\begin{equation}
?T^{\mu_{1}\dots\mu_{m}}{}_{\lambda_{1}\dots\lambda_{n}}?=T\left(\beta^{\mu_{1}},\dotsb,\beta^{\mu_{m}},e_{\lambda_{1}},\dotsb,e_{\lambda_{n}}\right).
\end{equation}

A general tensor is a sum of such pure tensor terms, so that any tensor
$T$ can be represented by a $\left(m+n\right)$-dimensional array
of scalars. For example, any tensor of order 2 is a matrix, and type
$(1,1)$ tensors are linear mappings operating on vectors or forms
via ordinary matrix multiplication if they are all expressed in terms
of components in the same basis. Basis-independent quantities from
linear algebra such as the trace and determinant are then well defined
on such tensors.

\noindent %
\begin{framed}%
\noindent $\triangle$ It is important to remember that a tensor $T^{\mu\nu}$
or $T_{\mu\nu}$ can be written as a matrix of scalars, but linear
algebra operations only are valid for linear operators $?T^{\mu}{}_{\nu}?$.
A similar source of potential confusion is that the (anti-)symmetry
of $T^{\mu\nu}$ or $T_{\mu\nu}$ is basis independent, while that
of $?T^{\mu}{}_{\nu}?$ is not.\end{framed}

\noindent %
\begin{framed}%
\noindent $\triangle$ A potentially confusing aspect of component
notation is the basis vectors $e_{\mu}$, which are not components
of a 1-form but rather vectors, with $\mu$ a label, not an index.
Similarly, the basis 1-forms $\beta^{\lambda}$ should not be confused
with components of a vector.\end{framed}

The Latin letters of abstract index notation (e.g. $?T^{ab}{}_{cd}?$)
can thus be viewed as placeholders for what would be indices in a
particular basis, while the Greek letters of component notation represent
an actual array of scalars that depend on a specific basis. The reason
for the different notations is to clearly distinguish tensor identities,
true in any basis, from equations true only in a specific basis.

\noindent %
\begin{framed}%
\noindent $\triangle$ It is common in general relativity and other
topics to abuse both abstract and index notation to represent objects
that are non-tensorial (see Section \ref{sec:Parallel-transport}). \end{framed}

\noindent %
\begin{framed}%
\noindent $\triangle$ Note that if abstract index notation is not
being used, Latin and Greek indices are often used to make other distinctions,
a common one being between indices ranging over three space indices
and indices ranging over four space-time indices. \end{framed}

\noindent %
\begin{framed}%
\noindent $\triangle$ Note that ``rank\index{rank}” and ``dimension\index{dimension}”
are overloaded terms across these constructs: ``rank'' is sometimes
used to refer to the order of the tensor, which is the dimensionality
of the corresponding multi-dimensional array; the dimension of a tensor
is that of the underlying vector space, and so is the length of a
side of the corresponding array (also sometimes called the dimension
of the array). However, the rank of a order 2 tensor coincides with
the rank of the corresponding matrix. \end{framed}

\subsection{\label{subsec:Exterior-forms-as-multilinear-mappings}Exterior forms
as multilinear mappings }

An \textbf{exterior form}\index{exterior form:and the exterior product}
(AKA $k$-form\index{k-form}, alternating form\index{alternating form})
is defined to be an element of $\Lambda^{k}V^{*}$. Just as we formed
the isomorphism $\otimes\varphi_{i}\mapsto\Pi\varphi_{i}$ to view
tensors as multilinear mappings on $V$, we can view $k$-forms as
alternating multilinear mappings on $V$. Restricting attention to
the exterior product of $k$ 1-forms $\bigwedge\varphi_{i}$, we define
the isomorphism

\begin{equation}
\begin{aligned}\bigwedge_{i=1}^{k}\varphi_{i} & \mapsto\sum_{\pi}\textrm{sign}\left(\pi\right)\prod_{i=1}^{k}\varphi_{\pi\left(i\right)}\\
 & =\sum_{i_{1},i_{2},\dotsc,i_{k}}\varepsilon^{i_{1}i_{2}\dots i_{k}}\varphi_{i_{1}}\varphi_{i_{2}}\dotsm\varphi_{i_{k}},
\end{aligned}
\end{equation}
where $\pi$ is any permutation of the $k$ indices, sign$(\pi)$
is the sign of the permutation, and $\varepsilon$ is the \textbf{permutation
symbol}\index{permutation symbol} (AKA completely anti-symmetric
symbol\index{completely anti-symmetric symbol}, Levi-Civita symbol\index{Levi-Civita symbol},
alternating symbol\index{alternating symbol}, $\varepsilon$-symbol\index{varepsilon-symbol@$\varepsilon$-symbol}\index{epsilon symbol}),
defined to be $+1$ for even index permutations, $-1$ for odd, and
$0$ otherwise. 

\noindent %
\begin{framed}%
\noindent \sun{} The above isomorphism extends the interpretation
of forms acting on vectors as yielding a projection. Specifically,
if the parallelepiped $\varphi^{\sharp}=\bigwedge\varphi_{i}^{\sharp}$
has volume $V$, then $\varphi(v_{1},\ldots v_{k})/V$ is the volume
of the projection of the parallelepiped $v=\bigwedge v_{i}$ onto
$\varphi^{\sharp}$.\end{framed}Extending this to arbitrary forms $\varphi\in\Lambda^{j}V^{*}$ and
$\psi\in\Lambda^{k}V^{*}$, we have

\begin{equation}
\begin{aligned} & \left(\varphi\wedge\psi\right)\left(v_{1},\dotsc,v_{j+k}\right)\\
 & \mapsto\cfrac{1}{j!k!}\sum_{\pi}\textrm{sign}\left(\pi\right)\varphi\left(v_{\pi\left(1\right)},\dotsc,v_{\pi\left(j\right)}\right)\psi\left(v_{\pi\left(j+1\right)},\dotsc,v_{\pi\left(j+k\right)}\right).
\end{aligned}
\end{equation}

Just as with tensors, this isomorphism is canonical but not unique;
but in the case of exterior forms, other isomorphisms are in common
use. The main alternative isomorphism inserts a term $1/k!$ in the
first relation above, which results in $1/j!k!$ being replaced by
$1/\left(j+k\right)!$ in the second. Note that this alternative is
inconsistent with the interpretation of exterior products as parallelepipeds.

\noindent %
\begin{framed}%
\noindent $\triangle$ It is important to understand which convention
a given author is using. The first convention above is common in physics,
and we will adhere to it here. \end{framed}

\subsection{Exterior forms as completely anti-symmetric tensors }

An immediate result of this view of forms as multilinear mappings
is that we can also view forms as completely anti-symmetric tensors
under the identification of $\prod\varphi_{i}$ with $\bigotimes\varphi_{i}$.
For example, for a 2-form we have the equivalent expressions 
\begin{equation}
\begin{aligned}\left(\varphi\wedge\psi\right)\left(v,w\right) & \leftrightarrow\left(\varphi\otimes\psi-\psi\otimes\varphi\right)\left(v,w\right)\\
 & \leftrightarrow\varphi\left(v\right)\psi\left(w\right)-\psi\left(v\right)\varphi\left(w\right).
\end{aligned}
\end{equation}
Note however that this identification does not lead to equality of
the inner products defined on tensors and exterior forms; instead
for two $k$-forms we have

\begin{equation}
\left\langle \bigwedge\varphi_{i},\bigwedge\psi_{j}\right\rangle _{\textrm{form}}=\textrm{det}\left(\left\langle \varphi_{i},\psi_{j}\right\rangle \right),
\end{equation}
while as tensors we have

\begin{equation}
\left\langle \bigwedge\varphi_{i},\bigwedge\psi_{j}\right\rangle _{\textrm{tensor}}=\left\langle \varepsilon^{I}\varphi_{I},\varepsilon^{J}\varphi_{J}\right\rangle _{\textrm{tensor}}=k!\textrm{det}\left(\left\langle \varphi_{i},\psi_{j}\right\rangle \right).
\end{equation}
Fortunately, the tensor inner product is almost always expressed explicitly
in terms of index contractions, so we will continue to use the $\left\langle \;,\,\right\rangle $
notation for the inner product of $k$-forms.

Also note that this isomorphism between the exterior product and the
tensor product can be similarly used to identify the exterior product
of vectors with a completely anti-symmetric contravariant tensor.
In the following section we identify exterior forms with lower index
anti-symmetric arrays; we can similarly identify the exterior product
of vectors with upper index anti-symmetric arrays.

\subsection{Exterior forms as anti-symmetric arrays }

In terms of a basis $\beta^{\mu}$ of $V^{*}$, we can write a $k$-form
$\varphi$ as

\begin{equation}
\varphi=\frac{1}{k!}\sum_{\mu_{1},\dotsc,\mu_{k}}\varphi_{\mu_{1}\dots\mu_{k}}\beta^{\mu_{1}}\wedge\dotsb\wedge\beta^{\mu_{k}}.
\end{equation}

\noindent %
\begin{framed}%
\noindent $\triangle$ The above way of writing the components is
not unique, and others are in common use, the main alternative omitting
the factorial.\end{framed}The advantage of the expression above is that, with our isomorphism
convention, the component array can be identified with the anti-symmetric
covariant tensor component array in the same basis:

\begin{equation}
\varphi\mapsto\frac{1}{k!}\varphi_{\mu_{1}\dots\mu_{k}}\sum_{\pi}\textrm{sign}\left(\pi\right)\bigotimes_{i}\beta^{\pi\left(i\right)}=\varphi_{\mu_{1}\dots\mu_{k}}\beta^{\mu_{1}}\otimes\cdots\otimes\beta^{\mu_{k}}
\end{equation}
Here we have dropped the summation sign in favor of the Einstein summation
convention, and the last equality follows from the anti-symmetry of
the component array. This means that as with tensors, in terms of
the $k$-form as a multilinear mapping we have

\begin{equation}
\varphi_{\mu_{1}\dots\mu_{k}}=\varphi\left(e_{\mu_{1}},\dotsb,e_{\mu_{k}}\right).
\end{equation}

\subsection{\label{subsec:Algebra-valued-exterior-forms}Algebra-valued exterior
forms }

We can extend the view of exterior forms as real-valued linear mappings
to define \textbf{algebra-valued forms}\index{algebra-valued form}\index{exterior form:algebra-valued}.
These follow the same construction as in Section \ref{subsec:Exterior-forms-as-multilinear-mappings}
above, starting from an algebra-valued 1-form 
\begin{equation}
\check{\Theta}\colon V\to\mathfrak{a},
\end{equation}
so that general forms are alternating multilinear maps from $k$ vectors
to a real algebra $\mathfrak{a}$ whose vector multiplication takes
the place of multiplication in $\mathbb{R}$. Since this vector multiplication
may not be commutative, we need to be more careful in terms of ordering
in the isomorphism to ensure antisymmetry, i.e. for two algebra-valued
1-forms we define 
\begin{equation}
(\check{\Theta}\wedge\check{\Psi})(v,w)\equiv\check{\Theta}(v)\check{\Psi}(w)-\check{\Theta}(w)\check{\Psi}(v).
\end{equation}

An algebra-valued form whose values are defined by matrices is a \textbf{matrix-valued
form}\index{matrix-valued form}\index{exterior form:matrix-valued}.
Exterior forms that take values in a matrix group can also be considered
as matrix-valued forms, but it must be understood that under addition
the values may no longer be in the group. One can also form the exterior
product between a matrix-valued form and a \textbf{vector-valued form}\index{vector-valued form}\index{exterior form:vector-valued}.
To reduce confusion when dealing with algebra- and vector-valued forms,
we will indicate them with (non-standard) decorations, for example
in the case of a matrix-valued 1-form acting on a vector-valued 1-form,
\begin{equation}
(\check{\Theta}\wedge\vec{\varphi})(v,w)\equiv\check{\Theta}(v)\vec{\varphi}(w)-\check{\Theta}(w)\vec{\varphi}(v).
\end{equation}
\begin{framed}%
\noindent $\triangle$ Since the elements of an algebra are vectors,
algebra-valued forms may be considered as vector-valued forms whose
values can be multiplied. We will reserve the term vector-valued forms
for forms whose values are acted on by matrix-valued forms.\end{framed}%
\begin{framed}%
\noindent $\triangle$ An additional distinction can be made between
forms that take values which are concrete matrices and column vectors
(and thus depend upon the basis of the underlying vector space), and
forms that take values which are abstract linear transformations and
abstract vectors (and thus are basis-independent). We will attempt
to distinguish between these by referring to the specific matrix or
abstract group, and by only using ``vector-valued'' when the value
is an abstract vector.\end{framed}

A notational issue arises in the particular case of Lie algebra valued
form\index{Lie algebra valued form}s, where the related associative
algebra in the relation $[\check{\Theta},\check{\Psi}]=\check{\Theta}\check{\Psi}-\check{\Psi}\check{\Theta}$
could also be in use. In this case multiplication of values could
use either the Lie commutator or that of the related associative algebra.
We will denote the exterior product using the Lie commutator by $\check{\Theta}[\wedge]\check{\Psi}$.
Some authors use $[\check{\Theta},\check{\Psi}]$ or $[\check{\Theta}\wedge\check{\Psi}]$,
but both can be ambiguous, motivating us to introduce our (non-standard)
notation. The expression $\check{\Theta}\wedge\check{\Psi}$ is then
reserved for the exterior product using the underlying associative
algebra (e.g. that of matrix multiplication if the associative algebra
is defined this way). For two Lie algebra-valued 1-forms we then have

\begin{equation}
\begin{aligned}(\check{\Theta}[\wedge]\check{\Psi})\left(v,w\right) & =[\check{\Theta}\left(v\right),\check{\Psi}\left(w\right)]-[\check{\Theta}\left(w\right),\check{\Psi}\left(v\right)]\\
 & =\check{\Theta}\left(v\right)\check{\Psi}\left(w\right)-\check{\Psi}\left(w\right)\check{\Theta}\left(v\right)-\check{\Theta}\left(w\right)\check{\Psi}\left(v\right)+\check{\Psi}\left(v\right)\check{\Theta}\left(w\right)\\
 & =(\check{\Theta}\wedge\check{\Psi}+\check{\Psi}\wedge\check{\Theta})\left(v,w\right).
\end{aligned}
\end{equation}
Note that $[\check{\Theta},\check{\Psi}](v,w)=\check{\Theta}(v)\check{\Psi}(w)-\check{\Psi}(v)\check{\Theta}(w)$
is a distinct construction, as is $[\check{\Theta}(v),\check{\Psi}(w)]=\check{\Theta}(v)\check{\Psi}(w)-\check{\Psi}(w)\check{\Theta}(v)$;
neither are in general anti-symmetric and thus do not yield forms.
Also note that e.g. for two 1-forms $\check{\Theta}[\wedge]\check{\Psi}\neq\check{\Theta}\wedge\check{\Psi}-\check{\Psi}\wedge\check{\Theta}$,
and $(\check{\Theta}[\wedge]\check{\Theta})\left(v,w\right)=2[\check{\Theta}\left(v\right),\check{\Theta}\left(w\right)]$
does not in general vanish, so $[\wedge]$ does not act like a Lie
commutator in these respects. However, for algebra-valued $j$- and
$k$-forms $\check{\Theta}$ and $\check{\Psi}$, the operation $[\wedge]$
does in fact follow a graded commutativity rule 

\begin{equation}
\check{\Theta}[\wedge]\check{\Psi}=\left(-1\right)^{jk+1}\check{\Psi}[\wedge]\check{\Theta},
\end{equation}
and with an algebra-valued $m$-form $\check{\Xi}$ we find a graded
Jacobi identity of

\begin{equation}
\left(-1\right)^{jm}(\check{\Theta}[\wedge]\check{\Psi})\left[\wedge\right]\check{\Xi}+\left(-1\right)^{kj}(\check{\Psi}[\wedge]\check{\Xi})\left[\wedge\right]\check{\Theta}+\left(-1\right)^{mk}(\check{\Xi}[\wedge]\check{\Theta})\left[\wedge\right]\check{\Psi}=0.
\end{equation}

Algebra-valued forms also introduce potentially ambiguous index notation.
If a basis is chosen for the algebra $\mathfrak{a}$, the value of
an algebra-valued form may be expressed using component notation as
$\Theta^{\mu}$; or if the algebra is defined in terms of matrices,
an element might be written $?\Theta^{\alpha}{}_{\beta}?$, an expression
that has nothing to do with the basis of $\mathfrak{a}$. Then for
example an algebra-valued 1-form might be written $?\Theta^{\mu}{}_{\gamma}?$
or $?\Theta^{\alpha}{}_{\beta\gamma}?$. 

\noindent %
\begin{framed}%
\noindent $\triangle$ In considering algebra-valued forms expressed
in index notation, extra care must be taken to identify the type of
form in question, and to match each index with the aspect of the object
it was meant to represent. \end{framed}

\subsection{\label{subsec:The-Hodge-star}The Hodge star }

A pseudo inner product determines orthonormal bases for $V$, among
which we can choose a specific one $\hat{e}_{\mu}$. The ordering
of the $\hat{e}_{\mu}$ determines a choice of orientation. This orientation
uniquely determines an orthonormal basis (i.e. a unit ``length''
vector) for the one-dimensional vector space $\Lambda^{n}V$, namely
the \textbf{unit $n$-vector}\index{unit n-vector@unit \textit{n}-vector}
(AKA orientation $n$-vector, volume element) 
\begin{equation}
\Omega\equiv\hat{e}_{1}\wedge\dotsb\wedge\hat{e}_{n}.
\end{equation}

\noindent %
\begin{framed}%
\noindent $\triangle$ Many symbols are used in the literature for
the unit $n$-vector and related quantities, including $\varepsilon$,
$i$, $I$, and $\omega$; to avoid confusion with the other common
uses of these symbols, we will use the (non-standard) symbol $\Omega$. \end{framed}

Since $\Lambda^{n}V$ is one-dimensional, every element of $\Lambda^{n}V$
is a real multiple of $\Omega$. Thus $\Omega$ sets up a bijection
(dependent upon the inner product and choice of orientation) between
$\Lambda^{n}V$ and $\Lambda^{0}V$ = $\mathbb{R}$. In general, $\Lambda^{k}V$
and $\Lambda^{n-k}V$ are vector spaces of equal dimension, and thus
we can also set up a bijection between them.

The \textbf{Hodge star operator}\index{Hodge star operator}\index{star operator}
(AKA Hodge dual\index{Hodge dual}\index{dual}) is defined to be
the linear map 
\begin{equation}
*\colon\Lambda^{k}V\to\Lambda^{n-k}V
\end{equation}
that acts on any $A,B\in\Lambda^{k}V$ such that 
\begin{equation}
A\wedge*B=\left\langle A,B\right\rangle \Omega.
\end{equation}
In particular, we immediately obtain
\begin{equation}
A\wedge*A=\left\langle A,A\right\rangle \Omega.
\end{equation}
\begin{framed}%
\noindent \sun{} These relations allow one to think of the Hodge star
$*$ as an operator that that yields the ``orthogonal complement
with the same magnitude,'' or alternatively that ``swaps the exterior
and inner products.''\end{framed}

The Hodge star operator is dependent upon a choice of inner product
and orientation, but beyond that is independent of any particular
basis. However, if we choose an orthonormal basis $\hat{e}_{\mu}$
oriented with $\Omega$, we can take $\hat{A}\equiv\hat{e}_{1}\wedge\dotsb\wedge\hat{e}_{k}$
and $\hat{C}\equiv\hat{e}_{k+1}\wedge\dotsb\wedge\hat{e}_{n}$, in
which case $*\hat{A}=\left\langle \hat{A},\hat{A}\right\rangle \hat{C}$,
i.e. $*\hat{A}$ is constructed from an orthonormal basis for the
orthogonal complement of $\hat{A}$; in fact, this relation can be
used as an equivalent definition of the Hodge star, and for a pseudo
inner product of signature $(r,s)$ results in 
\begin{equation}
\left\langle A,B\right\rangle =(-1)^{s}\left\langle *A,*B\right\rangle .
\end{equation}

Below we list some easily derived facts about the Hodge star operator,
where $V$ is $n$-dimensional with unit $n$-vector $\Omega$ and
a pseudo inner product of signature $(r,s)$:
\begin{itemize}
\item $*\Omega=\left(-1\right)^{s}\Rightarrow\left(*C\right)\Omega=\left(-1\right)^{s}C$
if $C\in\Lambda^{n}V$
\item $*1=\Omega\Rightarrow\left\langle *a,\Omega\right\rangle =\left(-1\right)^{s}a$
if $a\in\Lambda^{0}V$
\item $**A=\left(-1\right)^{k\left(n-k\right)+s}A=\left(-1\right)^{k\left(n-1\right)+s}A$,
where $A\in\Lambda^{k}V$
\item $A\wedge*B=B\wedge*A$ if $A,B\in\Lambda^{k}V$
\item $*\left(A\wedge*B\right)=\left\langle A\wedge*B,\Omega\right\rangle =(-1)^{s}\left\langle A,B\right\rangle $
if $A,B\in\Lambda^{k}V$ 
\end{itemize}
\noindent %
\begin{framed}%
\noindent $\triangle$ Some texts (including earlier versions of this
paper and the first edition of the book based upon it) instead define
the Hodge star by the relation $A\wedge C=\left\langle *A,C\right\rangle \Omega$
for $A\in\Lambda^{k}V$, $C\in\Lambda^{n-k}V$, which prefixes our
current Hodge star by the factor $(-1)^{s}$.\end{framed}

Note that $*A$ is not a basis-independent object, since it reverses
sign upon changing the chosen orientation. Such an object is prefixed
by the word \textbf{pseudo-}\index{pseudo-}, e.g. $*v$ is called
a \textbf{pseudo-vector}\index{pseudo-vector} (AKA axial vector\index{axial vector},
in which case $v$ is called a polar vector\index{polar vector})
and $\Omega$ itself is a \textbf{pseudo-scalar}\index{pseudo-scalar}.

\noindent %
\begin{framed}%
\noindent $\triangle$ The use of ``pseudo'' to indicate a quantity
that reverses sign upon a change of orientation should not be confused
with the use of ``pseudo'' to indicate an inner product that is
not positive-definite. There are also other uses of ``pseudo'' in
use. In particular, in general relativity the term ``pseudo-tensor\index{pseudo-tensor}''
is sometimes used, where neither of the above meanings are implied;
instead this signifies that the tensor (to be defined in \ref{subsec:Tensors})
is not in fact a tensor.\end{framed}

\section{\label{sec:Differentiable-manifolds}Differentiable manifolds}

Differentiable manifolds allow us to graft calculus onto a topological
manifold, which we can think of as a ``rubber sheet.'' The constructions
of coordinates and tangent vectors enable us to define a family of
derivatives associated with the concept of how vector fields change
on the manifold. The challenge is in defining all these objects without
an ambient space, which our intuitive picture normally depends upon.

\noindent %
\begin{framed}%
\noindent $\triangle$ Note that a differentiable manifold includes
no concept of length or distance (a metric), and no structure that
allows tangent vectors at different points to be compared or related
to each other (a connection). It is important to remember that nothing
in this section depends upon these two extra structures. \end{framed}

When dealing with manifolds, there are two main approaches one can
take: express everything in terms of coordinates, or strive to express
everything in a coordinate-free fashion. In keeping with our attempt
to focus on concepts rather than calculations, we will take the latter
approach, but will take pains to carefully express fundamental concepts
in terms of coordinates in order to derive a picture of what these
coordinate-free tools do. 

\subsection{Coordinates }

A key feature of a topological manifold $M^{n}$ is that every point
has an open neighborhood homeomorphic to an open subset of $\mathbb{R}^{n}$.
To make this precise we define the following terms.
\begin{itemize}
\item \textbf{Coordinate chart}\index{coordinate chart} (AKA parameterization\index{parameterization},
patch\index{patch}, system of coordinates\index{system of coordinates}):
a homeomorphism $\alpha\colon U\to\mathbb{R}^{n}$ from an open set
$U\subset M^{n}$ to an open subset of $\mathbb{R}^{n}$
\item \textbf{Coordinate functions}\index{coordinate function} (AKA coordinates\index{coordinates}):
the maps $a^{\mu}\colon U\to\mathbb{R}$ that project $\alpha$ down
to one of the canonical Cartesian components 
\item \textbf{Atlas}\index{atlas}: a collection of coordinate charts that
cover the manifold 
\item \textbf{Coordinate transformation}\index{coordinate transformation}
(AKA change of coordinates\index{change of coordinates}, transition
function\index{transition function}): in a region covered by two
charts, we can construct the map $\alpha_{2}\circ\alpha_{1}^{-1}\colon\mathbb{R}^{n}\to\mathbb{R}^{n}$
\end{itemize}
\begin{figure}[H]
\noindent \begin{centering}
\includegraphics[width=1\columnwidth]{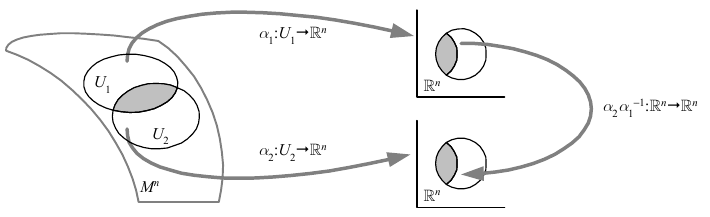}
\par\end{centering}
\caption{In the intersection of two coordinate charts we can construct the
coordinate transformation, a homeomorphism on $\mathbb{R}^{n}$.}
\end{figure}

\noindent %
\begin{framed}%
\noindent $\triangle$ A coordinate chart is sometimes defined to
be the inverse map $\alpha^{-1}\colon\mathbb{R}^{n}\to M$ valid on
an open subset of $\mathbb{R}^{n}$, with similar changes to related
definitions such as coordinate functions.\end{framed}

The coordinate transformations are simply maps on Euclidean space,
so we can require them to be infinitely differentiable (AKA smooth\index{smooth},
$C^{\infty}$). An atlas whose charts all have smooth coordinate transformations
determines a \textbf{differentiable structure}\index{differentiable structure},
which turns the topological manifold into a \textbf{differentiable
manifold}\index{manifold!differentiable} (AKA smooth manifold\index{smooth manifold}).
Two differential structures are considered to be equivalent if the
union of their atlases still results in smooth coordinate transformations.
Unless otherwise noted, from this point forward ``manifold\index{manifold}”
will mean differentiable manifold.

A \textbf{complex manifold}\index{complex manifold} is defined to
have an atlas of charts to $\mathbb{C}^{n}$ whose coordinate transformations
are analytic. Complex $n$-manifolds are a subset of real $2n$-manifolds,
but atlases are highly constrained since complex analytic functions
are much more constrained than smooth functions. By ``manifold''
we will always mean a real manifold.

With the addition of a differentiable structure, one can define the
various tools of calculus on manifolds in a straightforward way. Differentiable
functions $f\colon U\to\mathbb{R}$ require the map $f\circ\alpha^{-1}\colon\mathbb{R}^{n}\to\mathbb{R}$
to be differentiable, and differentials $\partial/\partial a^{\mu}$
are defined at a point $p\in U$ by

\begin{equation}
\left.\frac{\partial}{\partial a^{\mu}}\left(f\right)\right|_{p}\equiv\left.\frac{\partial}{\partial x^{\mu}}\left(f\circ\alpha^{-1}\left(x\right)\right)\right|_{x=\alpha(p)}.
\end{equation}
where $x\in\mathbb{R}^{n}$. All of the usual relations of calculus
hold with these definitions.

\noindent %
\begin{framed}%
\noindent $\triangle$ To avoid clutter, a common abuse of notation
is to use $x^{\mu}$ to denote any or all of three quantities: the
point $p\in M$, the coordinate functions $a^{\mu}\colon M\to\mathbb{R}$,
and the $\mathbb{R}^{n}$ $n$-tuplet $x^{\mu}=a^{\mu}\left(p\right)$.
Similarly, the differential $\partial/\partial a^{\mu}$ is often
denoted $\partial/\partial x^{u}$. We will follow these conventions
going forward, but when dealing with fundamental definitions or pictures,
it is important to distinguish these very different quantities from
each other. Another shortcut is to denote differentials by $\partial_{\mu}$;
as with basis vectors, it is important to remember that these are
labels, not component indices. \end{framed}

\subsection{\label{subsec:Tangent-vectors-and-differential-forms}Tangent vectors
and differential forms }

The \textbf{tangent space}\index{tangent space} $T_{p}U$ at a point
$p\in U$ is defined to be the vector space spanned by the differential
operators $\partial/\partial a^{\mu}\mid_{p}$. A \textbf{tangent
vector}\index{tangent vector} $v\in T_{p}U$ can then be expressed
in tensor component notation as $v=v^{\mu}\partial/\partial a^{\mu}$,
so that $v\left(a^{\mu}\right)=v^{\mu}$. The tangent vector $\partial/\partial a^{\mu}\mid_{p}$
applied to a function $f$ can be thought of as ``the change of $f$
in the direction of the $\mu^{\textrm{th}}$ coordinate line at $p$.''

\begin{figure}[H]
\noindent \begin{centering}
\includegraphics[width=0.9\columnwidth]{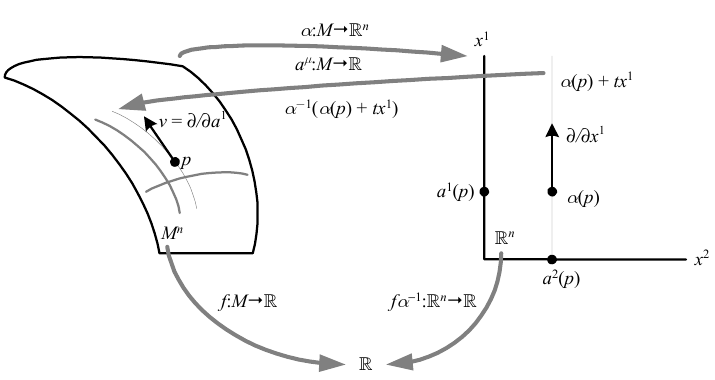}
\par\end{centering}
\caption{In a particular coordinate chart, a tangent vector $v$ operates on
a function by taking the derivative of the composite function in $\mathbb{R}^{n}$
in the direction of $v^{\mu}\partial/\partial a^{\mu}$.}
\end{figure}
Thus at a point $p$, we have 
\begin{equation}
v^{\mu}\frac{\partial}{\partial a^{\mu}}\left(f\right)=v^{\mu}\frac{\partial}{\partial x^{\mu}}\left(f\circ\alpha^{-1}\left(x\right)\right),
\end{equation}
where $x=\alpha(p)$. The coordinate line $\alpha^{-1}\left(a^{\mu}\left(p\right)+tv^{\mu}x^{\mu}\right)$
is a parameterized curve on $M$, and thus it and the tangent vector
itself are coordinate-independent objects. In this chart, any parametrized
curve is defined to have tangent $v$ at $t=0$ if its coordinates
are $C^{\mu}\left(t\right)\equiv a^{\mu}+tv^{\mu}$ to first order
in $t$; therefore the coordinates of the tangent vector to $C$ at
any point may be obtained by 
\begin{equation}
v^{\mu}=\frac{\mathrm{d}C^{\mu}}{\mathrm{d}t}.
\end{equation}
In another coordinate chart, the coordinate line that yields the same
operator on functions near $p$ can be seen to correspond to the familiar
transformation of vector components

\begin{equation}
v=v^{\mu}\frac{\partial}{\partial a^{\mu}}=\left(v^{\lambda}\frac{\partial b^{\mu}}{\partial a^{\lambda}}\right)\frac{\partial}{\partial b^{\mu}}.
\end{equation}
We can consider the point ``$p$ moved in the direction $v$” by
abusing notation to write $p^{\mu}+tv^{\mu}$ in place of $\alpha^{-1}\left(a^{\mu}\left(p\right)+tv^{\mu}x^{\mu}\right)$;
this is a coordinate-dependent expression, but in the limit $\varepsilon\rightarrow0$
we can unambiguously write $p+\varepsilon v$ to refer to the concept
``$p$ moved infinitesimally in the direction $v$,'' which is coordinate-independent.
This allows us to write 
\begin{equation}
v(f)=\underset{\varepsilon\rightarrow0}{\textrm{lim}}\frac{1}{\varepsilon}\left[f_{p+\varepsilon v}-f_{p}\right].
\end{equation}

\begin{figure}[H]
\noindent \begin{centering}
\includegraphics[width=0.8\columnwidth]{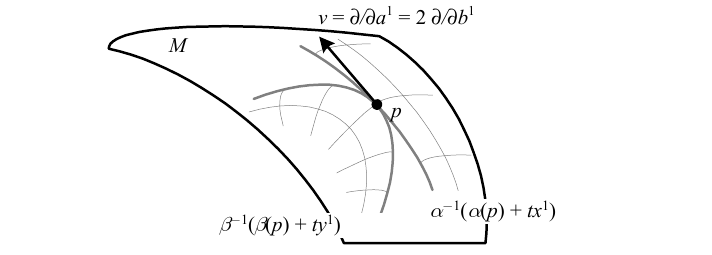}
\par\end{centering}
\caption{A tangent vector $v$ in terms of two different coordinate charts.
$v^{\mu}=(1,0)$ in chart $\alpha$ with coordinate functions $a^{\mu}\left(p\right)=x^{\mu}$,
and $v^{\mu}=\left(2,0\right)$ in chart $\beta$ with coordinate
functions $b^{\mu}\left(p\right)=y^{\mu}$. The divergent coordinate
lines show that the concept of moving a point \textquotedblleft in
the direction of $v$” can only be coordinate-independent in the infinitesimal
limit.}
\end{figure}

The set of all tangent spaces in a region $U$ is called the \textbf{tangent
bundle}\index{tangent bundle}, and is denoted $TU$. A (smooth, contravariant)
\textbf{vector field}\index{vector field} on $U$ is then a tangent
vector defined at each point such that its application to a smooth
function on $U$ is again smooth. Similarly, a \textbf{covariant vector
field}\index{covariant vector field} is a $1$-form defined at each
point such that its value on a vector field is a smooth function,
and a \textbf{tensor field}\index{tensor field} is the tensor product
of vector fields and covariant vector fields.

\noindent %
\begin{framed}%
\noindent $\triangle$ Tensor fields (including vector fields and
covariant vector fields) are written using the same notation as tensors,
making it important to distinguish the two situations. In particular,
one can define a (pseudo) metric tensor field, which is then usually
referred to as simply a metric. \end{framed}

Note that a tensor field must remain a tensor locally at any point
$p$, i.e. it must be a multi-linear mapping. For example, a covariant
tensor field can only depend upon the values of its vector field arguments
at $p$, since otherwise one could add a vector field that vanishes
at $p$ and obtain a different result. This means that operators such
as the derivatives on manifolds we will see in Sections \ref{sec:Derivatives-on-manifolds}
and \ref{sec:Parallel-transport} cannot usually be viewed as tensors,
since they measure the difference between arguments at different points. 

Since vectors are operators on functions, we can apply one vector
field to another. Following the practice of using $\partial/\partial x^{u}$
to refer to $\partial/\partial a^{\mu}$, this can be used to define
the \textbf{Lie bracket of vector fields}\index{Lie bracket:of vector fields}\index{Vector field:Lie bracket of}

\begin{equation}
\begin{aligned}\left[v,w\right](f) & \equiv v(w(f))-w(v(f))\\
\Rightarrow\left[v,w\right] & =\left(v^{\mu}\frac{\partial w^{\lambda}}{\partial x^{\mu}}-w^{\mu}\frac{\partial v^{\lambda}}{\partial x^{\mu}}\right)\frac{\partial}{\partial x^{\lambda}}.
\end{aligned}
\end{equation}
Here we have used the equality of mixed partials, and can easily verify
that $\left[v,w\right]$ is anti-commuting and satisfies the Jacobi
identity. Since this expression is coordinate-independent, $\left[v,w\right]$
is a vector field and we can thus view $\textrm{vect}\left(M\right)$,
the set of all vector fields on $M$, as the infinite-dimensional
\textbf{Lie algebra of vector fields}\index{Lie algebra:of vector fields}
on $M$, with vector multiplication defined by the Lie bracket.

Having defined vector and tensor fields on manifolds, we can now define
a \textbf{differential form}\index{differential form} as an alternating
covariant tensor field, i.e. an exterior form in $\Lambda\left(T_{p}U\right)$
smoothly defined for every point $p$.

\noindent %
\begin{framed}%
\noindent $\triangle$ Just as tensor fields are usually referred
to as simply tensors, differential forms are usually referred to as
simply \textbf{forms}, and a $k$-form is written simply $\varphi\in\Lambda^{k}M$.
It is important to remember that in the context of manifolds, a $k$-form
is an exterior form smoothly defined on $k$ elements of the tangent
space at each point, i.e. an anti-symmetric covariant $k$-tensor
field. \end{framed}

On a differentiable manifold, the existence of $k$-forms makes possible
a more concrete definition of orientability: a manifold $M^{n}$ is
orientable\index{orientable} if there exists a non-vanishing $n$-form.
Such a form is called a \textbf{volume form}\index{volume form} (AKA
volume element\index{volume element}), since it gains a Jacobian-like
determinant factor under invertible linear transformations.

\noindent %
\begin{framed}%
\noindent $\triangle$ The term ``volume form'' or ``volume element''
is sometimes defined in physics to reflect the intuitive idea of a
form which returns the volume spanned by its argument vectors; however,
volume is always positive, so that in this usage we are more accurately
referring to a \textbf{volume pseudo-form}\index{volume pseudo-form}
whose value is the absolute value of the volume form as we have defined
it, and which exists on any differentiable manifold, including those
which are non-orientable.\end{framed}

\subsection{Frames }

A \textbf{frame}\index{frame} $e_{\mu}$ on $U\subset M^{n}$ is
defined to be a tensor field of bases for the tangent spaces at each
point, i.e. $n$ linearly independent smooth vector fields $e_{\mu}$. 

\begin{figure}[H]
\noindent \begin{centering}
\includegraphics[width=0.8\columnwidth]{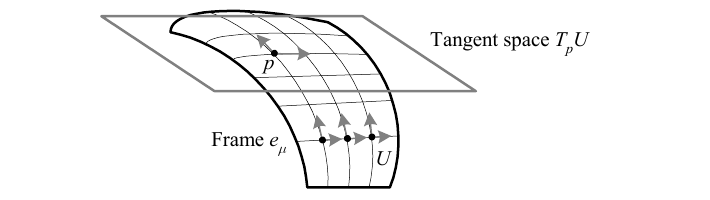}
\par\end{centering}
\caption{A frame $e_{\mu}$ is $n$ smooth vector fields that together provide
a basis for the tangent space at every point.}
\end{figure}

The concept of frame has a particularly large number of synonyms,
including comoving frame\index{comoving frame}, repère mobile\index{repère mobile},
vielbein\index{vielbein}, $n$-frame\index{n-frame}, and $n$-bein\index{n-bein}
(where $n$ is the dimension). The \textbf{dual frame}\index{frame!dual},
the $1$-forms $\beta^{\mu}$ corresponding to a frame $e_{\mu}$,
is also often simply called the frame.

When using particular coordinates $x^{\mu}$, the frame $e_{\mu}=\partial/\partial x^{\mu}$
is called the \textbf{coordinate frame}\index{coordinate frame} (AKA
coordinate basis\index{coordinate basis} or associated basis\index{associated basis});
any other frame is then called a \textbf{non-coordinate frame}\index{non-coordinate frame}.
A \textbf{holonomic frame}\index{holonomic frame} is a coordinate
frame in some coordinates (though perhaps not the ones being used);
this condition is equivalent to requiring that $\left[e_{\mu},e_{\nu}\right]=0$,
a result which is sometimes called \textbf{Frobenius' theorem}\index{Frobenius' theorem}.
An \textbf{anholonomic frame}\index{anholonomic frame} is then a
frame that cannot be derived from any coordinate chart in its region
of definition. Using a non-coordinate frame suited to a specific problem
is sometimes called the \textbf{method of moving frames}\index{method of moving frames}. 

\noindent %
\begin{framed}%
\noindent $\triangle$ Note that the distinction between holonomic
and coordinate frames as defined here is often not made.\end{framed}

\begin{figure}[H]
\noindent \begin{centering}
\includegraphics[width=0.8\columnwidth]{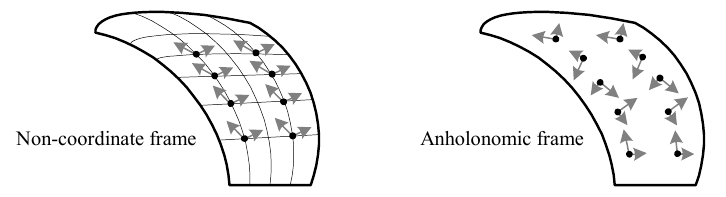}
\par\end{centering}
\caption{A non-coordinate frame is not tangent to the coordinate functions
being used, while an anholonomic frame cannot be derived from any
coordinate chart.}
\end{figure}

A frame cannot usually be globally defined on a manifold. A simple
way to see this is by the example of the 2-sphere $S^{2}$. Any drawing
of coordinate functions on a globe will have singularities, such as
the north and south poles when using latitude and longitude; these
are points where the associated coordinate frame will either be undefined
or will vanish. In general, there is no non-zero smooth vector field
that can be defined on $S^{n}$ for even $n$ (this is sometimes called
the \textbf{hedgehog theorem}\index{hedgehog theorem}, AKA \index{hairy ball theorem}hairy
ball theorem, coconut theorem\index{coconut theorem}). 

\begin{figure}[H]
\begin{centering}
\includegraphics[width=1\columnwidth]{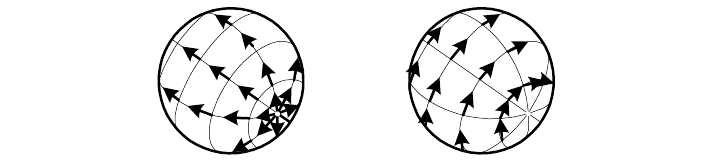}
\par\end{centering}
\caption{The hedgehog theorem for $S^{2}$, showing that any attempt to \textquotedblleft comb
the hair of a hedgehog\textquotedblright{} yields bald spots, in this
case at the poles.}
\end{figure}

A manifold that can have a global frame defined on it is called \textbf{parallelizable}\index{parallelizable}.
Some facts regarding parallelizable manifolds include:
\begin{itemize}
\item All parallelizable manifolds are orientable (and therefore have a
volume form), but as we saw with $S^{2}$ the converse is not in general
true
\item Any orientable 3-manifold $M^{3}$ is parallelizable $\Rightarrow$
any 4-manifold $M^{3}\times\mathbb{R}$ is parallelizable (important
in the case of the spacetime manifold) 
\item Of the $n$-spheres, only $S^{1}$, $S^{3}$, and $S^{7}$ are parallelizable
(this can be seen to be related to $\mathbb{C}$, $\mathbb{H}$, and
$\mathbb{O}$ being the only normed finite-dimensional real division
algebras beyond $\mathbb{R}$) 
\item The torus (with any number of holes) is the only closed orientable
surface with a non-zero smooth vector field 
\end{itemize}

\subsection{Tangent vectors in terms of frames }

It is important to remember that in following our intuitive picture
of a Euclidean surface, our central definitions were manifolds $M$
and tangent vectors $v$. These are the ``real'' intrinsic objects,
while their expressions in terms of a particular coordinate chart
and frame are arbitrary. Coordinates and frames are ``temporary''
tools we use to ``componentize'' points and tangents on a manifold.

In particular, if a manifold is defined in terms of a set of coordinate
functions that feature a singularity, this singularity may be due
to the coordinates extending outside of their valid chart, telling
us nothing about whether the manifold itself has a singularity. Every
point of a well-defined differentiable manifold always has a local
coordinate chart and tangent vectors.

For example, given the typical spherical coordinate chart for $S^{2}$
the associated frame will be singular at the poles, since they are
outside of $U$ for that chart; nevertheless, tangent vectors are
well-defined at these points, and can be expressed perfectly normally
in a different chart.

\begin{figure}[H]
\noindent \begin{centering}
\includegraphics[width=0.8\columnwidth]{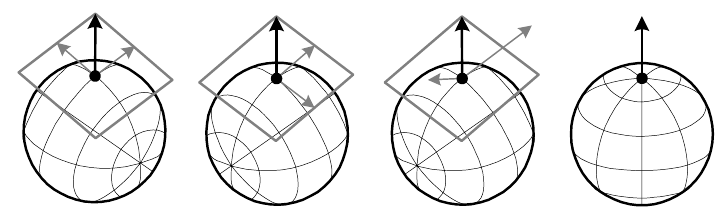}
\par\end{centering}
\caption{A manifold and tangent vector expressed in terms of different coordinate
functions and frames.}
\end{figure}

In the above figure, we see the following situations depicted:
\begin{itemize}
\item $v=e_{1}+e_{2}=\partial/\partial x^{1}+\partial/\partial x^{2}$ (expressed
in a coordinate frame) 
\item $v=e_{1}^{\prime}-e_{2}^{\prime}=\partial/\partial x^{\prime1}-\partial/\partial x^{\prime2}$
(using a different coordinate frame) 
\item $v=e_{1}^{\prime\prime}+3e_{2}^{\prime\prime}=\partial/\partial x^{\prime1}-\partial/\partial x^{\prime2}$
(in a non-coordinate frame) 
\end{itemize}
The final figure depicts coordinate functions that are singular at
the point of interest; the manifold and vector are still well-defined,
but the tangent space at this point cannot be expressed in terms of
this coordinate chart. 

\noindent %
\begin{framed}%
\noindent $\triangle$ In general, when working with objects on manifolds,
it is important to keep clearly in mind whether a given symbol represents
a vector, form, or function (0-form); whether any given index is a
label, an abstract index or a component index in a particular frame
or coordinates; and whether the object is a field with a value at
each point, or is only valid at a particular point. Any calculation
can always be made explicit by expressing everything in terms of functions
and differential operators on them.\end{framed}

\subsection{Diffeomorphisms }

In the same way that spaces or topological manifolds are equivalent
if they are related by a homeomorphism, differentiable manifolds are
equivalent if they are related by a \textbf{diffeomorphism}\index{diffeomorphism},
a homeomorphism that is differentiable along with its inverse. As
usual we define differentiability by moving the mapping to $\mathbb{R}^{n}$,
e.g. $\Phi\colon M\to N$ is differentiable if $\alpha_{N}\circ\Phi\circ\alpha_{M}^{-1}\colon\mathbb{\mathbb{R}}^{m}\to\mathbb{R}^{n}$
is, where $\alpha_{M}$ and $\alpha_{N}$ are charts for $M$ and
$N$. Intuitively, a diffeomorphism like a homeomorphism can be thought
of as arbitrary stretching and bending, but it is ``nicer'' in that
it preserves the differentiable structure.

\noindent %
\begin{framed}%
\noindent $\triangle$ It is important to distinguish between coordinate
transformations, which are locally defined and so may have singularities
outside of a given region; and diffeomorphisms, which are globally
defined and form a group. One can define a coordinate transformation
on a region of a manifold that avoids any resulting singularities,
but a diffeomorphism must be smooth on the entire manifold. \end{framed}

\subsection{\label{subsec:The-differential-and-pullback}The differential and
pullback }

If we consider a general mapping between manifolds $\Phi\colon M^{m}\to N^{n}$,
we can choose charts $\alpha_{M}\colon M\to\mathbb{R}^{m}$ and $\alpha_{N}\colon N\to\mathbb{R}^{n}$,
with coordinate functions $x^{\mu}$ and $y^{\nu}$, so that the mapping
$\alpha_{N}\circ\Phi\colon M\to\mathbb{R}^{n}$ can be represented
by $n$ functions $\Phi^{\nu}\colon M\to\mathbb{R}$. This allows
us to write down an expression for the induced \textbf{tangent mapping}\index{tangent mapping}
or \textbf{differential}\index{differential} (aka pushforward\index{pushforward},
derivative\index{derivative}) $\mathrm{d}\Phi\colon TM\to TN$ (also
denoted $T\Phi$ or $\Phi_{*}$ or sometimes simply $\Phi$ if it
is clear the argument is a tangent vector). For a tangent vector $v=v^{\mu}\partial/\partial x^{\mu}$
at a point $p\in M$ we define

\begin{equation}
\left.\mathrm{d}\Phi\left(v\right)\right|_{p}\equiv\left.v^{\mu}\frac{\partial\Phi^{\nu}}{\partial x^{\mu}}\frac{\partial}{\partial y^{\nu}}\right|_{\Phi\left(p\right)}.
\end{equation}
This definition can be shown to be coordinate-independent and to follow
our intuitive expectation that mapped tangent vectors stay tangent
to mapped curves. If $M=N$ and $\Phi$ is the identity, $\mathrm{d}\Phi$
is just the vector component transformation in Section \ref{subsec:Tangent-vectors-and-differential-forms}.
The matrix 
\begin{equation}
J_{\Phi}(x)\equiv\frac{\partial\Phi^{\nu}}{\partial x^{\mu}}
\end{equation}
is called the \textbf{Jacobian matrix}\index{Jacobian matrix} (AKA
Jacobian). For the parametrized curve $C\colon\mathbb{R}\to N^{n}$,
we define the tangent to the curve at $t\in\mathbb{R}$ to be
\begin{equation}
\begin{aligned}\dot{C}\left(t\right) & \equiv\left.\mathrm{d}C\left(\frac{\partial}{\partial x}\right)\right|_{t}\\
 & =\left.\frac{\partial C^{\lambda}}{\partial x}\frac{\partial}{\partial y^{\lambda}}\right|_{C\left(t\right)},
\end{aligned}
\end{equation}
which is also denoted $\frac{\mathrm{d}C\left(t\right)}{\mathrm{d}t}$
and coincides with the Euclidean tangent to a curve if $N=\mathbb{R}^{n}$.

If $\Phi$ is a diffeomorphism, $\mathrm{d}\Phi$ is an isomorphism
between the tangent spaces at every point in $M$. The \textbf{inverse
function theorem}\index{inverse function theorem} says that the converse
is true locally: if $\mathrm{d}\Phi_{p}$ is an isomorphism at $p\in M$,
then $\Phi$ is locally a diffeomorphism. In particular, this means
that if in some coordinates the Jacobian matrix is nonsingular, then
$\alpha_{N}\circ\Phi\circ\alpha_{M}^{-1}$ represents a locally valid
coordinate transformation and $\Phi^{\nu}=y^{\nu}$. 

A mapping between manifolds $\Phi\colon M^{m}\to N^{n}$ also can
be used to naturally define the \textbf{pullback}\index{pullback}
of a form $\Phi^{*}\colon\Lambda^{k}N\to\Lambda^{k}M$ by 
\begin{equation}
\Phi^{*}\varphi\left(v_{1},\dotsc,v_{k}\right)=\varphi\left(\mathrm{d}\Phi\left(v_{1}\right),\dotsc,\mathrm{d}\Phi\left(v_{k}\right)\right),
\end{equation}
where the name indicates that a form on $N$ can be ``pulled back”
to $M$ using $\Phi$. Note that the composition of pullbacks is then
\begin{equation}
\Psi^{*}\Phi^{*}\varphi=(\Phi\Psi)^{*}\varphi.
\end{equation}

\begin{figure}[H]
\noindent \begin{centering}
\includegraphics[width=0.8\columnwidth]{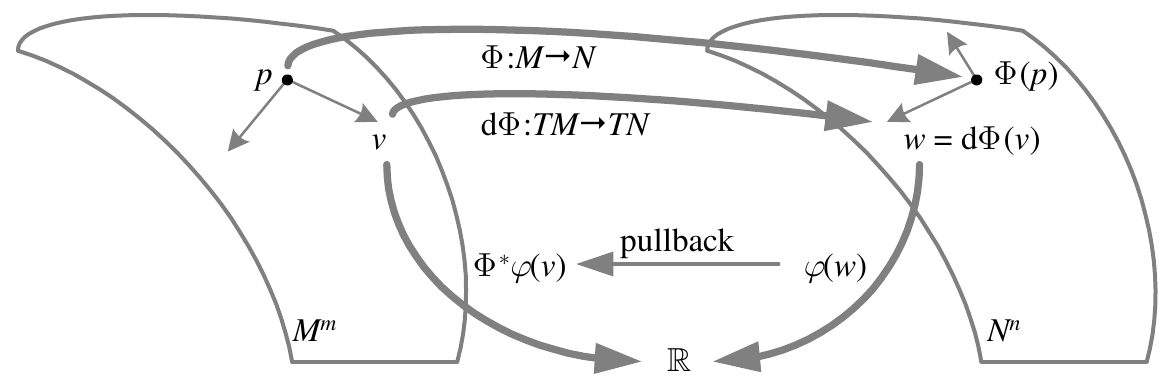}
\par\end{centering}
\caption{Forms $\varphi$ on $N$ are pulled back to $M$ by sending argument
vectors to $N$ using $\mathrm{d}\Phi$.}
\end{figure}

Note that for a mapping $f\colon M\to\mathbb{R}$, we have $\mathrm{d}f\colon TM\to T\mathbb{R}\cong\mathbb{R}$,
so that $\mathrm{d}f\left(v\right)=v^{\mu}\partial f/\partial x^{\mu}=v\left(f\right)$,
the directional derivative of $f$. Let us apply this to the coordinate
function $x^{1}\colon M\to\mathbb{R}$. Then we have $\mathrm{d}x^{1}\left(v\right)=v^{\mu}\partial x^{1}/\partial x^{\mu}=v^{1}$,
so that in particular $\mathrm{d}x^{\nu}\left(\partial/\partial x^{\mu}\right)=?\delta^{\nu}{}_{\mu}?$,
i.e. $\mathrm{d}x^{\mu}$ is in fact the dual frame to $\partial/\partial x^{\mu}$.
Thus in a given coordinate system, we can write a general tensor of
type $(m,n)$ as

\begin{equation}
T=?T^{\mu_{1}\dots\mu_{m}}{}_{\nu_{1}\dots\nu_{n}}?\frac{\partial}{\partial x^{\mu_{1}}}\otimes\dotsb\otimes\frac{\partial}{\partial x^{\mu_{m}}}\otimes\mathrm{d}x^{\nu_{1}}\otimes\dotsb\otimes\mathrm{d}x^{\nu_{n}}.
\end{equation}
In particular, the metric tensor is often written 
\begin{equation}
\mathrm{d}s^{2}\equiv g=g_{\mu\nu}\mathrm{d}x^{\mu}\mathrm{d}x^{\nu},
\end{equation}
where the Einstein summation convention is used and the tensor symbol
omitted. A general $k$-form $\varphi\in\Lambda^{k}M$ can then be
written as

\begin{equation}
\varphi=\underset{\mu_{1}<\dotsb<\mu_{k}}{\sum}\varphi_{\mu_{1}\dots\mu_{k}}\mathrm{d}x^{\mu_{1}}\wedge\dotsb\wedge\mathrm{d}x^{\mu_{k}}.
\end{equation}
From either the tangent mapping definition or the behavior of the
exterior product under a change of basis, we can see that under a
change of coordinates we have
\begin{equation}
\mathrm{d}y^{\mu_{1}}\wedge\dotsb\wedge\mathrm{d}y^{\mu_{k}}=\textrm{det}\left(\frac{\partial y^{\nu}}{\partial x^{\mu}}\right)\mathrm{d}x^{\mu_{1}}\wedge\dotsb\wedge\mathrm{d}x^{\mu_{k}}.
\end{equation}
This is the familiar \textbf{Jacobian determinant}\index{Jacobian determinant}
(like the Jacobian matrix, also often called the Jacobian) that appears
in the change of coordinates rule for integrals from calculus, and
explains the name of the volume form as defined previously in terms
of the exterior product. 

In summary, the differential $\mathrm{d}$ has a single definition,
but is used in several different settings that are not related in
an immediately obvious way.

\begin{table}[H]
\begin{tabular*}{1\columnwidth}{@{\extracolsep{\fill}}|l|>{\raggedright}p{0.25\columnwidth}|>{\raggedright}p{0.25\columnwidth}|>{\raggedright}p{0.2\columnwidth}|}
\hline 
Construct & Argument & Other names & Other symbols\tabularnewline
\hline 
\hline 
$\mathrm{d}\Phi\colon TM\to TN$ & $\Phi\colon M\to N$ & Tangent mapping & $T\Phi$, $\Phi_{*}$, $\Phi$\tabularnewline
\hline 
$\mathrm{d}f\colon TM\to\mathbb{R}$ & $f\colon M\to\mathbb{R}$ & Directional derivative & $v\left(f\right)$, $\mathrm{d}_{v}f$, $\nabla_{v}f$\tabularnewline
\hline 
$\mathrm{d}x^{\mu}\colon TM\to\mathbb{R}$ & $x^{\mu}\colon M\to\mathbb{R}$ & Dual frame to $\partial/\partial x^{\mu}$ & $\beta^{\mu}$\tabularnewline
\hline 
\end{tabular*}

\caption{Various uses of the differential on manifolds.}
\end{table}

\subsection{Immersions and embeddings }

We can generalize and make precise the concept of a surface embedded
in 3-dimensional space with the following definitions concerning a
differentiable map $\Phi\colon M^{m}\to N^{n}$:
\begin{itemize}
\item \textbf{Immersion}\index{immersion}: $\mathrm{d}\Phi$ is injective
for all $p\in M$; intuitively, a smooth mapping that doesn't collapse
the tangent spaces 
\item \textbf{Submanifold}\index{submanifold}: an immersion with $\Phi$
injective; intuitively, an immersion that doesn't intersect itself 
\item \textbf{Embedding}\index{manifold!embedding}\index{embedding} (AKA
imbedding\index{imbedding}): a submanifold with $\Phi$ a homeomorphism
onto $\Phi\left(M\right)$; intuitively, a submanifold that doesn't
have intersecting limit points 
\end{itemize}
\begin{figure}[H]
\noindent \begin{centering}
\includegraphics[width=0.8\columnwidth]{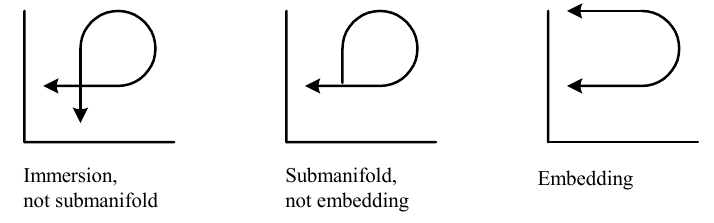}
\par\end{centering}
\caption{$\mathbb{R}$ immersed in $\mathbb{R}^{2}$; the second immersion
approaches a self-intersection in the limit as the line approaches
infinity.}
\end{figure}

The difference in dimension $\left(n-m\right)$ is called the \textbf{codimension}\index{codimension}
of the embedding. The \textbf{Whitney embedding theorem}\index{Whitney embedding theorem}
states that for positive codimension, any $M^{m}$ can be immersed
in $\mathbb{R}{}^{(2m-1)}$ and embedded in $\mathbb{R}^{2m}$. Thus
we can view differentiable manifolds as generalized surfaces that
we study without making reference to the enclosing Euclidean space.
The limiting dimension of this theorem is illustrated by noting that
the real projective space $\mathbb{R}\textrm{P}^{m}$ cannot be embedded
in $\mathbb{R}^{(2m-1)}$.

\section{\label{sec:Derivatives-on-manifolds}Derivatives on manifolds }

In this section we will introduce various objects that in some way
measure how vectors or forms change from point to point on a manifold.

\subsection{\label{subsec:Derivations}Derivations}

In general, we define a \textbf{derivation}\index{derivation} to
be a linear map $\mathcal{D}\colon\mathfrak{a}\to\mathfrak{a}$ on
an algebra $\mathfrak{a}$ that follows the \textbf{Leibniz rule}\index{Leibniz rule}
(AKA product rule\index{product rule})

\begin{flushleft}
\begin{equation}
\mathcal{D}(AB)=(\mathcal{D}A)B+A(\mathcal{D}B).
\end{equation}
As noted previously in Section \ref{subsec:Tangent-vectors-and-differential-forms},
the set $\mathrm{vect}(M)$ of vector fields on a manifold form a
Lie algebra; the Lie bracket operation with a fixed vector field $\left[u,\;\right]$
is then a derivation on this algebra, since the Leibniz rule
\par\end{flushleft}

\begin{equation}
\left[u,\left[v,w\right]\right]=\left[\left[u,v\right],w\right]+\left[v,\left[u,w\right]\right]
\end{equation}
is just the Jacobi identity.

For a graded algebra, e.g. the exterior algebra, the \textbf{degree}\index{degree}
of a derivation is the integer $c$ where $\mathcal{D}\colon\Lambda^{k}M\to\Lambda^{k+c}M$.
A \textbf{graded derivation}\index{graded derivation} is defined
to follow the \textbf{graded Leibniz rule}\index{graded Leibniz rule},
e.g. for a $k$-form $\varphi$,

\begin{flushleft}
\begin{equation}
\mathcal{D}\left(\varphi\wedge\psi\right)=\mathcal{D}\varphi\wedge\psi+\left(-1\right)^{kc}\varphi\wedge\mathcal{D}\psi.
\end{equation}
If $c$ is odd, a graded derivation is sometimes called an \textbf{anti-derivation}\index{anti-derivation}
(AKA skew-derivation\index{skew-derivation}).
\par\end{flushleft}

\subsection{\label{subsec:The-Lie-derivative-of-a-vector-field}The Lie derivative
of a vector field }

Without some kind of additional structure, there is no way to ``transport''
vectors, or compare them at different points on a manifold, and therefore
no way to construct a vector derivative. The simplest way to introduce
this structure is via another vector field, which leads us to the
\textbf{Lie derivative}\index{derivative!Lie} 
\begin{equation}
L_{v}w\equiv\left[v,w\right].
\end{equation}
As noted above, $L_{v}$ is a derivation due to the Jacobi identity.
In this section we define the Lie derivative in terms of infinitesimal
vector transport, and explore its geometrical meaning.

Given any vector field $v$ on $M$, it can be shown (\cite{FrankelGeom}
pp. 125-127) that there exists a parameterized curve $v_{p}(t)$ at
every point $p\in M$ such that $v_{p}(0)=p$ and $\dot{v}_{p}(t)$
is the value of the vector field $v$ at the point $v_{p}(t)$ (the
dot indicates the derivative with respect to $t$, which as usual
is calculated on the curve mapped to $\mathbb{R}^{n}$ by the coordinate
chart). Each curve in this family is in general only well-defined
locally, i.e. for $-\varepsilon<t<\varepsilon$, and is thus called
the \textbf{local flow}\index{local flow} of $v$.

\begin{figure}[H]
\noindent \begin{centering}
\includegraphics[width=0.8\columnwidth]{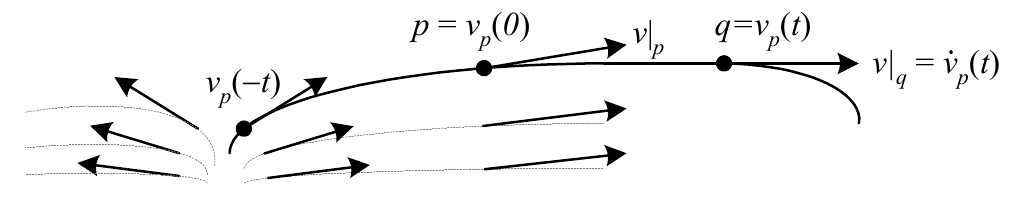}
\par\end{centering}
\caption{A depiction of the local flow of a vector field $v$, with details
on the local parameterized curve $v_{p}(t)$ at a point $p$.}
\end{figure}

For a fixed value of $t$, there is some region $U\subset M$ where
the map $\Phi_{t}\colon U\to U$ defined by $p\mapsto v_{p}\left(t\right)$
is a diffeomorphism, and within the valid domain of $t$ the maps
$\Phi_{t}$ satisfy the abelian group law $\Phi_{t}\circ\Phi_{s}=\Phi_{t+s}$;
thus the $\Phi_{t}$ are called a \textbf{local one-parameter group
of diffeomorphisms}\index{local one-parameter group of diffeomorphisms}.
This name is somewhat misleading, since due to the limited domain
of $t$ the maps $\Phi_{t}$ do not actually form a group; the ``local''
reflects the fact that the diffeomorphisms are not on all of $M$.
In the case that these maps are in fact valid for all of $t$ and
$M$, $v$ is called a \textbf{complete vector field}\index{complete vector field},
and the $\Phi_{t}$ are called a\textbf{ one-parameter group of diffeomorphisms}\index{one-parameter group of diffeomorphisms}.
If $M$ is compact, then every vector field is complete; if not, then
a vector field is complete if it has \textbf{compact support}\index{compact support}
(is non-zero on a compact subset of $M$). 

The tangent map $\mathrm{d}\Phi$ defined by the vector field $v$
is then the extra structure we need to ``transport'' vectors. $\mathrm{d}\Phi$
maps a vector tangent to the curve $C$ to a vector tangent to the
curve $\Phi\left(C\right)$; it ``pushes vectors along the flow of
$v$.'' We can now define the Lie derivative as a limit

\begin{equation}
\begin{aligned}L_{v}w & \equiv\underset{\varepsilon\rightarrow0}{\textrm{lim}}\frac{1}{\varepsilon}\left[\mathrm{d}\Phi_{-\varepsilon}\left(w\left|_{v_{p}\left(\varepsilon\right)}\right.\right)-w\left|_{p}\right.\right]\\
 & =\underset{\varepsilon\rightarrow0}{\textrm{lim}}\frac{1}{\varepsilon}\left[w\left|_{v_{p}\left(\varepsilon\right)}\right.-\mathrm{d}\Phi_{\varepsilon}\left(w\left|_{p}\right.\right)\right]\\
 & =\underset{\varepsilon\rightarrow0}{\textrm{lim}}\frac{1}{\varepsilon}\left[w\left|_{p}\right.-\mathrm{d}\Phi_{\varepsilon}\left(w\left|_{v_{p}\left(-\varepsilon\right)}\right.\right)\right].
\end{aligned}
\end{equation}

\begin{figure}[H]
\noindent \begin{centering}
\includegraphics[width=0.8\columnwidth]{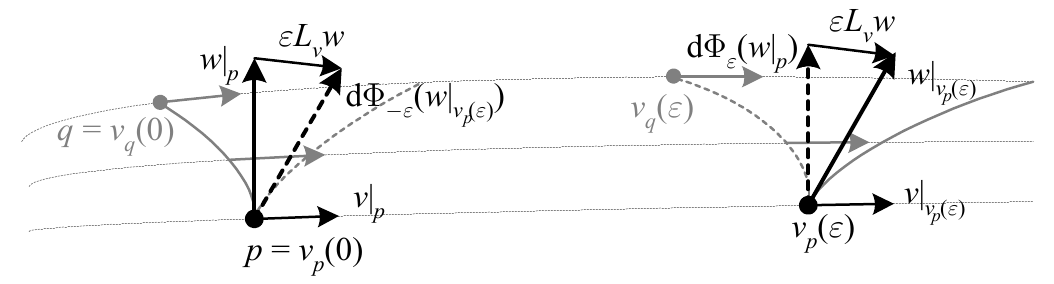}
\par\end{centering}
\caption{\label{fig:The-Lie-derivative}The Lie derivative $L_{v}w$ is \textquotedblleft the
difference between $w$ and its transport by the local flow of $v$.\textquotedblright{} }
\end{figure}

\noindent %
\begin{framed}%
\noindent \sun{} In this and future depictions of vector derivatives,
the situation is simplified by focusing on the change in the vector
field $w$ while showing the ``transport'' of $w$ as a parallel
displacement. This has the advantage of highlighting the equivalency
of defining the derivative at either 0 or $\varepsilon$ in the limit
$\varepsilon\rightarrow0$. Depicting $L_{v}w$ as a non-parallel
vector at $v_{p}\left(t\right)$ would be more accurate, but would
obscure this fact. We also will follow the picture here in using words
to characterize derivatives: namely, ``the difference'' is short
for ``the difference per unit $\varepsilon$ to order $\varepsilon$
in the limit $\varepsilon\rightarrow0$.''\end{framed}

This definition can be shown to be equivalent to $L_{v}w\equiv\left[v,w\right]$.
Another way of depicting the Lie derivative that highlights the anti-commutativity
of the Lie bracket is to consider $L_{v}w$ in terms of a loop defined
by the flows of $v$ and $w$. 

\begin{figure}[H]
\noindent \begin{centering}
\includegraphics[width=0.8\columnwidth]{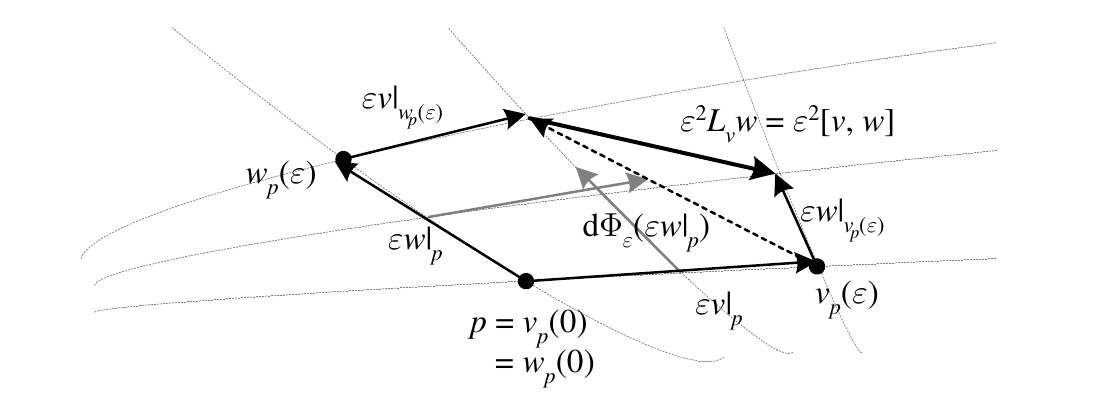}
\par\end{centering}
\caption{The Lie derivative $L_{v}w$ can also be pictured as the vector field
whose local flow is the \textquotedblleft commutator of the flows
of $v$ and $w$,\textquotedblright{} i.e. it is the difference between
the local flow of $v$ followed by $w$ and that of $w$ followed
by $v$. Thus $L_{v}w$ \textquotedblleft completes the parallelogram\textquotedblright{}
formed by the flow lines.}
\end{figure}

\subsection{The Lie derivative of an exterior form }

The Lie derivative $L_{v}$ can be applied to a $k$-form $\varphi$
by using the pullback of $\varphi$ by the diffeomorphism $\Phi$
associated with the flow of $v$, i.e. applied to $k$ vectors $w_{1},\ldots,w_{k}$
we define

\begin{equation}
L_{v}\varphi\left(w_{1},\ldots,w_{k}\right)\equiv\underset{\varepsilon\rightarrow0}{\textrm{lim}}\frac{1}{\varepsilon}\left[\varphi\left(\mathrm{d}\Phi_{\varepsilon}\left(w_{1},\ldots,w_{k}\right)\right)-\varphi\left(w_{1},\ldots,w_{k}\right)\right].
\end{equation}
$L_{v}\varphi$ thus measures the change in $\varphi$ as its arguments
are transported by the local flow of $v$. In the case of a 0-form
$f$, this is just the differential or directional derivative $L_{v}f=v(f)=\mathrm{d}f(v)$.

\begin{figure}[H]
\noindent \begin{centering}
\includegraphics[width=0.8\columnwidth]{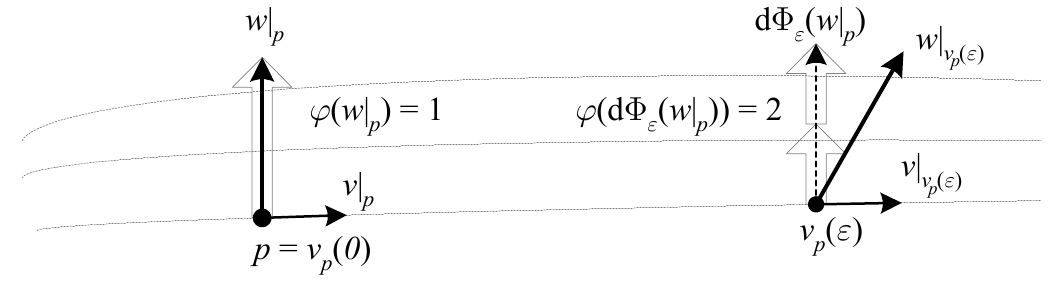}
\par\end{centering}
\caption{The Lie derivative illustrated for a 1-form $\varphi$ with $\varepsilon=1$.
$L_{v}\varphi$ is \textquotedblleft the difference between $\varphi$
applied to $w$ and $\varphi$ applied to $w$ transported by the
local flow of $v$,\textquotedblright{} so above we have $L_{v}\varphi(w)=2-1=1$
(valid in the limit $\varepsilon\rightarrow0$ if $\varphi$ changes
linearly in the range shown).}
\end{figure}

By using the above definitions of the Lie derivative applied to vectors
and 1-forms, and noting that we can derive a Leibniz rule over contraction
$L_{v}(\varphi\left(w\right))=\left(L_{v}\varphi\right)(w)+\varphi\left(L_{v}w\right)$,
we arrive at an expression for the Lie derivative applied to general
tensors, viewed as real-valued mappings on vectors and 1-forms:
\begin{equation}
\begin{aligned}L_{v}T(\varphi_{1},\ldots,\varphi_{m},w_{1},\ldots,w_{n}) & =v\left(T(\varphi_{1},\ldots,\varphi_{m},w_{1},\ldots,w_{n})\right)\\
 & -\sum_{j=1}^{m}T\left(\varphi_{1},\ldots,L_{v}\varphi_{j},\ldots,\varphi_{m},w_{1},\ldots,w_{n}\right)\\
 & -\sum_{j=1}^{n}T\left(\varphi_{1},\ldots,\varphi_{m},w_{1},\ldots,L_{v}w_{j},\ldots,w_{n}\right)
\end{aligned}
\end{equation}
In a holonomic frame, this yields an expression for the Lie derivative
of a tensor in terms of coordinates 
\begin{equation}
\begin{aligned}L_{v}?T^{\mu_{1}\dots\mu_{m}}{}_{\sigma_{1}\dots\sigma_{n}}? & =v^{\lambda}\frac{\partial}{\partial x^{\lambda}}?T^{\mu_{1}\dots\mu_{m}}{}_{\sigma_{1}\dots\sigma_{n}}?\\
 & -\sum_{j=1}^{m}\left(\frac{\partial v^{\mu_{j}}}{\partial x^{\lambda}}\right)?T^{\mu_{1}\dots\mu_{j-1}\lambda\mu_{j+1}\dots\mu_{m}}{}_{\sigma_{1}\dots\sigma_{n}}?\\
 & +\sum_{j=1}^{n}\left(\frac{\partial v^{\lambda}}{\partial x^{\sigma_{j}}}\right)?T^{\mu_{1}\dots\mu_{m}}{}_{\sigma_{1}\dots\sigma_{j-1}\lambda\sigma_{j+1}\dots\sigma_{n}}?.
\end{aligned}
\end{equation}
From this we can confirm that the Lie derivative satisfies the Leibniz
rule over the tensor product, and therefore is a derivation of degree
0 on both the tensor algebra and the exterior algebra.

\subsection{\label{subsec:The-exterior-derivative-of-a-1-form}The exterior derivative
of a 1-form }

The Lie derivative $L_{v}\varphi$ is defined in terms of a vector
field $v$, and its value as a ``change in $\varphi$'' is computed
by using $v$ to transport the arguments of $\varphi$. In contrast,
recall that the differential $\mathrm{d}$ takes a 0-form $f\colon M\to\mathbb{R}$
to a 1-form $\mathrm{d}f\colon TM\to\mathbb{R}$ with 
\begin{equation}
\mathrm{d}f(v)=v(f).
\end{equation}
Thus $\mathrm{d}$ is a derivation of degree +1 on 0-forms, whose
value as a ``change in $f$'' is computed using the vector field
argument of the resulting 1-form.

We would like to generalize $\mathrm{d}$ to $k$-forms by extending
this idea of including the ``direction argument'' by increasing
the degree of the form. It turns out that if we also require the property
\begin{equation}
\mathrm{d}\left(\mathrm{d}\left(\varphi\right)\right)=0
\end{equation}
(or ``$\mathrm{d}^{2}=0$''), there is a unique graded derivation
of degree +1 that extends $\mathrm{d}$ to general $k$-forms; this
derivation is called the \textbf{exterior derivative}\index{exterior derivative}.
We first explore the exterior derivative of a 1-form.

The exterior derivative of a 1-form is defined by
\begin{equation}
\mathrm{d}\varphi\left(v,w\right)\equiv v\left(\varphi\left(w\right)\right)-w\left(\varphi\left(v\right)\right)-\varphi\left(\left[v,w\right]\right),
\end{equation}
where e.g. 

\begin{equation}
v\left(\varphi\left(w\right)\right)=\underset{\varepsilon\rightarrow0}{\textrm{lim}}\frac{1}{\varepsilon}\left[\varphi\left(w\left|_{v_{p}\left(\varepsilon\right)}\right.\right)-\varphi\left(w\left|_{p}\right.\right)\right]
\end{equation}
measures the change in $\varphi\left(w\right)$ in the direction $v$,
so that

\begin{equation}
\begin{aligned}\mathrm{d}\varphi\left(v,w\right) & =\underset{\varepsilon\rightarrow0}{\textrm{lim}}\frac{1}{\varepsilon^{2}}\left[\left(\varphi\left(\varepsilon w\left|_{v_{p}\left(\varepsilon\right)}\right.\right)-\varphi\left(\varepsilon w\left|_{p}\right.\right)\right)\right.\\
 & \phantom{{=\underset{\varepsilon\rightarrow0}{\textrm{lim}}\frac{1}{\varepsilon^{2}}\left[-\right.}}-\left(\varphi\left(\varepsilon v\left|_{w_{p}\left(\varepsilon\right)}\right.\right)-\varphi\left(\varepsilon v\left|_{p}\right.\right)\right)\\
 & \phantom{{=\underset{\varepsilon\rightarrow0}{\textrm{lim}}\frac{1}{\varepsilon^{2}}\left[-\right.}}\left.-\varphi\left(\varepsilon^{2}\left[v,w\right]\right)\right].
\end{aligned}
\end{equation}
The term involving the Lie bracket ``completes the parallelogram''
formed by $v$ and $w$, so that $\mathrm{d}\varphi\left(v,w\right)$
can be viewed as the ``sum of $\varphi$ on the boundary of the surface
defined by its arguments.''

\begin{figure}[H]
\begin{centering}
\includegraphics[width=0.5\columnwidth]{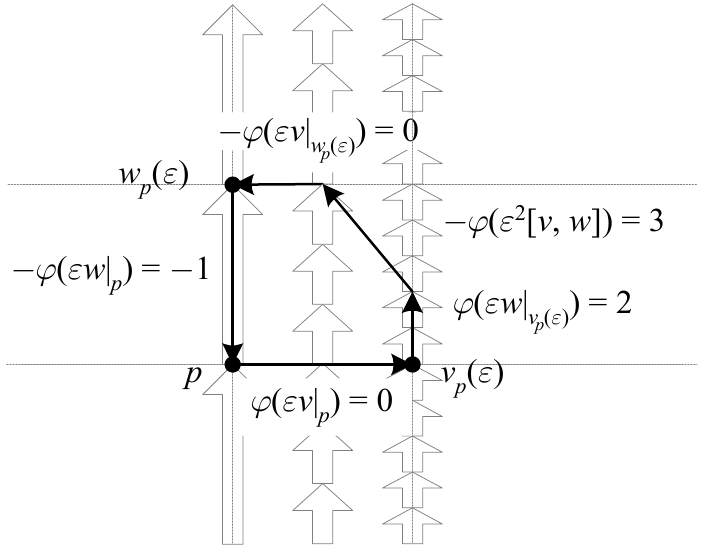}
\par\end{centering}
\caption{The exterior derivative of a 1-form $\mathrm{d}\varphi\left(v,w\right)$
is the sum of $\varphi$ along the boundary of the completed parallelogram
defined by $v$ and $w$. So if in the diagram $\varepsilon=1$, we
have $\mathrm{d}\varphi\left(v,w\right)=\left(2-1\right)-\left(0-0\right)+3=4$.
This value is valid in the limit $\varepsilon\rightarrow0$ if the
sum varies like $\varepsilon^{2}$ as depicted in the figure.}
\end{figure}

The identity $\mathrm{d}^{2}=0$ can then be seen as stating the intuitive
fact that the boundary of a boundary is zero. If $\varphi=\mathrm{d}f$,
then $\varphi\left(v\right)=\mathrm{d}f\left(v\right)=v\left(f\right)$,
the change in $f$ along $v$. Thus e.g. $\varepsilon\varphi\left(v\left|_{p}\right.\right)=f\left(v_{p}\left(\varepsilon\right)\right)-f\left(p\right)$,
so that the value of $\varphi$ on $v$ is the difference between
the values of $f$ on the two points which are the boundary of $v$.
Each endpoint will be cancelled by a starting point as we add up values
of $\varphi$ along a sequence of vectors, resulting in the difference
between the values of $f$ at the boundary of the total path defined
by these vectors. $\mathrm{d}\varphi$ is the value of $\varphi$
over the boundary path of the surface defined by its arguments, which
has no boundary points and so vanishes.

\begin{figure}[H]
\noindent \begin{centering}
\includegraphics[width=0.8\columnwidth]{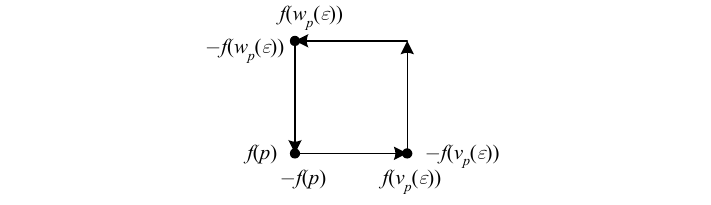}
\par\end{centering}
\caption{$\mathrm{d}^{2}=0$ corresponds to the boundary of a boundary is zero:
each term $\varphi(v)=\mathrm{d}f(v)$ is the difference between the
values of $f$ on the boundary points of $v$, which cancel as we
traverse the boundary of the surface defined by the arguments of $\mathrm{d}\varphi(v,w)$.
In the figure we assume a vanishing Lie bracket for simplicity.}
\end{figure}

Note that $\mathrm{d}\varphi\left(v,w\right)$ measures the interaction
between $\varphi$ and the vector fields $v$ and $w$, thus avoiding
the need to ``transport'' vectors. In particular, a non-zero exterior
derivative can be pictured as resulting from either the vector fields
or $\varphi$ ``changing,'' i.e. changing with regard to the implied
coordinates of our pictures.

\begin{figure}[H]
\noindent \begin{centering}
\includegraphics[width=0.7\columnwidth]{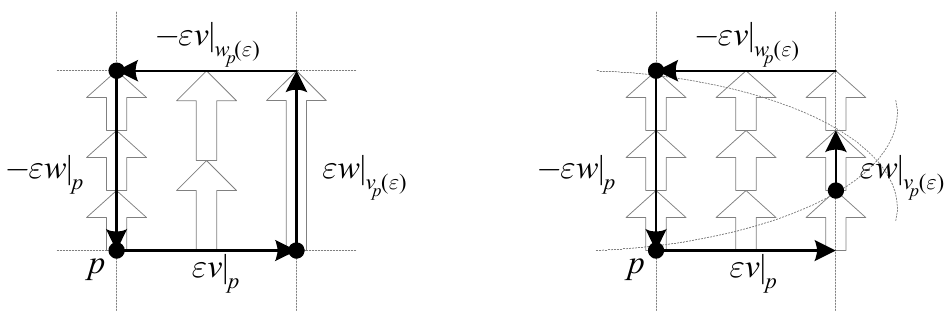}
\par\end{centering}
\caption{A non-zero exterior derivative $\mathrm{d}\varphi\left(v,w\right)$
results from changes in $\varphi\left(v\right)$ or $\varphi\left(w\right)$,
not changes in either $\varphi$ or the vector fields alone as compared
to some transport.}
\end{figure}

If we calculate $\mathrm{d}\varphi\left(e_{1},e_{2}\right)$ explicitly
in a holonomic frame in two dimensions, $\mathrm{d}\left(\varphi_{1}\mathrm{d}x^{1}+\varphi_{2}\mathrm{d}x^{2}\right)=\mathrm{d}\varphi_{1}\wedge\mathrm{d}x^{1}+\mathrm{d}\varphi_{2}\wedge\mathrm{d}x^{2}$,
so applying this to the basis vector fields $e_{1}$ and $e_{2}$
we have
\begin{equation}
\begin{aligned}\mathrm{d}\varphi\left(e_{1},e_{2}\right) & =\mathrm{d}\varphi_{1}\left(e_{1}\right)\cdot\mathrm{d}x^{1}\left(e_{2}\right)-\mathrm{d}\varphi_{1}\left(e_{2}\right)\cdot\mathrm{d}x^{1}\left(e_{1}\right)\\
 & \phantom{{}=}+\mathrm{d}\varphi_{2}\left(e_{1}\right)\cdot\mathrm{d}x^{2}\left(e_{2}\right)-\mathrm{d}\varphi_{2}\left(e_{2}\right)\cdot\mathrm{d}x^{2}\left(e_{1}\right)\\
 & =e_{1}\left(\varphi_{2}\right)-e_{2}\left(\varphi_{1}\right)\\
 & =\frac{\partial\varphi_{2}}{\partial x^{1}}-\frac{\partial\varphi_{1}}{\partial x^{2}}.
\end{aligned}
\end{equation}
Note that a holonomic dual frame $\beta^{\mu}=\mathrm{d}x^{\mu}$
satisfies $\mathrm{d}\beta^{\mu}=\mathrm{dd}x^{\mu}=0$.

\subsection{\label{subsec:The-exterior-derivative-of-a-k-form}The exterior derivative
of a k-form }

The extension of the coordinate-free definition of $\mathrm{d}$ to
general $k$-forms gives the expression

\begin{equation}
\begin{aligned} & \mathrm{d}\varphi\left(v_{0},\dotsc,v_{k}\right)\\
 & \equiv\sum_{j=0}^{k}\left(-1\right)^{j}v_{j}\left(\varphi\left(v_{0},\dotsc,v_{j-1},v_{j+1},\dotsc,v_{k}\right)\right)\\
 & \phantom{{}=}+\sum_{i<j}\left(-1\right)^{i+j}\varphi\left(\left[v_{i},v_{j}\right],v_{0},\dotsc,v_{i-1},v_{i+1},\dotsc,v_{j-1},v_{j+1},\dotsc,v_{k}\right).
\end{aligned}
\end{equation}

Our picture of $\mathrm{d}^{2}=0$ for 1-forms then can be extended
to higher dimensions. For example, assuming vanishing Lie brackets
to simplify the picture, the exterior derivative of a 2-form $\mathrm{d}\varphi\left(u,v,w\right)$
can be viewed as the ``sum of $\varphi$ on the boundary faces of
the cube defined by its arguments.'' If $\varphi=\mathrm{d}\psi\left(v,w\right)$
is the boundary of a face, $\mathrm{d}\varphi=\mathrm{d}^{2}\psi$
is the sum of the boundaries of the faces; each edge is then counted
by two faces with opposite signs, thus canceling and confirming that
$\mathrm{d}^{2}=0$.

\begin{figure}[H]
\noindent \begin{centering}
\includegraphics[width=0.8\columnwidth]{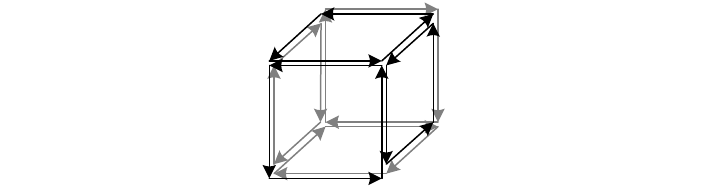}
\par\end{centering}
\caption{\label{fig:The-3-form}The 3-form $\mathrm{d}\varphi=\mathrm{d}^{2}\psi$
sums $\psi$ over the edges of the faces of a cube. The sum vanishes
since each edge is counted twice with opposite signs.}
\end{figure}

In a holonomic frame, we can obtain an expression for $\mathrm{d}\varphi$
in terms of coordinates
\begin{equation}
\begin{aligned}\mathrm{d}\varphi & =\sum_{\mu_{0}<\dotsb<\mu_{k}}\left(\sum_{j=0}^{k}\left(-1\right)^{j}\frac{\partial}{\partial x^{\mu_{j}}}\varphi_{\mu_{0}\dots\mu_{j-1}\mu_{j+1}\dots\mu_{k}}\right)\mathrm{d}x^{\mu_{0}}\wedge\dotsb\wedge\mathrm{d}x^{\mu_{k}}\\
 & =\frac{\partial}{\partial x^{\mu_{0}}}\varphi_{\mu_{1}\dots\mu_{k}}\mathrm{d}x^{\mu_{0}}\wedge\dotsb\wedge\mathrm{d}x^{\mu_{k}}\\
 & =\frac{\partial\varphi_{I}}{\partial x^{\mu_{0}}}\mathrm{d}x^{\mu_{0}}\wedge\mathrm{d}x^{I},
\end{aligned}
\end{equation}
so that in terms of array components we have

\begin{equation}
\begin{aligned}\left(\mathrm{d}\varphi\right)_{\mu_{0}\dots\mu_{k}} & =\sum_{j=0}^{k}\left(-1\right)^{j}\frac{\partial}{\partial x^{\mu_{j}}}\varphi_{\mu_{0}\dots\mu_{j-1}\mu_{j+1}\dots\mu_{k}}.\end{aligned}
\end{equation}
It is not hard to see that the exterior derivative commutes with the
pullback, i.e. $\Phi^{*}\mathrm{d}\varphi=\mathrm{d}\Phi^{*}\varphi$. 

\noindent %
\begin{framed}%
\noindent $\triangle$ Despite a convenient description using coordinates
associated with a holonomic frame, it is important to keep in mind
that the exterior derivative of a form is frame- and coordinate-independent. \end{framed}

If we include an inner product, vector calculus can be seen to correspond
to exterior calculus on $\mathbb{R}^{3}$, and can thus be generalized
to arbitrary dimensions:
\begin{itemize}
\item For a function (0-form) $f$, the components of the 1-form $\mathrm{d}f$
correspond to those of the gradient\index{gradient} of $f$, i.e.
$\left(\mathrm{d}f\right)_{\mu}=(\nabla f)^{\mu}$ or $\nabla f=(\mathrm{d}f)^{\sharp}$;
a generalization of the gradient is then the 1-form $\mathrm{d}f$
\item For a 1-form with components equal to those of a vector $\varphi_{\mu}=v^{\mu}$,
the components of $\mathrm{d}\varphi$ correspond to those of the
curl\index{curl} of $v$, i.e. $(\mathrm{d}\varphi)_{\mu}=(\nabla\times v)^{\mu}$
or $(\nabla\times v)=(*\mathrm{d}(v^{\flat}))^{\sharp}$; a generalization
of the curl is then the 2-form $\mathrm{d}\varphi$ 
\item For a 2-form with components equal to those of a vector $\psi_{\mu}=(*\varphi)_{\mu}=v^{\mu}$,
the value of $\mathrm{d}\psi$ corresponds to the value of the divergence\index{divergence}
of $v$, i.e. $\mathrm{d}\psi=\nabla\cdot v$ or $\nabla\cdot v=*\mathrm{d}(*(v^{\flat}))$;
a generalization of the divergence is then the value $*\mathrm{d}(*\varphi)$ 
\end{itemize}
In $\mathbb{R}^{3}$ the relations curl grad = div curl = 0 thus correspond
to the property $\mathrm{d}^{2}=0$. Note that we have used the musical
isomorphisms on $\mathbb{R}^{3}$, which imply an inner product, as
does the Hodge star. The generalizations can be extended to a pseudo
inner product with signature $\left(r,s\right)$ by defining the divergence
as 
\begin{equation}
\mathrm{div}\left(v\right)\equiv(-1)^{s}*\mathrm{d}(*v^{\flat}),
\end{equation}
which is then independent of both signature and orientation.

Finally, the classical gradient, curl, and divergence integral theorems\index{divergence theorem}
in vector calculus are generalized to \textbf{Stokes' theorem}\index{Stokes' theorem}:
for an $(n-1)$-form $\varphi$ on a compact oriented manifold $M^{n}$
with boundary $\partial M$,

\begin{equation}
\int_{M}\,\mathrm{d}\varphi=\int_{\partial M}\varphi.
\end{equation}
This is essentially the integral form of the property $\mathrm{d}^{2}=0$:
summing $\mathrm{d}\varphi$ over $M$ can be pictured as summing
$\varphi$ over the boundaries of infinitesimal volumes, so that all
internal boundaries cancel and what is left is $\varphi$ over the
outer boundary $\partial M$.

\begin{figure}[H]
\begin{centering}
\includegraphics[width=0.5\columnwidth]{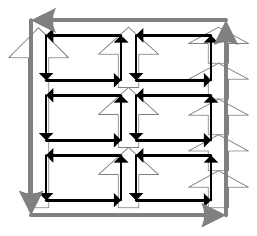}
\par\end{centering}
\caption{The integral of $\mathrm{d}\varphi$ over $M$ can be pictured as
summing $\varphi$ over the boundaries of infinitesimal volumes, so
that all internal boundaries cancel and what is left is $\varphi$
over the outer boundary $\partial M$.}
\end{figure}

We will not address the details of defining integration on manifolds
here, but the basic idea is relatively straightforward: a coordinate
chart maps an $n$-dimensional sub-manifold $U$ of $M^{n}$ to $S\in\mathbb{R}^{n}$;
an $n$-form $\varphi$ can then be written $f(x_{I})\,\mathrm{d}x^{I}$,
and its integral over $U$ is defined to be $\int_{S}f(x_{I})\,\mathrm{d}x^{I}$,
which can be shown to be coordinate-independent. Note that without
additional structure on the manifold, we cannot integrate functions
or other forms over $M^{n}$ besides $n$-forms (or pseudo $n$-forms
if $U$ is non-orientable).

We can further generalize the divergence to $k$-forms $\varphi$
by defining the \textbf{codifferential}\index{codifferential} (AKA
coderivative\index{coderivative}, exterior coderivative) 
\begin{align}
\begin{aligned}\delta\varphi & \equiv(-1)^{k}*^{-1}\mathrm{d}(*\varphi)\\
 & =(-1)^{n(k+1)+s+1}*\mathrm{d}(*\varphi)\\
\Rightarrow\mathrm{div}\left(v\right) & =-\delta v^{\flat}.
\end{aligned}
\end{align}
The map $\delta:\Lambda^{k}M\to\Lambda^{k-1}M$ does not follow the
Leibniz rule and so is not a derivation. However, we do have $\delta^{2}=0$,
so that we may write $\Delta\equiv\left(\mathrm{d}+\delta\right)^{2}=\mathrm{d}\delta+\delta\mathrm{d}$,
which (usually for $s=0$) is called the \textbf{Laplace-Beltrami
operator}\index{Laplace-Beltrami operator} (AKA Laplace operator,
Laplacian, Laplace–de Rham operator\index{Laplace–de Rham operator});
a form on a Riemannian manifold for which it vanishes is called a
\textbf{harmonic form}\index{harmonic form}. For $f\in\Lambda^{0}M$
we have $\delta f=0$ and $\Delta f=-\mathrm{div}\left(\nabla f\right)$;
despite the sign, this last is often itself written $\Delta f\equiv\nabla^{2}f\equiv\nabla\cdot\nabla f$,
except on a flat Lorentzian manifold where it is denoted $\square f=\partial^{\mu}\partial_{\mu}f\equiv-\partial_{t}^{2}f+\Delta f$,
where $\square$ is called the \textbf{d'Alembertian}\index{d'Alembertian}
(AKA d'Alembert operator\index{d'Alembert operator}, wave operator\index{wave operator},
box operator\index{box operator}), whose spatial part $\Delta$ is
called the \textbf{Laplacian}\index{Laplacian} (AKA Laplace operator\index{Laplace operator}).
$\Delta f=0$ is then called \textbf{Laplace's equation}\index{Laplace's equation},
while a fixed $\rho\in\Lambda^{0}M$ defines \textbf{Poisson's equation}\index{Poisson's equation}
$\Delta f=\rho$.%
\begin{framed}%
\noindent $\triangle$ In the mostly minuses signature on a Lorentzian
manifold, the operators above may be defined with a negative sign.
It is also important to remember that the sign of $\Delta$ may change
depending upon whether it is applied to a function or a form.\end{framed}

If $\varphi\in\Lambda^{k}M$ and $\psi\in\Lambda^{k+1}M$ so that
$\mathrm{d}\left(\varphi\wedge*\psi\right)\in\Lambda^{n}M$, it is
not hard to see that
\begin{align}
\int_{\partial M}\left(\varphi\wedge*\psi\right) & =\int_{M}\left\langle \mathrm{d}\varphi,\psi\right\rangle \Omega-\left\langle \varphi,\delta\psi\right\rangle \Omega,
\end{align}
which means that if $\varphi\wedge*\psi$ vanishes on $\partial M$
(or $\partial M=0$) we have $\int_{M}\left\langle \mathrm{d}\varphi,\psi\right\rangle \Omega=\int_{M}\left\langle \varphi,\delta\psi\right\rangle \Omega$.
In particular, for $f\in\Lambda^{0}M$ and $v^{\flat}\in\Lambda^{1}M$,
we have
\begin{align}
\int_{\partial M}f\left(*v^{\flat}\right) & =\int_{M}\left\langle \nabla f,v\right\rangle \Omega+f\mathrm{div}\left(v\right)\Omega,
\end{align}
or for $f=1$ and recalling that $i_{v}\Omega=*(v^{\flat})$,
\begin{align}
\int_{\partial M}*v^{\flat} & =\int_{M}\mathrm{div}\left(v\right)\Omega\\
\Rightarrow\int_{M}\mathrm{div}\left(v\right)\Omega & =\int_{\partial M}i_{v}\Omega\\
 & =\int_{\partial M}\left\langle v,\hat{n}\right\rangle i_{\hat{n}}\Omega,
\end{align}
where $\hat{n}$ is the unit normal vector to $\partial M$, the classical
\textbf{divergence theorem}\index{divergence theorem} (AKA Gauss's
theorem, Ostrogradsky's theorem).

\subsection{\label{subsec:Relationships-between-derivations}Relationships between
derivations }

We can define one other derivation on $k$-forms, the \textbf{interior
derivative}\index{interior derivative} (AKA inner derivative\index{inner derivative},
inner multiplication\index{inner multiplication}), which is the generalization
of the interior product to forms on manifolds, i.e. for a given vector
$v$ it is the graded degree $-1$ derivation 
\begin{equation}
\left(i_{v}\varphi\right)\left(w_{2},\dotsc,w_{k}\right)\equiv\varphi\left(v,w_{2},\dotsc,w_{k}\right)
\end{equation}
on $k$-forms $\varphi$, which follows the graded Leibniz rule 
\begin{equation}
i_{v}\left(\varphi\wedge\psi\right)=(i_{v}\varphi)\wedge\psi+\left(-1\right)^{k}\varphi\wedge(i_{v}\psi).
\end{equation}
The graded commutativity of forms immediately gives the property $i_{v}i_{w}+i_{w}i_{v}=i_{v}^{2}=0$.
We define $i_{v}f\equiv0$ for a 0-form $f$ and note that $i_{v}\Omega=*(v^{\flat})$. 

The interior, exterior, and Lie derivatives then form an infinite-dimensional
graded Lie algebra with the following relations:
\begin{itemize}
\item $\left[L_{v},L_{w}\right]\equiv L_{v}L_{w}-L_{w}L_{v}=L_{\left[v,w\right]}$
\item $\left[i_{v},i_{w}\right]\equiv i_{v}i_{w}+i_{w}i_{v}=0$
\item $\left[\mathrm{d},\mathrm{d}\right]\equiv\mathrm{d}^{2}+\mathrm{d}^{2}=0$
\item $\left[L_{v},i_{w}\right]\equiv L_{v}i_{w}-i_{w}L_{v}=i_{\left[v,w\right]}$
\item $\left[L_{v},\mathrm{d}\right]\equiv L_{v}\mathrm{d}-\mathrm{d}L_{v}=0$
\item $\left[i_{v},\mathrm{d}\right]\equiv i_{v}\mathrm{d}+\mathrm{d}i_{v}=L_{v}$
\end{itemize}
This last relation is sometimes called \textbf{Cartan's formula}\index{Cartan's formula}
(AKA Cartan's magic formula).

\section{\label{sec:The-divergence-currents-and-tensor-densities}The divergence,
currents, and tensor densities }

\subsection{Coordinate and tensor divergences}

Using the results from Section \ref{subsec:The-divergence-and-conserved-quantities},
we can derive many useful coordinate dependent relations. Adopting
the common abbreviation 
\begin{equation}
\sqrt{g}\equiv\sqrt{\left|\det\left(g_{\mu\nu}\right)\right|}
\end{equation}
and including torsion for completeness, we expand both sides of the
coordinate divergence expression to get
\begin{align}
\partial_{\lambda}\sqrt{g}=\sqrt{g}\left(?\Gamma^{\mu}{}_{\lambda\mu}?-?T^{\mu}{}_{\mu\lambda}?\right),
\end{align}
which along with the expressions for the metric derivative from Section
\ref{subsec:The-Levi-Civita-connection} yields
\begin{align}
\begin{aligned}\partial_{\lambda}\left(\sqrt{g}g^{\mu\nu}\right) & =\sqrt{g}\left(g^{\mu\nu}?\Gamma^{\sigma}{}_{\lambda\sigma}?-g^{\mu\nu}?T^{\sigma}{}_{\sigma\lambda}?-?\Gamma^{\mu\nu}{}_{\lambda}?-?\Gamma^{\nu\mu}{}_{\lambda}?\right)\\
\Rightarrow\partial_{\nu}\left(\sqrt{g}g^{\mu\nu}\right) & =-\sqrt{g}\left(?\Gamma^{\mu\nu}{}_{\nu}?-?T^{\nu\mu}{}_{\nu}?\right).
\end{aligned}
\end{align}

From $\det\left(\exp\left(g\right)\right)=\exp\left(\mathrm{tr}\left(g\right)\right)\Rightarrow\ln\left(\det\left(g\right)\right)=\mathrm{tr}\left(\ln\left(g\right)\right)$,
we can take the derivative of the components upon which it turns out
that

\begin{equation}
\begin{aligned}\frac{1}{\det\left(g\right)}\partial_{\lambda}\left(\det\left(g\right)\right) & =\mathrm{tr}\left(g^{-1}\partial_{\lambda}g\right)\\
 & =g^{\mu\nu}\partial_{\lambda}g_{\mu\nu}\\
\Rightarrow\partial_{\lambda}\sqrt{g} & =\frac{1}{2}\sqrt{g}g^{\mu\nu}\partial_{\lambda}g_{\mu\nu}\\
\Rightarrow?\Gamma^{\mu}{}_{\lambda\mu}?-?T^{\mu}{}_{\mu\lambda}? & =\frac{1}{2}g^{\mu\nu}\partial_{\lambda}g_{\mu\nu}.
\end{aligned}
\end{equation}
By considering the inverse matrix, we see that these expressions are
also valid with $g^{\mu\nu}\partial_{\lambda}g_{\mu\nu}\rightarrow-g_{\mu\nu}\partial_{\lambda}g^{\mu\nu}$.
The first line above actually applies to any variation; applying it
to the Lie derivative and using its coordinate expression gives us

\begin{align}
\begin{aligned}L_{u}\sqrt{g} & =\sqrt{g}\mathrm{div}\left(u\right)\\
\Rightarrow\mathrm{div}\left(u\right) & =\frac{1}{2}g^{\mu\nu}L_{u}g_{\mu\nu}.
\end{aligned}
\end{align}

If we consider an anti-symmetric tensor $F^{\mu\nu}$ and a symmetric
tensor $G^{\mu\nu}$, it is not hard to see that
\begin{align}
\begin{aligned}\nabla_{\nu}F^{\mu\nu}-?T^{\lambda}{}_{\lambda\nu}?F^{\mu\nu} & =\frac{1}{\sqrt{g}}\partial_{\nu}\left(\sqrt{g}F^{\mu\nu}\right)-\frac{1}{2}?T^{\mu}{}_{\lambda\nu}?F^{\lambda\nu},\\
\nabla_{\nu}G^{\mu\nu}-?T^{\lambda}{}_{\lambda\nu}?G^{\mu\nu} & =\frac{1}{\sqrt{g}}\partial_{\nu}\left(\sqrt{g}G^{\mu\nu}\right)+?\Gamma^{\mu}{}_{\lambda\nu}?G^{\lambda\nu},\\
\nabla_{\nu}?G_{\mu}{}^{\nu}?-?T^{\lambda}{}_{\lambda\nu}??G_{\mu}{}^{\nu}? & =\frac{1}{\sqrt{g}}\partial_{\nu}\left(\sqrt{g}?G_{\mu}{}^{\nu}?\right)-?\Gamma^{\lambda}{}_{\mu\nu}??G_{\lambda}{}^{\nu}?\\
 & =\frac{1}{\sqrt{g}}\partial_{\nu}\left(\sqrt{g}?G_{\mu}{}^{\nu}?\right)-\frac{1}{2}\partial_{\mu}g_{\lambda\nu}G^{\lambda\nu}+?T^{\lambda}{}_{\mu\nu}??G_{\lambda}{}^{\nu}?.
\end{aligned}
\end{align}
The above expressions are more commonly presented with zero torsion,
under which condition we denote the covariant derivative $\overline{\nabla}_{\nu}$;
this then defines the ``divergence'' of the tensor. It can also
be shown (\cite{FrankelGeom} p. 365) that the ``divergence'' of
an exterior $k$-form expressed as an anti-symmetric tensor can be
written in terms of the hodge star as

\begin{align}
\begin{aligned}\overline{\nabla}^{\nu}F_{\nu\mu_{2}\cdots\mu_{k}} & \equiv g^{\nu\mu_{1}}\overline{\nabla}_{\nu}F_{\mu_{1}\cdots\mu_{k}}\\
 & =-\left(\delta F\right)_{\mu_{2}\cdots\mu_{k}}\\
 & =\left(-1\right){}^{n(k+1)+s}\left(*\mathrm{d}\left(*F\right)\right)_{\mu_{2}\cdots\mu_{k}}.
\end{aligned}
\end{align}

\subsection{\label{subsec:Coordinate-and-tensor-divergence-theorems}Coordinate
and tensor divergence theorems}

The expression for the divergence theorem from Section \ref{subsec:The-divergence-and-conserved-quantities}
is
\begin{equation}
\begin{aligned}\int_{V}\mathrm{div}(u)\mathrm{d}V & =\int_{\partial V}i_{u}\mathrm{d}V\\
 & =\int_{\partial V}\left\langle u,\hat{n}\right\rangle \mathrm{d}S,
\end{aligned}
\end{equation}
where $V$ is an $n$-dimensional compact submanifold of $M^{n}$,
$\hat{n}$ is the unit normal vector to $\partial V$, and $\mathrm{d}S\equiv i_{\hat{n}}\mathrm{d}V$
is the induced volume element (``surface element\index{surface element}'')
for $\partial V$. If we choose an orthonormal frame with $e_{1}=\hat{n}$
on $\partial V$, the divergence theorem can be written
\begin{equation}
\begin{aligned}\int_{V}\mathrm{div}(u)\mathrm{d}V & =\int_{\partial V}u^{1}\mathrm{d}S,\end{aligned}
\end{equation}
and if we can choose coordinates with $x^{1}$ constant on $\partial V$
and normal to it, the divergence theorem can be written
\begin{equation}
\begin{aligned}\int_{V}\partial_{\lambda}\left(\sqrt{g}u^{\lambda}\right)\mathrm{d}^{n}x & =\int_{\partial V}\sqrt{g}\mathrm{d}x^{1}\left(u\right)\mathrm{d}^{n-1}x\\
 & =\int_{\partial V}u^{1}\sqrt{g}\mathrm{d}^{n-1}x,
\end{aligned}
\end{equation}
where $\mathrm{d}^{n}x\equiv\mathrm{d}x^{1}\wedge\cdots\wedge\mathrm{d}x^{n}$
and $\mathrm{d}^{n-1}x\equiv\mathrm{d}x^{2}\wedge\cdots\wedge\mathrm{d}x^{n}$. 

Since the ``divergence'' of a tensor $T$ with order greater than
1 is tensor-valued, and the parallel transport of tensors is path-dependent,
we cannot in general integrate to get a divergence theorem for tensors.
In the case of a flat metric and zero torsion however, we can choose
coordinates whose coordinate frame is orthonormal, so that the frame
is its own parallel transport, i.e. $\nabla_{v}\left(\beta^{\mu}\right)=0$.
For e.g. a tensor $T^{ab}$, we can then define a coordinate-dependent
vector $J^{\mu}$ for each index $\mu$
\begin{equation}
\begin{aligned}J^{\mu} & \equiv T\left(\beta^{\mu},\quad\right)\\
\Rightarrow\left(J^{\mu}\right)^{b} & =T^{\mu b}\\
\Rightarrow\overline{\nabla}_{v}J^{\mu} & \overset{\cancel{R}}{=}\beta^{\mu}\overline{\nabla}_{v}T\\
\Rightarrow\overline{\nabla}_{b}\left(J^{\mu}\right)^{b} & \overset{\cancel{R}}{=}\overline{\nabla}_{b}T^{\mu b}\\
\Rightarrow\int_{V}\overline{\nabla}_{b}T^{\mu b}\mathrm{d}V & \overset{\cancel{R}}{=}\int_{V}\overline{\nabla}_{b}\left(J^{\mu}\right)^{b}\mathrm{d}V\\
 & =\int_{\partial V}T^{\mu}{}_{b}\hat{n}^{b}\mathrm{d}S.
\end{aligned}
\end{equation}
For arbitrary coordinates, the components of the coordinate frame
are by definition constant, i.e. $\partial_{v}\left(\mathrm{d}x^{\mu}\right)=0$;
we can therefore write 
\begin{equation}
\begin{aligned}\sqrt{g}J^{\mu} & \equiv\sqrt{g}T\left(\mathrm{d}x^{\mu},\quad\right)\\
\Rightarrow\partial_{\nu}\left(\sqrt{g}J^{\mu}\right)^{\nu} & =\partial_{\nu}\left(\sqrt{g}T^{\mu\nu}\right)\\
\Rightarrow\int_{V}\partial_{\nu}\left(\sqrt{g}T^{\mu\nu}\right)\mathrm{d}^{n}x & =\int_{V}\partial_{\nu}\left(\sqrt{g}J^{\mu}\right)^{\nu}\mathrm{d}^{n}x\\
 & =\int_{V}\nabla_{b}\left(J^{\mu}\right)^{b}\mathrm{d}V\\
 & =\int_{\partial V}T^{\mu}{}_{b}\hat{n}^{b}\mathrm{d}S.
\end{aligned}
\end{equation}
This relation remains true in the presence of both curvature and torsion,
however it is important to note that $\partial_{\nu}\left(\sqrt{g}T^{\mu\nu}\right)$
is not a ``divergence'' and $T^{\mu b}=\left(J^{\mu}\right)^{b}$
is coordinate-dependent. In the special case of an anti-symmetric
tensor under zero torsion, we can write 
\begin{equation}
\begin{aligned}\int_{V}\overline{\nabla}_{\nu}F^{\mu\nu}\mathrm{d}V & =\int_{V}\partial_{\nu}\left(\sqrt{g}F^{\mu\nu}\right)\mathrm{d}^{n}x\\
 & =\int_{\partial V}F^{\mu}{}_{b}\hat{n}^{b}\mathrm{d}S.
\end{aligned}
\end{equation}

\subsection{Current forms and densities}

In Section \ref{subsec:The-divergence-and-conserved-quantities} we
defined the current vector (AKA flux) $j\equiv\rho u$, where $\rho$
is the density of the physical quantity $Q$ and $u$ is a velocity
field, and then combined them into the four-current $J\equiv(\rho,j^{\mu})$.
There are a number quantities that can be defined around this concept:

{\footnotesize{}}
\begin{table}[H]
{\footnotesize{}}%
\begin{tabular*}{1\columnwidth}{@{\extracolsep{\fill}}>{\raggedright}m{0.2\columnwidth}l>{\raggedright}m{0.33\columnwidth}}
\toprule 
{\footnotesize{}Quantity} & {\footnotesize{}Definition} & {\footnotesize{}Meaning}\tabularnewline\addlinespace[0.01in]
\midrule
\midrule 
{\footnotesize{}Current vector\index{Current vector}} & {\footnotesize{}$j\equiv\rho u$} & {\footnotesize{}The vector whose length is the amount of $Q$ per
unit time crossing a unit area perpendicular to $j$}\tabularnewline
\midrule 
{\footnotesize{}Current form\index{Current form}} & {\footnotesize{}$\begin{aligned}\zeta & \equiv i_{j}\mathrm{d}V\\
 & =\left\langle j,\hat{n}\right\rangle \mathrm{d}S
\end{aligned}
$} & {\footnotesize{}The $(n-1)$-form which gives the amount of $Q$ per
unit time crossing the area defined by the argument vectors}\tabularnewline
\midrule 
{\footnotesize{}Current density\index{Current density}} & {\footnotesize{}$\begin{aligned}\mathfrak{j} & \equiv\sqrt{g}\,j\\
\Rightarrow\zeta & =\mathfrak{j}^{1}\mathrm{d}^{2}x
\end{aligned}
$} & {\footnotesize{}The vector whose coordinate length is the amount of
$Q$ per unit time crossing a unit coordinate area perpendicular to
$j$}\tabularnewline
\midrule 
{\footnotesize{}Current\index{Current}} & {\footnotesize{}$\begin{aligned}I_{S} & \equiv\int_{S}\zeta\\
 & =\int_{S}\left\langle j,\hat{n}\right\rangle \mathrm{d}S\\
 & =\int_{S}\mathfrak{j}^{1}\mathrm{d}^{2}x
\end{aligned}
$} & {\footnotesize{}The amount of $Q$ per unit time crossing $S$}\tabularnewline
\midrule 
{\footnotesize{}Four-current \index{Four-current}} & {\footnotesize{}$J\equiv(\rho,j^{\mu})$} & {\footnotesize{}Current vector on the spacetime manifold}\tabularnewline
\bottomrule
\end{tabular*}{\footnotesize\par}

{\footnotesize{}\caption{Quantities related to current.}
}{\footnotesize\par}

{\footnotesize{}Notes: $\rho$ is the density of the physical quantity
$Q$, $u$ is a velocity field, $\hat{n}$ is the unit normal to a
surface $S$, and $\mathrm{d}^{3}x$ are coordinates with $x^{1}$
constant on $S$ and normal to it. The four-current can be generalized
to other Lorentzian manifolds, and can also be turned into a form
$\xi\equiv i_{J}\mathrm{d}V$ or a density $\mathfrak{J}\equiv\sqrt{g}\,J$.}{\footnotesize\par}
\end{table}
{\footnotesize\par}

\noindent %
\begin{framed}%
\noindent $\triangle$ Note that the terms flux and current (as well
as flux density and current density) are not used consistently in
the literature.\end{framed}

\subsection{Tensor densities}

The current density $\mathfrak{j}$ defined in the previous section
is an example of a \textbf{tensor density}\index{tensor density},
which in general takes the form 
\begin{equation}
\mathfrak{T}\equiv\sqrt{g}^{W}T,
\end{equation}
where $T$ is a tensor and $W$ is called the \textbf{weight}\index{tensor density weight}.
Note that tensor densities are not coordinate-independent quantities,
and $\sqrt{g}$ itself can thus be called a scalar density. 

From the expressions in the preceding sections we get

\begin{equation}
\begin{aligned}\partial_{\lambda}\left(\mathfrak{T}\right) & =\sqrt{g}^{W}\partial_{\lambda}T+W\left(?\Gamma^{\mu}{}_{\lambda\mu}?-?T^{\mu}{}_{\mu\lambda}?\right)\mathfrak{T}\\
 & =\sqrt{g}^{W}\partial_{\lambda}T+\frac{W}{2}g^{\mu\nu}\partial_{\lambda}g_{\mu\nu}\mathfrak{T},\\
L_{u}\left(\mathfrak{T}\right) & =\sqrt{g}^{W}L_{u}T+W\mathrm{div}\left(u\right)\mathfrak{T}\\
 & =\sqrt{g}^{W}L_{u}T+\frac{W}{2}g^{\mu\nu}L_{u}g_{\mu\nu}\mathfrak{T},\\
\nabla_{\lambda}\left(\mathfrak{T}\right) & =\sqrt{g}^{W}\nabla_{\lambda}T,
\end{aligned}
\end{equation}
where the last is due to the covariant derivative of the metric vanishing.
In particular, this means that for zero torsion the divergence of
a vector density is

\begin{equation}
\begin{aligned}\overline{\nabla}_{\lambda}\mathfrak{J}^{\lambda} & =\sqrt{g}\overline{\nabla}_{\lambda}J^{\lambda}\\
 & =\sqrt{g}\mathrm{div}\left(J\right)\\
 & =\partial_{\lambda}\mathfrak{J}^{\lambda}.
\end{aligned}
\end{equation}
\begin{framed}%
\noindent $\triangle$ A potential source of confusion is the use
of the word ``density'' to indicate both an amount per unit area
or volume and the presence of the coordinate-dependent factor $\sqrt{g}$,
which as in the current density typically reflects the volume in question
being a unit coordinate volume instead of metric volume.\end{framed}

\subsection{Conserved currents and quantities}

In Section \ref{subsec:The-divergence-and-conserved-quantities} we
saw that a Lorentzian conserved current $\mathrm{div}(J)=0$ does
not imply a conserved quantity in the presence of curvature. If we
are willing to consider coordinate-dependent currents, at any given
point we can choose Riemann normal coordinates, which allows us to
recover a conserved quantity at that point in those coordinates.

In the integral form, we may also identify a coordinate-dependent
conserved quantity for a Lorentzian conserved current by integrating
over a space-like volume $S$ with coordinates such that $t\equiv x^{0}$
is constant on $S$ and normal to it, while $x^{1}$ is constant on
$\partial S$ and normal to it:
\begin{equation}
\begin{aligned}0 & =\int_{S}\sqrt{g}\mathrm{div}(J)\mathrm{d}^{3}x\\
 & =\int_{S}\partial_{\mu}\mathfrak{J}^{\mu}\mathrm{d}^{3}x\\
 & =\partial_{t}\left(\int_{S}\mathfrak{J}^{t}\mathrm{d}^{3}x\right)+\int_{S}\partial_{i}\mathfrak{J}^{i}\mathrm{d}^{3}x\\
 & =\partial_{t}\left(\int_{S}\mathfrak{J}^{t}\mathrm{d}^{3}x\right)+\int_{\partial S}\mathfrak{J}^{1}\mathrm{d}^{2}x
\end{aligned}
\end{equation}
Note that the coordinate-dependent factor $\sqrt{g}$ in $\mathfrak{J}=\sqrt{g}J$
cannot be absorbed into either $\mathrm{d}^{3}x$ or $\mathrm{d}^{2}x$
to yield a coordinate-independent quantity. More specifically, if
$\mathfrak{J}$ is either also normal to $S$ or vanishes on $\partial S$,
we have $\partial_{t}\left(\int_{S}\mathfrak{J}^{t}\mathrm{d}^{3}x\right)=0$.
This also holds if $S$ is infinite and $\mathfrak{J}$ vanishes rapidly
enough at spatial infinity. %
\begin{framed}%
\noindent $\triangle$ A conserved quantity as we have defined it
is a quantity whose amount in a volume of space changes in time by
the net amount that crosses the volume boundary. This concept is not
valid when $\mathrm{div}(J)=0$ in the presence of spacetime curvature,
but it is important to remember that this still means that $\int_{\partial V}\left\langle J,\hat{n}\right\rangle \mathrm{d}S=0$,
so that the same amount of the quantity enters and exits any finite
volume of spacetime; it is in this sense that the current is ``conserved.''\end{framed}

With regard to tensors, we can conclude from the divergence theorem
variants in Section \ref{subsec:Coordinate-and-tensor-divergence-theorems}
that in the case of an orthonormal coordinate frame under a flat metric
and the Levi-Civita covariant derivative, we have a coordinate-dependent
conserved quantity for each component of a tensor, corresponding to
a coordinate-dependent conserved current:
\begin{equation}
\begin{aligned}\overline{\nabla}_{\nu}T^{\mu\nu} & =0\\
\Rightarrow\partial_{0}T^{\mu0} & \overset{\cancel{R}}{=}-\overline{\nabla}_{j}T^{\mu j},\\
\int_{\partial V}T^{\mu}{}_{b}\hat{n}^{b}\mathrm{d}S & \overset{\cancel{R}}{=}0
\end{aligned}
\end{equation}
In the special case of an anti-symmetric tensor and the Levi-Civita
covariant derivative we also have a divergence theorem, and therefore
a coordinate-dependent conserved current for each component:
\begin{equation}
\begin{aligned}\overline{\nabla}_{\nu}F^{\mu\nu} & =0\\
\Rightarrow\int_{\partial V}F^{\mu}{}_{b}\hat{n}^{b}\mathrm{d}S & =0
\end{aligned}
\end{equation}

\end{document}